\documentclass[namedreferences]{SolarPhysics}
\usepackage[optionalrh,solaenum]{spr-sola-addons} 
\usepackage{solaheader} 
\usepackage{graphicx}                    
\usepackage{color}                       
\usepackage{url}                         

\begin{document}

\begin{article}

\begin{opening}

\title{Nonlinear force-free and potential field models of active-region and global coronal fields during the Whole Heliospheric Interval}

%
\author{G.J.D.~\surname{Petrie}$^{1}$\sep
        A.~\surname{Canou}$^{2}$\sep
        T.~\surname{Amari}$^{2}$      
       }

%

%
  \institute{$^{1}$ National Solar Observatory, Tucson, AZ 85719, USA
                     email: \url{gpetrie@noao.edu}\\ 
             $^{2}$ CNRS, Centre de Physique Th\'eorique de l'Ecole Polytechnique, F-91128 Palaiseau Cedex, France
                      email: \url{Aurelien.Canou@cpht.polytechnique.fr} email: \url{Tahar.Amari@cpht.polytechnique.fr} \\
             }

\begin{abstract}

Between 2008/3/24 and 2008/4/2, the three active regions NOAA ARs 10987, 10988 and 10989 were observed daily by the Synoptic Optical Long-term Investigations of the Sun (SOLIS) Vector Spectro-Magnetograph (VSM) while they traversed the solar disk.  We use these measurements and the nonlinear force-free magnetic field code XTRAPOL to reconstruct the coronal magnetic field for each active region and compare model field lines with images from the Solar Terrestrial RElations Observatory (STEREO)  and Hinode X-ray Telescope (XRT) telescopes.  Synoptic maps made from continuous, round-the-clock Global Oscillations Network Group (GONG) magnetograms provide information on the global photospheric field and potential-field source-surface models based on these maps describe the global coronal field during the Whole Heliospheric Interval (WHI) and its neighboring rotations.  Features of the modeled global field, such as the coronal holes and streamer belt locations, are discussed in comparison with extreme ultra-violet and coronagraph observations from STEREO.  The global field is found to be far from a minimum, dipolar state.  From the nonlinear models we compute physical quantities for the active regions such as the photospheric magnetic and electric current fluxes, the free magnetic energy and the relative helicity for each region each day where observations permit.  The interconnectivity of the three regions is addressed in the context of the potential-field source-surface model.  Using local and global quantities derived from the models, we briefly discuss the different observed activity levels of the regions.

\end{abstract}

%

\end{opening}


%
\section{Introduction}
\label{s:introduction} 

A large part of the low solar corona is dominated by the magnetic field.  This field is created inside the Sun and then emerges into the atmosphere.  The magnetic field plays a defining role in most coronal phenomena, quasi-static structures from cool, dense prominences to hot loops, and dynamical events such as flares and coronal mass ejections (CMEs).  The most spectacular events with the greatest impact on us
originate from the lower corona.  If we are to understand the various phenomena that occurred in the corona during WHI it is therefore crucial that we determine the 3D coronal magnetic field as well as possible.  Unfortunately, the coronal field is notoriously difficult to measure \cite{Tomczyketal2007,LiuLin2008}, significantly more difficult than in the much denser photosphere.  
It has therefore become common practice to reconstruct the coronal field from boundary conditions derived from photospheric field measurements \cite{Amarietal1997,Schrijveretal2006,Wiegelmann2008}.  Our best
estimate of coronal magnetic structure comes from such modeling efforts.

On the other hand, it is clear that there remain significant problems with the accuracy of the models.  The governing magnetohydrodynamics (MHD) equations are complex and even their equilibrium solutions are difficult to constrain.  Because the magnetic field so dominates the corona, most coronal fields can reasonably be assumed to be approximately force-free.   Models of the global lower coronal field usually do not include electric currents in the lower coronal domain either.  Such models, referred to as potential-field source-surface (PFSS) models, incorporate on their upper boundary a current "source surface" where the magnetic field is forced to be radial, mimicking the effect of the solar wind \cite{AltschulerNewkirk1969,Schattenetal1969,Hoeksema1984,WangSheeley1992}.  The global field produced by such models usually closely approximates much more expensive MHD models in practice \cite{Neugebaueretal1998,Rileyetal2006} and they have been successfully applied in studying basic properties of the global coronal field such as coronal holes, streamer belt locations and dipole axis tilts \cite{Zhaoetal2005,SchrijverLiu2008}.

Although \inlinecite{Schrijveretal2005} found that many quiescent active regions resemble potential field models, recent attempts to measure the accuracy of coronal magnetic field models extrapolated from photospheric field measurements have produced discouraging conclusions.  Since the launch of NASA's twin Solar Terrestrial Relations Observatory (STEREO) satellites it has been possible to estimate from simultaneous pairs of extreme ultra-violet (EUV) images from distinct vantage points the path through 3D space of any loop structure that can be correctly identified in both images.  \inlinecite{Sandmanetal2009} used this technique to test the accuracy of potential-field models of three active regions from line-of-sight MDI field measurements.  They measured the accuracy by calculating the misalignment angle between each observed loop and the model at different positions along the loop and noting the median value.  For the three regions studied they found averages and standard deviations $25^{\circ}\pm 8^{\circ}$, $19^{\circ}\pm 8^{\circ}$ and $36^{\circ}\pm 13^{\circ}$.  We remark that these observational tests do not involve direct comparison between observed and modeled coronal fields.  It is not known how closely the loop structures observed in EUV adhere to magnetic field trajectories.

As for NLFFF modeling, the picture is still more complicated.  There are several very different solution methods available and, unlike the potential-field problem, the nonlinear force-free problem cannot be immediately cast in a clear, mathematically well-posed way.  The force-free equations are nonlinear and of mixed mathematical (elliptic/hyperbolic) type.  Only two pieces of information, e.g., the normal components of the field $\bf B$ and the electric current $\bf J$, can be imposed on the boundary to give a mathematically well-posed boundary value problem, whereas observations of all three magnetic field vector components are available.  Different solution methods apply this information in different ways, often producing conflicting results.  To add to the complication, Grad-Rubin methods \cite{GradRubin1958} such as the one used here can incorporate data from either the positive or the negative magnetic polarity but not both in a single model.  Models from the two polarities can be very dissimilar, and often are in practice.  On the other hand, sometimes it is clear which polarity one must apply as boundary data \cite{CanouAmari2010}.  New approaches to take into account both polarities have been recently proposed by \inlinecite{WheatlandRegnier2009} and \inlinecite{AmariAly2010}.


\cite{Schrijveretal2008} and \cite{DeRosaetal2009} compared several NLFFF codes using common boundary data sets comprising Hinode vector magnetograms embedded within line-of-sight MDI magnetograms. They found major disparities between the various models, in their magnetic field and electric current structures.  \cite{Schrijveretal2008} find that the model fields differed significantly in geometry, energy content, and force-freeness.  \cite{DeRosaetal2009} concluded that while models can coincide well with observed features it remains difficult to determine whether a significant proportion of a modeled field accurately reproduces the coronal field over the entire coronal volume above an active region. \cite{CanouAmari2010} have re-analyzed the data set studied by DeRosa et al.~(2009), modeling the field of AR 10953 exploiting the fact that their Grad-Rubin method (also used here) takes its information on the photospheric electric current (via the force-free function $\alpha$) from only one of the magnetic polarities. They chose to integrate from the polarity most fully observed by Hinode, including a intense sunspot electric current flux, thereby avoiding the lack of information on the electric current in the MDI portion of the boundary data set. Finding evidence of a bald patch in the vector magnetogram they successfully modeled a magnetic flux rope that corresponded well with a Hinode XRT sigmoid and a filament seen in SMART H$\alpha$ data.

A further complication is that photospheric vector field measurements of a given solar phenomenon show significant disagreement from instrument to instrument and even between different inversions of a single Stokes data set\footnote{ftp://ftp.cora.nwra.com/pub/leka/papers/magcomp.pdf}.  In view of this, it may be optimistic to expect a coronal model extrapolated from photospheric data to agree with coronal observations.   \inlinecite{Wiegelmannetal2010} show encouraging evidence that
nonlinear force-free field models based on good
photospheric vector field measurements can give reasonable
estimates of the structure of the coronal field.

Nonlinear
force-free models give estimates of the free magnetic
energy and magnetic helicity of active region structures.
 Such modeling techniques can capture twisted coronal
magnetic structures, providing evidence that twisted flux
ropes exist in equilibrium in the solar corona (e.g.,
\inlinecite{Canouetal2009} and sometimes detailed knowledge of the
different twisted structures in an active region can
clarify discrepancies between the handedness of twisted
flux tubes ejected from the Sun (magnetic clouds) and the
handedness of twisted flux low in the corona \cite{RegnierAmari2004}.  Usually
electric currents flow in more than one direction in an
active region. Nonlinear
force-free fields can capture this current structure and
allow a more accurate and detailed estimate of active
region helicities.

In this paper we combine the two different types of coronal model discussed above, PFSS and NLFFF, to describe the global coronal field structure and the active region field properties during the WHI.  The PFSS models will be based on Global Oscillations Network Group (GONG) synoptic magnetograms and the NLFFF models extrapolated from Synoptic Optical Long-term Investigations of the Sun (SOLIS) vector magnetograms.  From the PFSS models for the WHI and its neighboring rotations we can study the coronal hole distributions and streamer belt locations, comparing them to synoptic EUV and coronagraph data from NASA's STEREO spacecraft.  Meanwhile NLFFF models of the active regions can provide information on the magnetic field including its basic structure, free magnetic energy and relative magnetic helicity.  We will compare the modeled field structures to Hinode/XRT and STEREO/SECCHI/EUVI images and interpret the PFSS and NLFFF models together to characterize the active regions.

The paper is organized as follows.  We will describe the modeling techniques in Section~\ref{s:pfssnlfff}.  Then, we will present PFSS models for Carrington rotations 2067-9 and associated observations of the global coronal structure in Section~\ref{s:globalstructure}, before presenting NLFFF models and observations of the active region structure in Section~\ref{s:activeregions} including a brief study of the interconnectedness of the regions and their temporal evolution.  We will conclude in Section~\ref{s:conclusion}.

\section{The PFSS and NLFFF models}
\label{s:pfssnlfff}

In the low corona, typical Alfv\`en speeds are so fast (1000~km/s) compared to the observed evolution time of structures that any persistent structure is static to a good approximation.  The magnetic field dominates the dynamics such that the Lorentz force ${\bf J}\times {\bf B}$, unopposed by significant plasma forces, is negligible such that almost all Maxwell stresses are contained within the field and the field and electric current are approximately parallel,

\begin{equation}
{\bf J} = \alpha ({\bf x}) {\bf B}.
\end{equation}

\noindent Because the quantity $\alpha $ is inversely proportional to the length scale, the current-free approximation $\alpha\approx 0$ is useful for studying large-scale structure.

Under the simplifying assumption that there are no significant electric currents in the corona, the coronal magnetic field may be represented by ${\bf B}=-{\bf\nabla}\psi(r,\theta ,\phi )$ in spherical coordinates where $\psi(r,\theta ,\phi )$ is a spherical harmonic series in $\theta$ and $\phi$.  The spherical harmonic coefficients are chosen such that the radial field component at the lower boundary of the model matches the photospheric magnetogram while the $r$-dependence of $\psi$ guarantees that the field is radial at a ``source-surface'', the outer surface of the model \cite{AltschulerNewkirk1969,Hoeksema1984}.  While line-of-sight boundary data can also be used, radial boundary conditions derived from line-of-sight data generally give better models because the photospheric field is approximately radial and because polar fields are poorly represented by line-of-sight measurements \cite{WangSheeley1992,PetriePatrikeeva2009}.  For a given magnetogram this method in theory provides a unique solution that is guaranteed to exist.  In practice the solution depends not only on the magnetogram, but also on the location and shape of the source surface and the number of spherical harmonics used.  For the PFSS models presented in this paper we use 36 spherical harmonics and we apply a spherical source surface of radius 2.5 solar radii.  The GONG synoptic maps used have $360\times 180$ pixels in longitude-sine(latitude) coordinates.

A force-free field obeys the equations

\begin{eqnarray}
{\bf\nabla}\times{\bf B} & = & \alpha {\bf B}, \label{ampere} \\
{\bf\nabla}\cdot {\bf B} & = & 0. \label{divergencefree}
\end{eqnarray}

The function $\alpha ({\bf x})$ must satisfy

\begin{equation}
{\bf B}\cdot{\bf\nabla}\alpha = 0. \label{propalpha}
\end{equation}
  
  Equations~(\ref{ampere}, \ref{propalpha}) have mixed elliptic-hyperbolic structure, consisting of an elliptic problem for $\bf B$ with fixed $\alpha$ and a hyperbolic problem for $\alpha$ with fixed $\bf B$.  The nonlinear force-free field calculation is of Grad-Rubin type, i.e. $\bf B$ and $\alpha$ are calculated iteratively with the elliptic and hyperbolic parts of the problem solved successively at each step.  To solve the elliptic problem one needs to impose the vertical field component $B_z$ on the boundary, whereas solving the hyperbolic problem requires information about $\alpha$ on either the part of the boundary where $B_z > 0$ or the part where $B_z < 0$.  Although all three components of the photospheric field are measured, only two pieces of information, e.g., the vertical components of the magnetic field and the electric current, may be imposed on the boundary to obtain a mathematically well-posed boundary-value problem for the mixed elliptic-hyperbolic NLFFF equations.  For this type of well-posed formulation, rigorous existence and partial uniqueness theorems have been proved \cite{Bineau1972,BoulAmari2000}.  In addition, the finite-difference XTRAPOL code used here works with a vector potential $A$ associated with the magnetic field, such that ${\bf B}={\bf\nabla}\times{\bf A}$, and with a staggered mesh such that the approximate solution is a member of the kernel of the ${\bf\nabla}\cdot$ so that the model field is constrained to be divergence-free to machine precision \cite{Amarietal1999}.  More recently, the code was extended to accommodate solutions with non-zero $\bf B$ and $\alpha$ on all six sides of the computational box \cite{Amarietal2006}.  In this paper we exploit all of these properties of the code.   All of the NLFF calculations in this paper were performed on $120\times 120\times 80$ grid points using a non-uniform grid where grid points are accumulated in region of strong current and strong magnetic field. In comparison a typical resolution of the SOLIS AR vector magnetograms is $300\times 300$ grid points.  For each magnetogram we calculate a model from each polarity.  If there is significant disagreement between them there is usually a simple explanation, as we discuss in later sections.  The NLFF calculation requires a potential field solution.  This is calculated independently of the global PFSS model discussed above, in the same Cartesian domain as the NLFFF model, using the SOLIS measurements for the vertical field component as boundary conditions.  The instrumental noise level of the VSM is a few G for the longitudinal field and about 40~G in the transverse direction (Jack harvey, private communication).  The results of the Stokes inversion calculation are not reliable for fields weaker than about 50~G.  In practice we include only fields stronger than about 50~G in our boundary data.  Because of the enormous differences in spatial resolution between the GONG synoptic map and the SOLIS magnetograms for the active regions, the question of consistency between these data sets and the associated coronal field models is not a trivial issue.  We will see as the paper progresses that, while the gross features of the active regions appear consistently in both data sets, the distinctive characteristics of the three regions are much more obvious in the vector data.

\section{The global structure}
\label{s:globalstructure}

\begin{figure} 
\begin{center}
\includegraphics[width=0.8\textwidth]{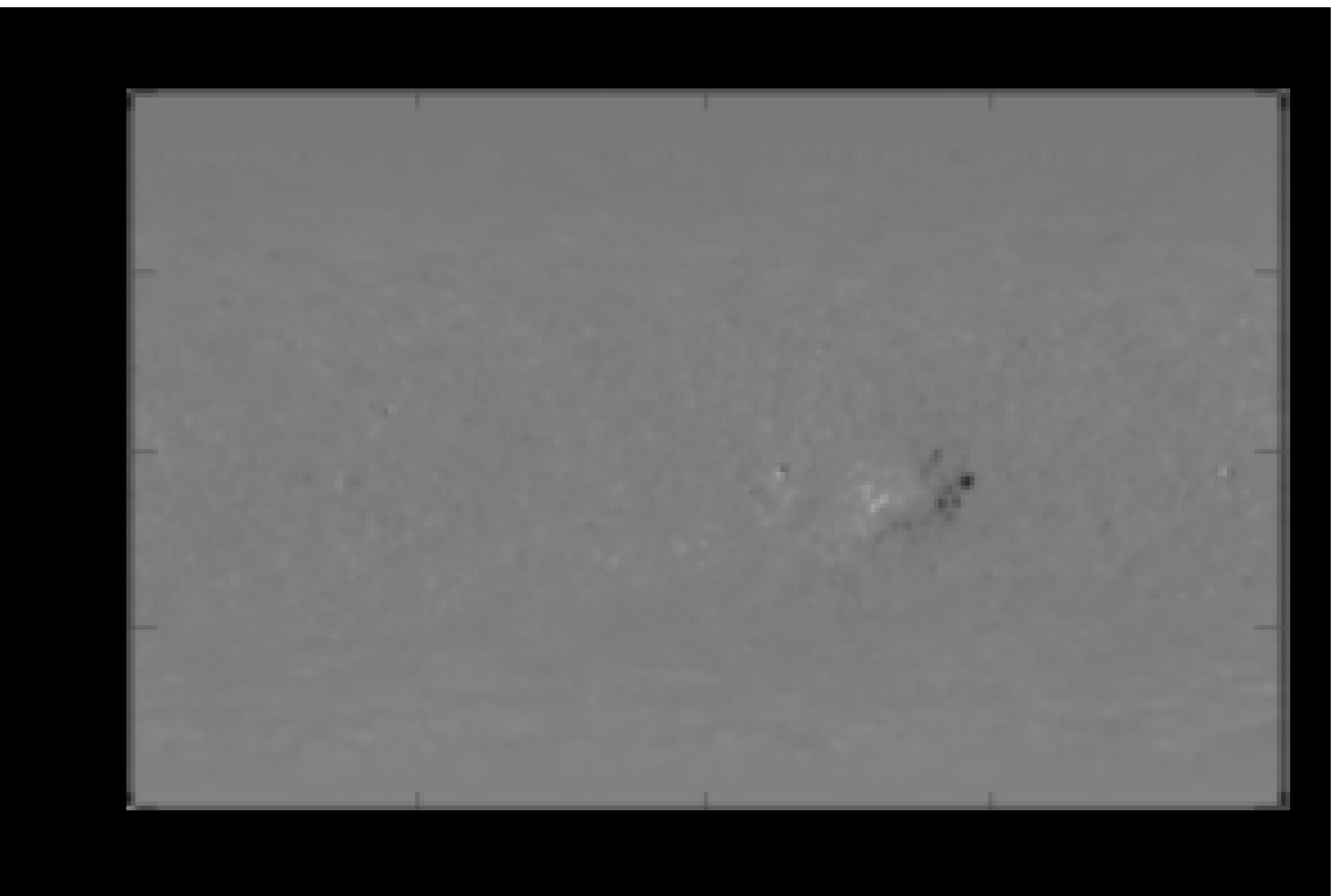}
\includegraphics[width=0.8\textwidth]{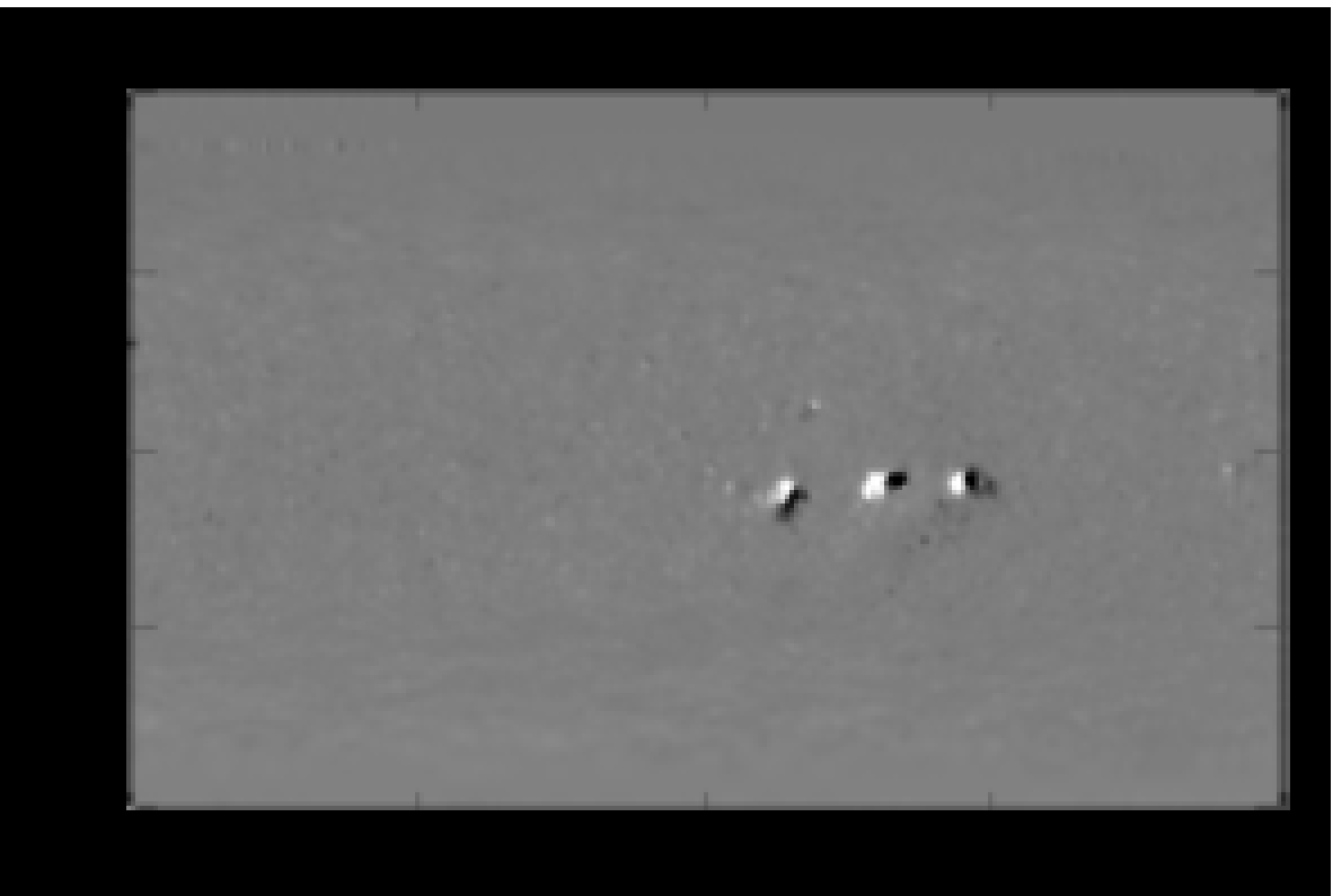}
\includegraphics[width=0.8\textwidth]{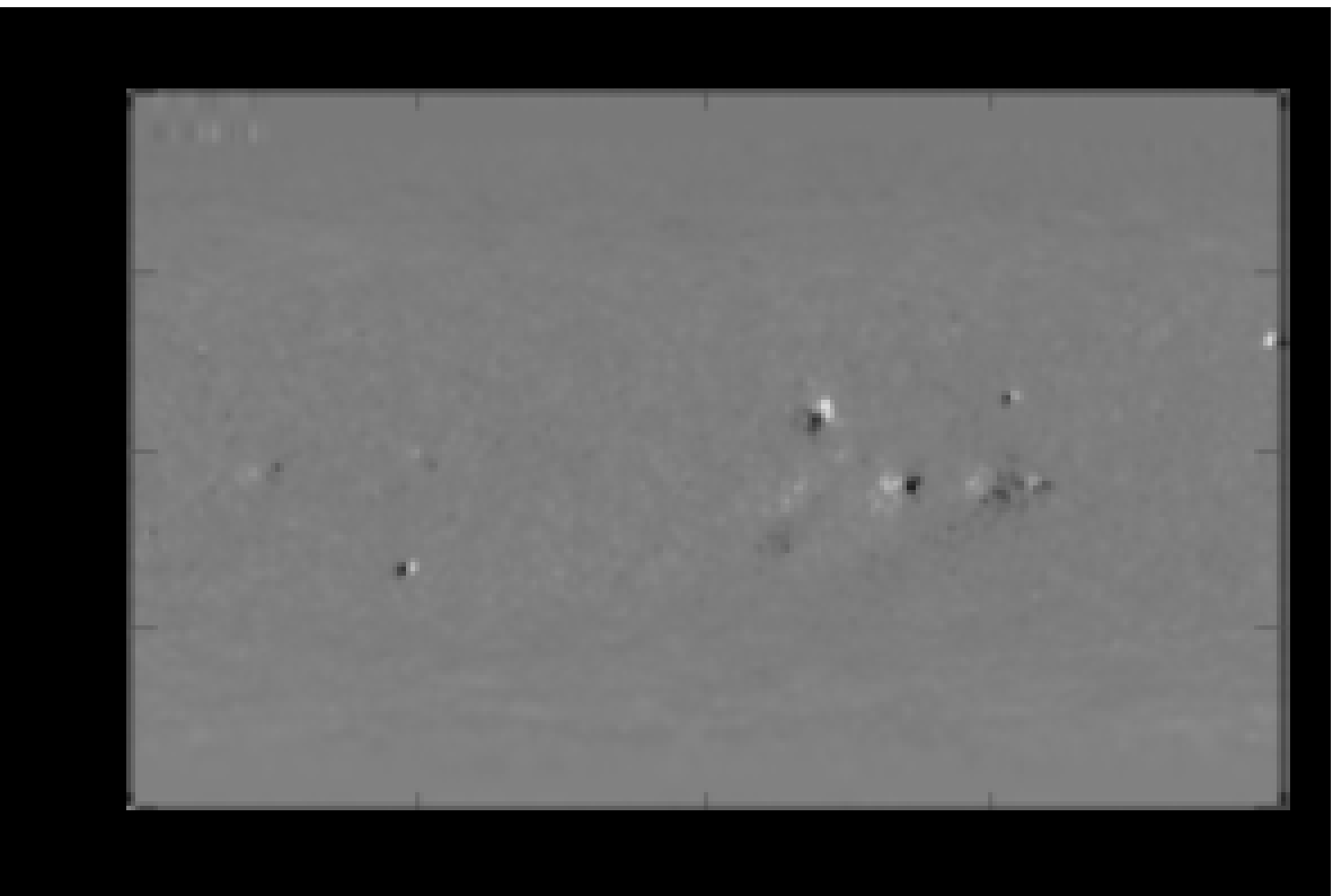}
\end{center}
\caption{GONG synoptic magnetograms of the radial magnetic field component for Carrington rotations 2067 (top), 2068 (WHI, middle) and 2069 (bottom).}
\label{fig:synmags}
\end{figure}

\begin{figure} 
\begin{center}
\includegraphics[width=0.8\textwidth]{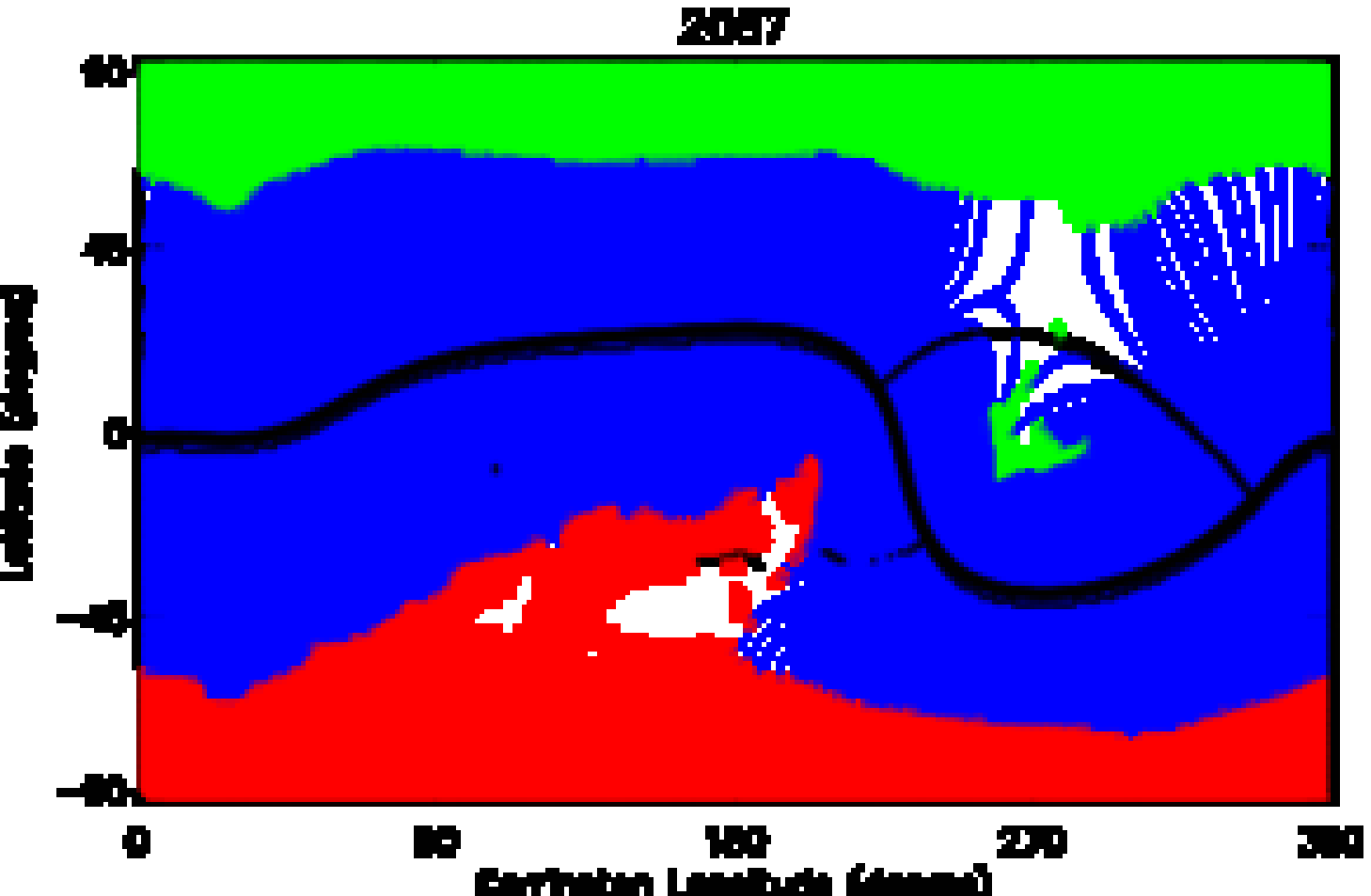}
\includegraphics[width=0.8\textwidth]{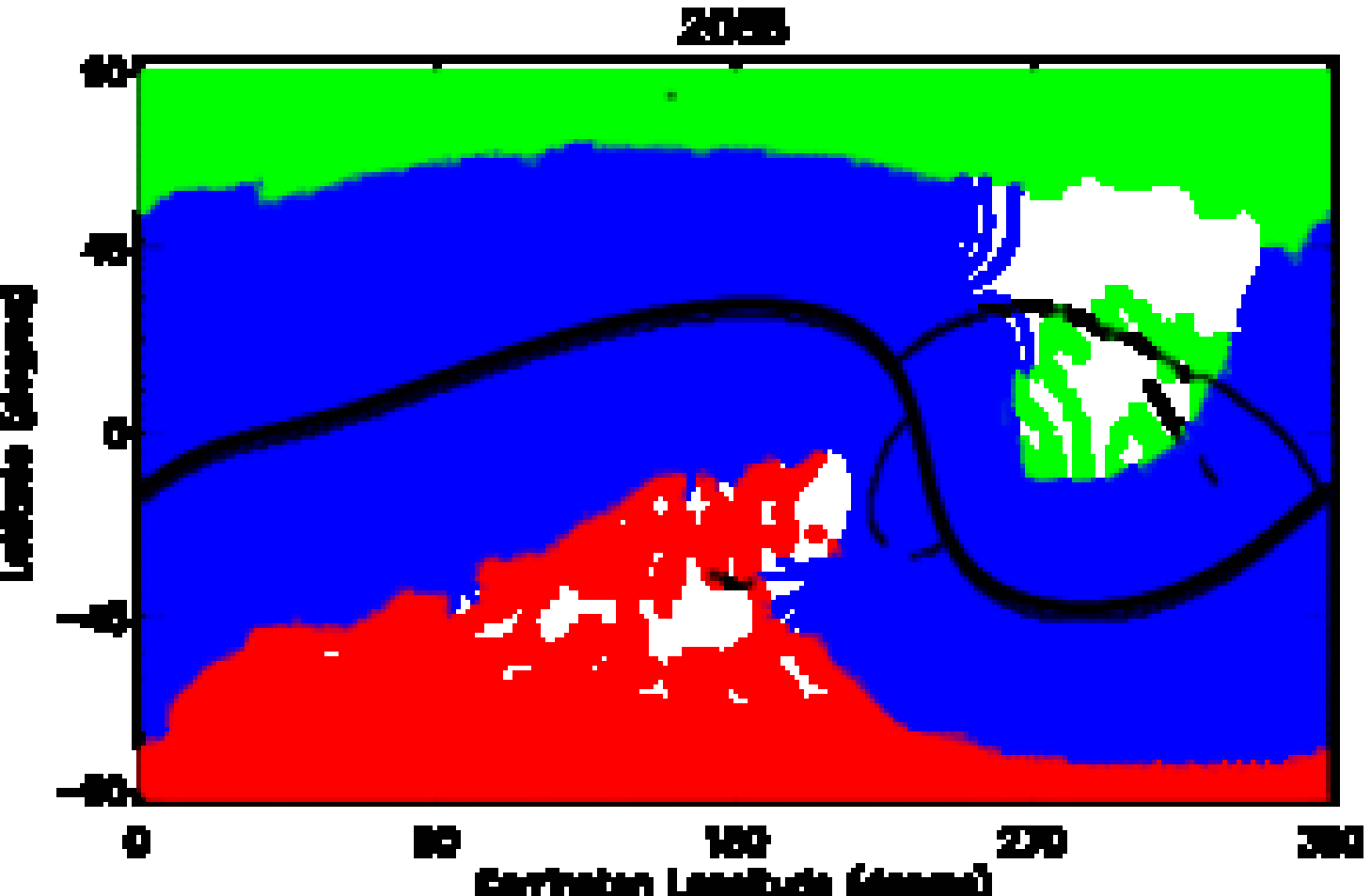}
\includegraphics[width=0.8\textwidth]{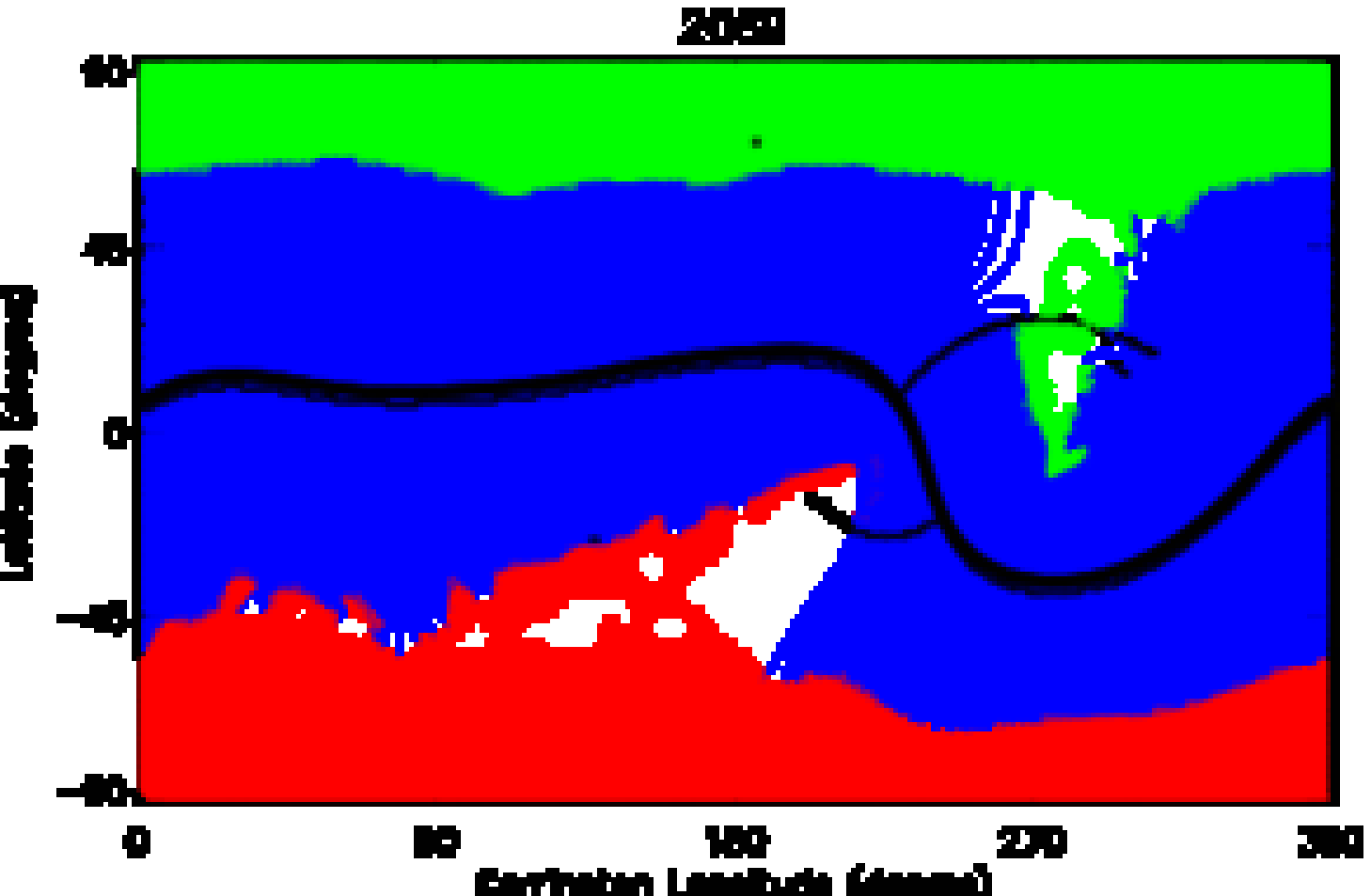}
\end{center}
\caption{PFSS models extrapolated from the GONG synoptic magnetograms of Figure~\ref{fig:synmags} for Carrington rotations 2067 (top), 2068 (WHI, middle) and 2069 (bottom).  Positive and negative coronal holes are colored red and green.  The streamer-belt neutral lines are represented by thick black lines and pseudo-streamer locations by thin black lines.  Streamer-belt fields are plotted in blue.}
\label{fig:synmodels}
\end{figure}

\begin{figure} 
\begin{center}
\includegraphics[width=0.49\textwidth]{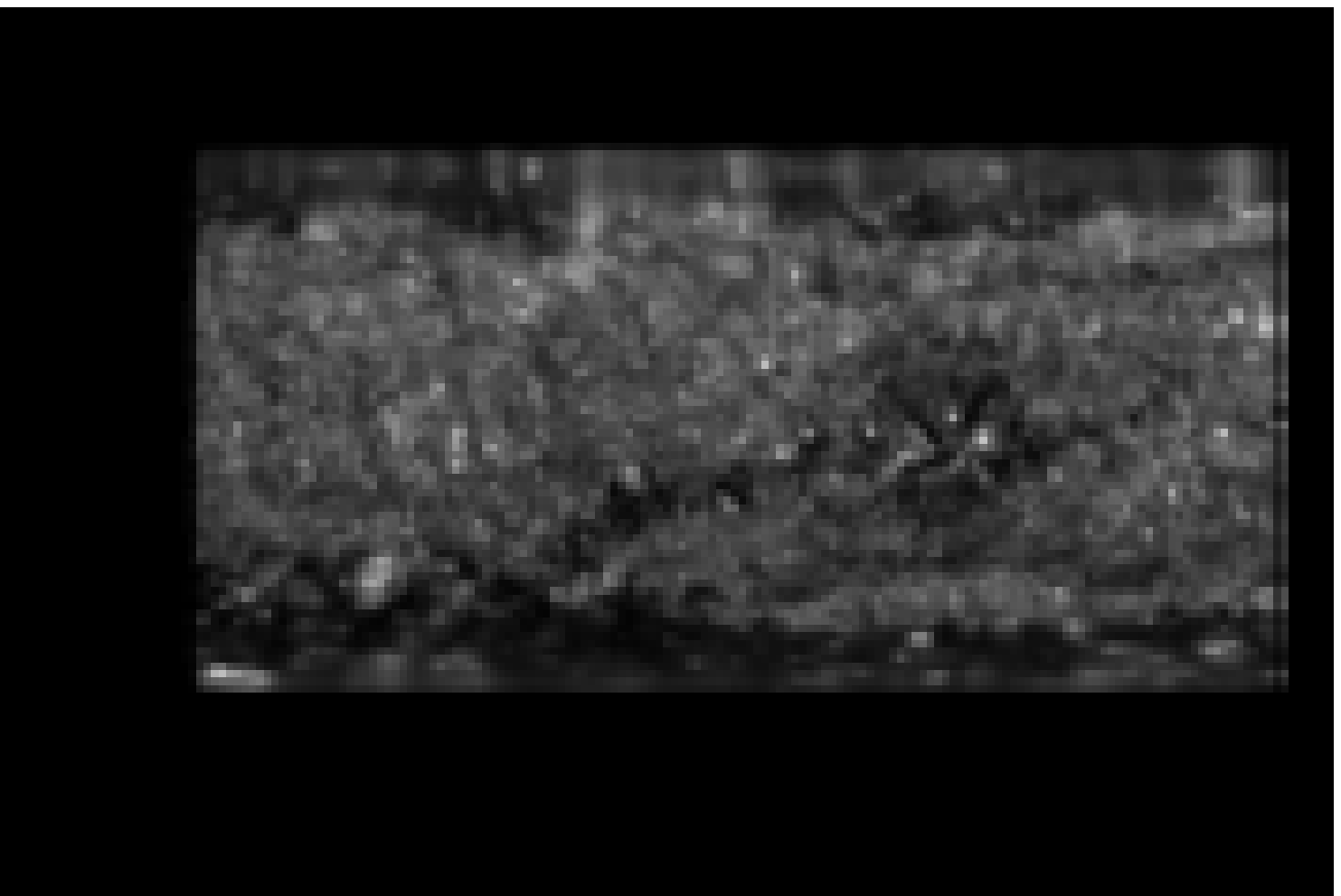}
\includegraphics[width=0.49\textwidth]{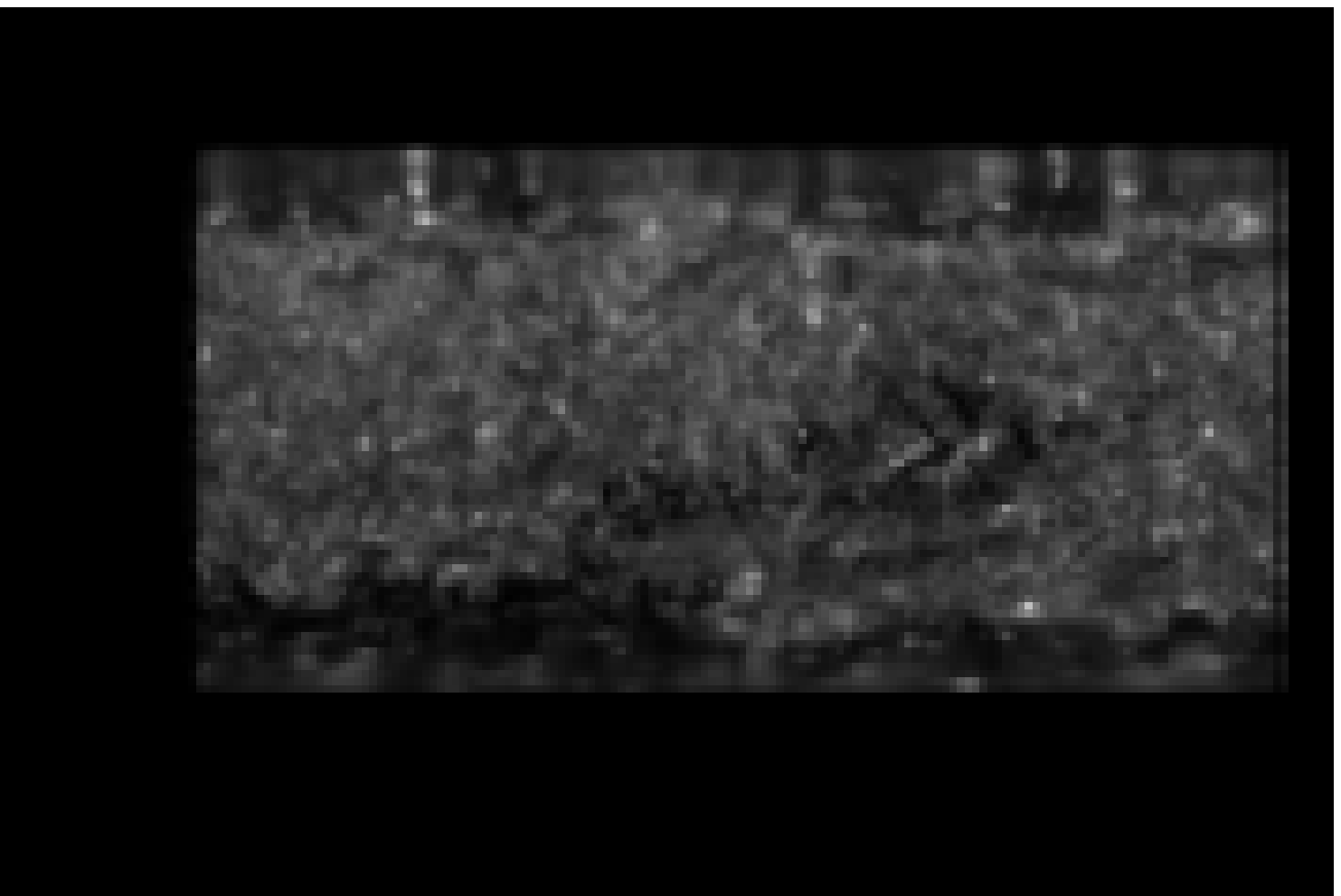}
\includegraphics[width=0.49\textwidth]{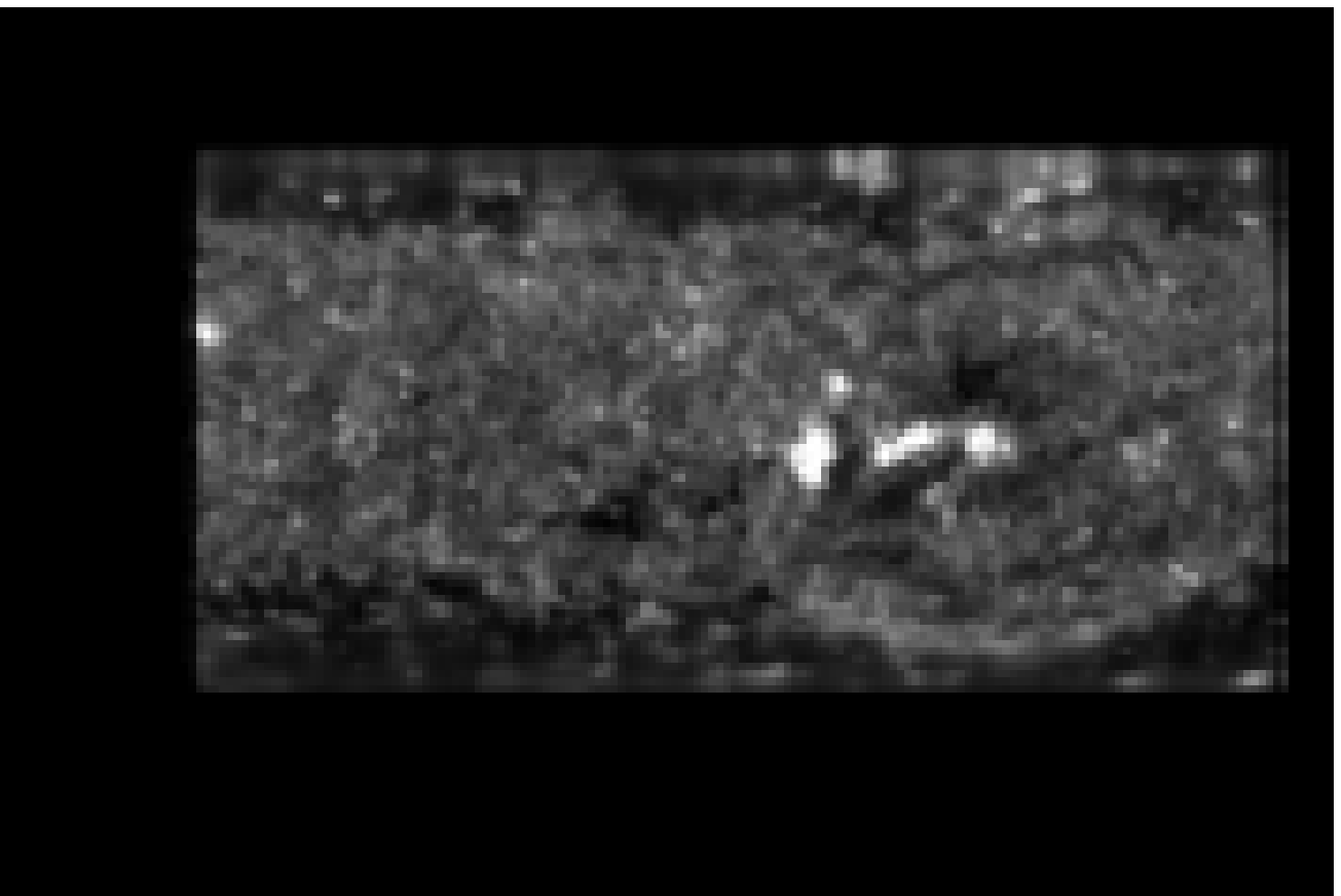}
\includegraphics[width=0.49\textwidth]{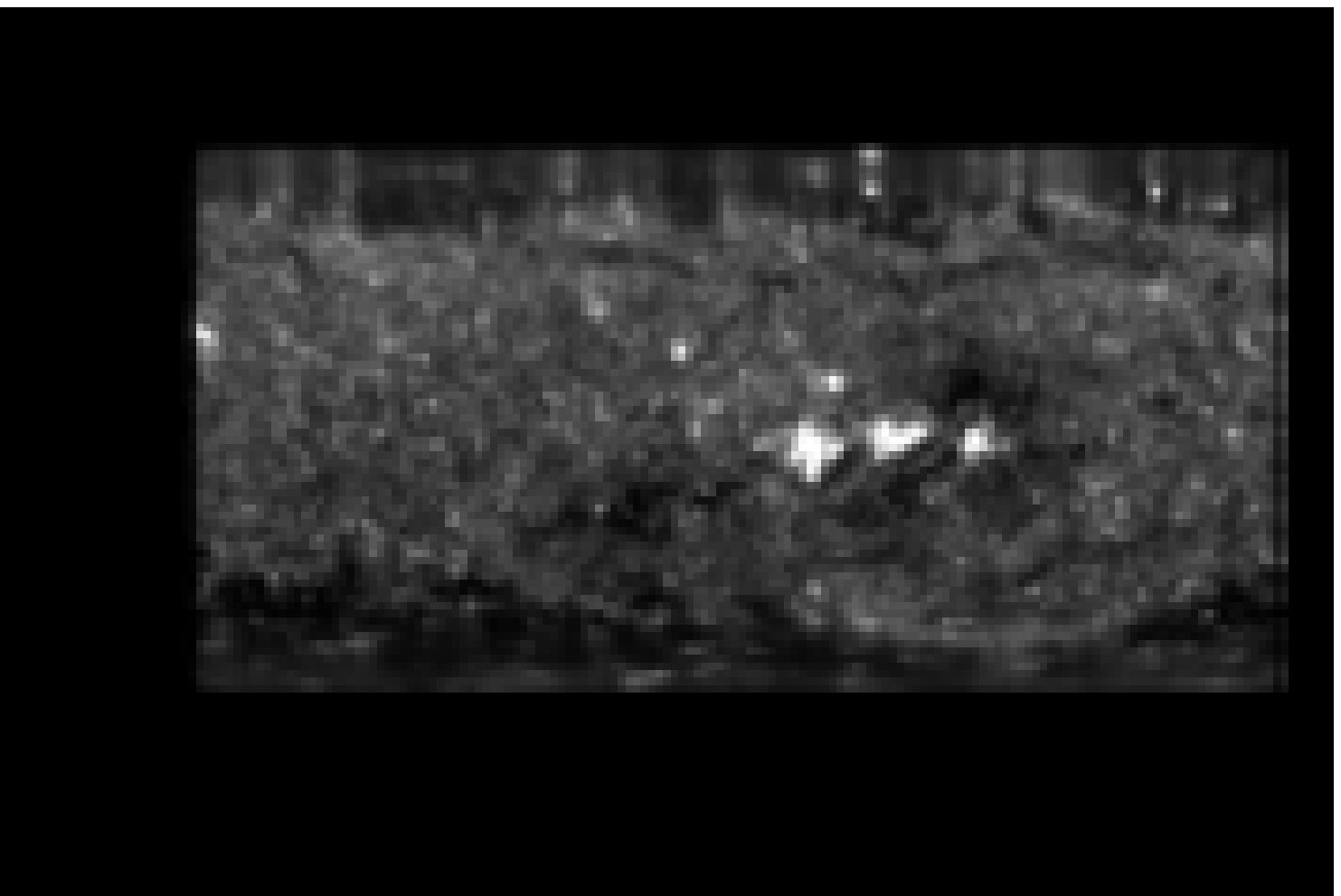}
\includegraphics[width=0.49\textwidth]{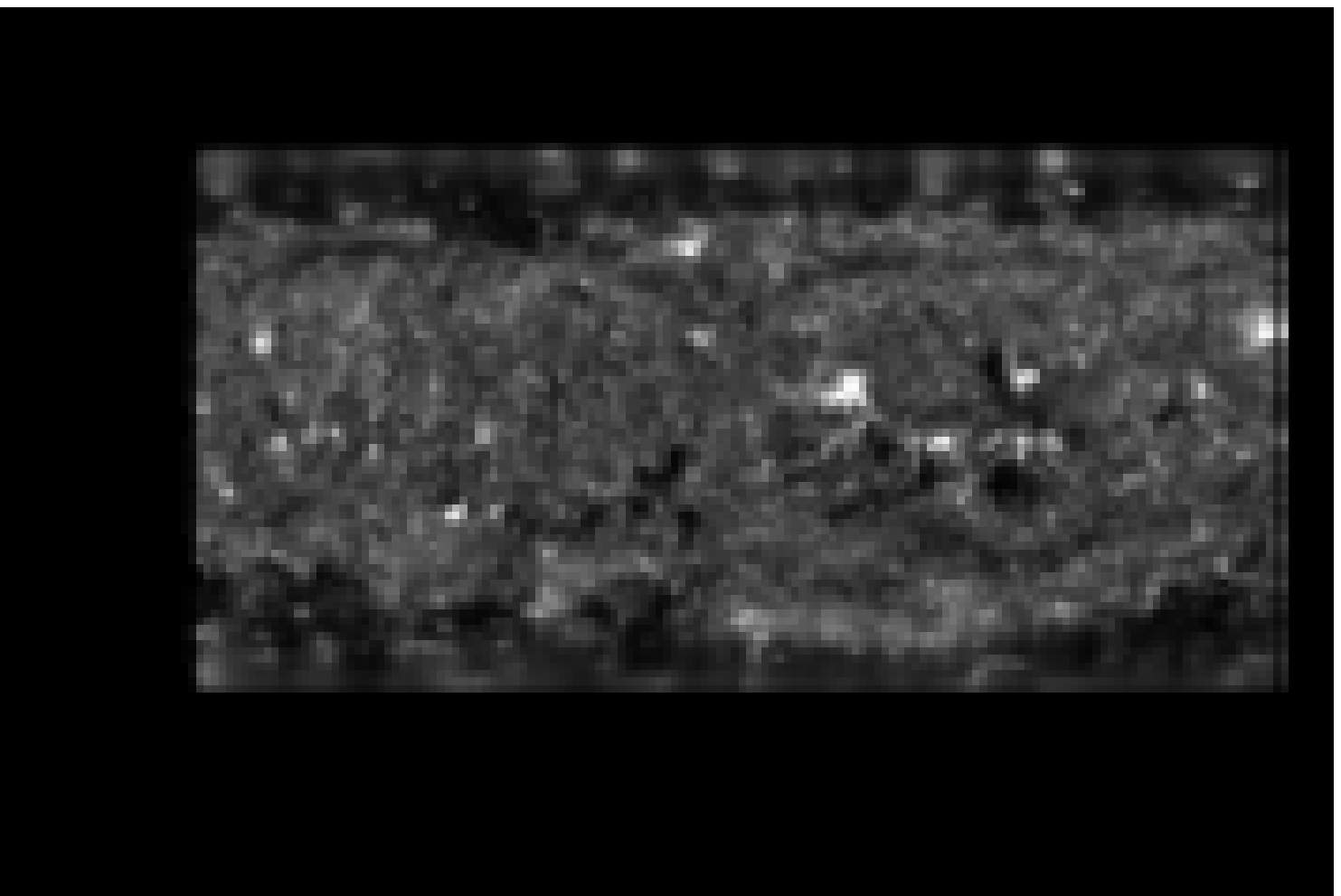}
\includegraphics[width=0.49\textwidth]{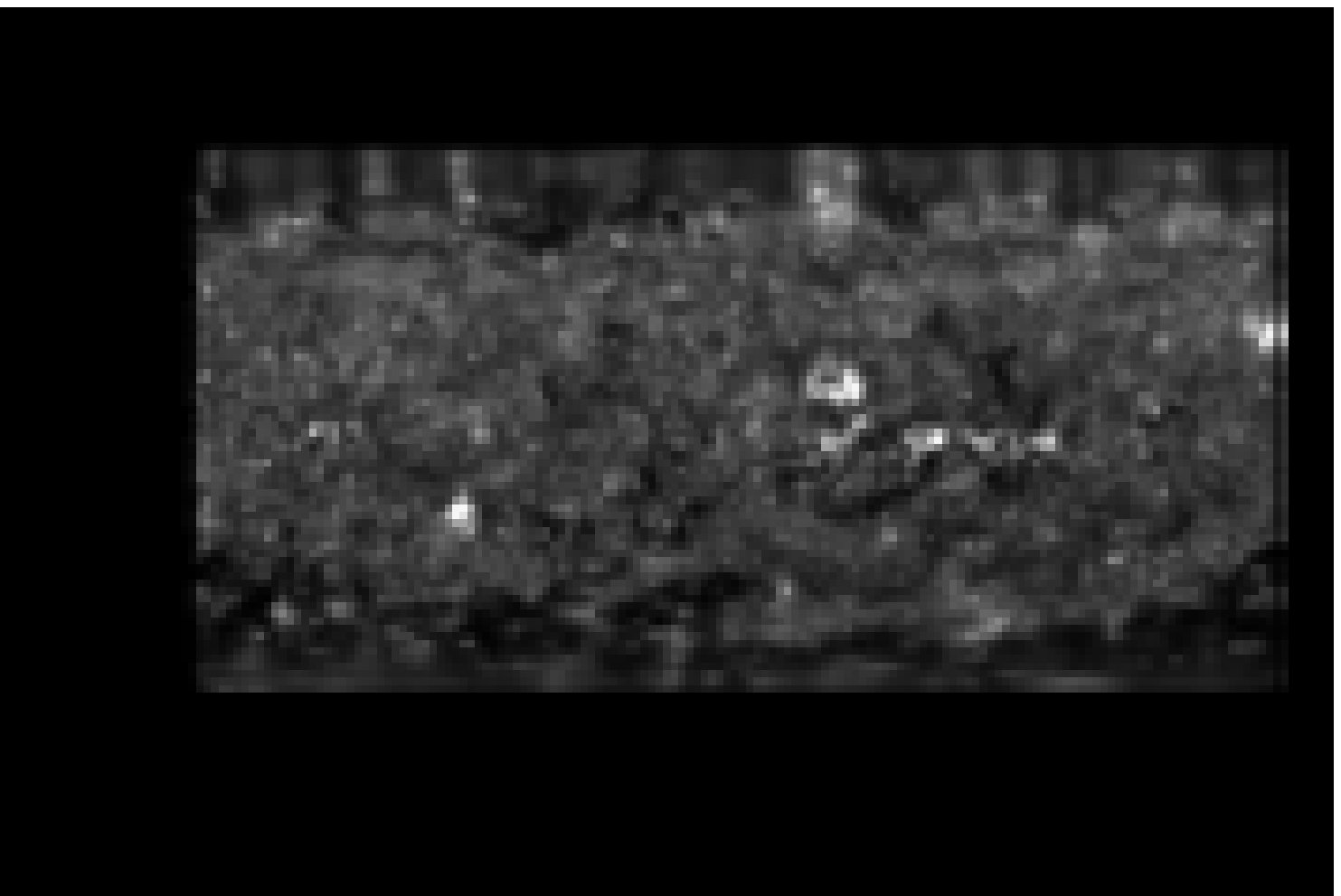}
\end{center}
\caption{STEREO/SECCHI/EUVI 171~\AA\  synoptic maps for CR 2067 (top pictures), CR 2068 (middle pictures) and CR 2069 (bottom pictures) from the ahead (left pictures) and behind (right pictures) spacecraft.  These show active regions as bright regions and coronal holes as dark regions.}
\label{fig:stereosyn171}
\end{figure}

\begin{figure} 
\begin{center}
\includegraphics[width=0.49\textwidth]{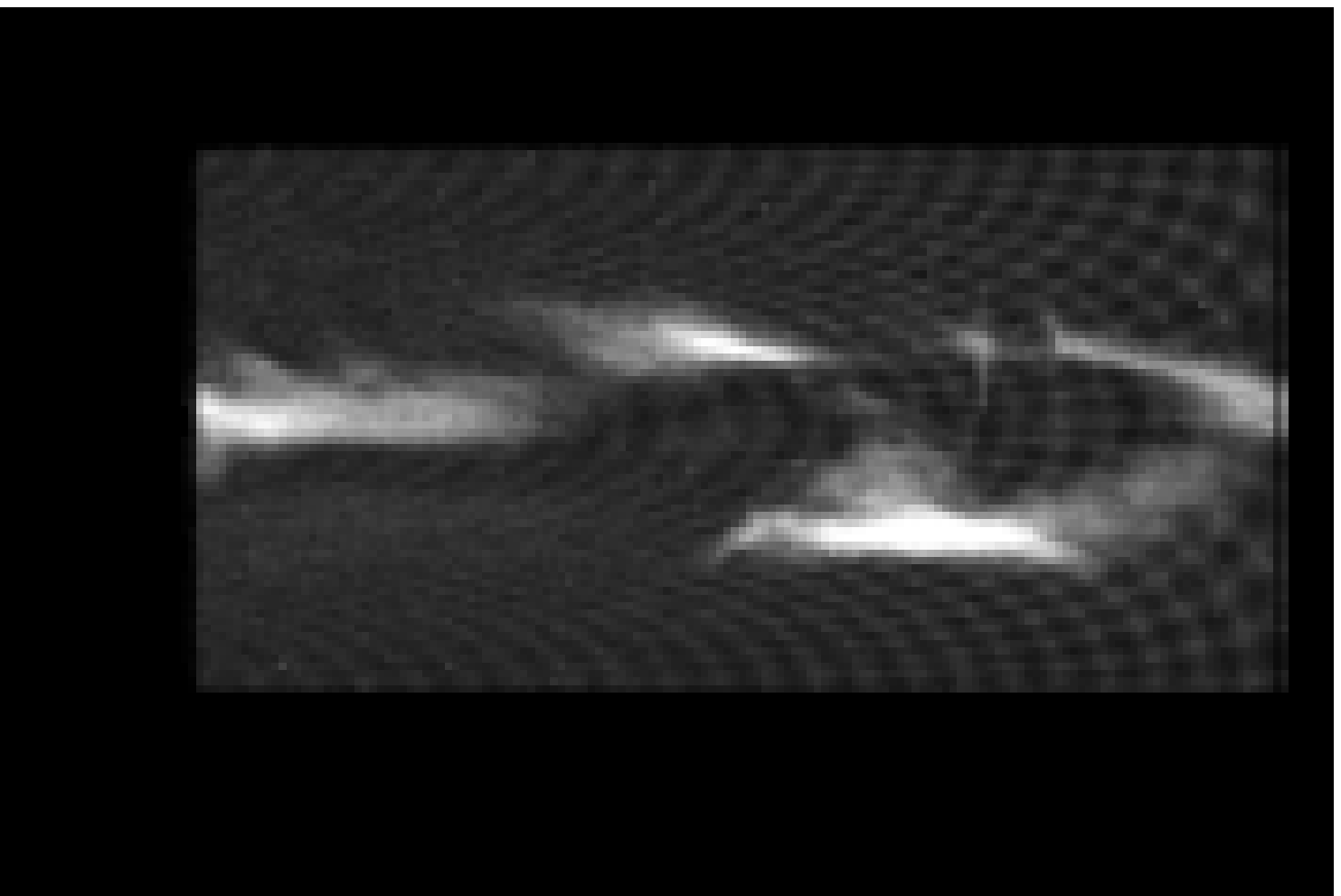}
\includegraphics[width=0.49\textwidth]{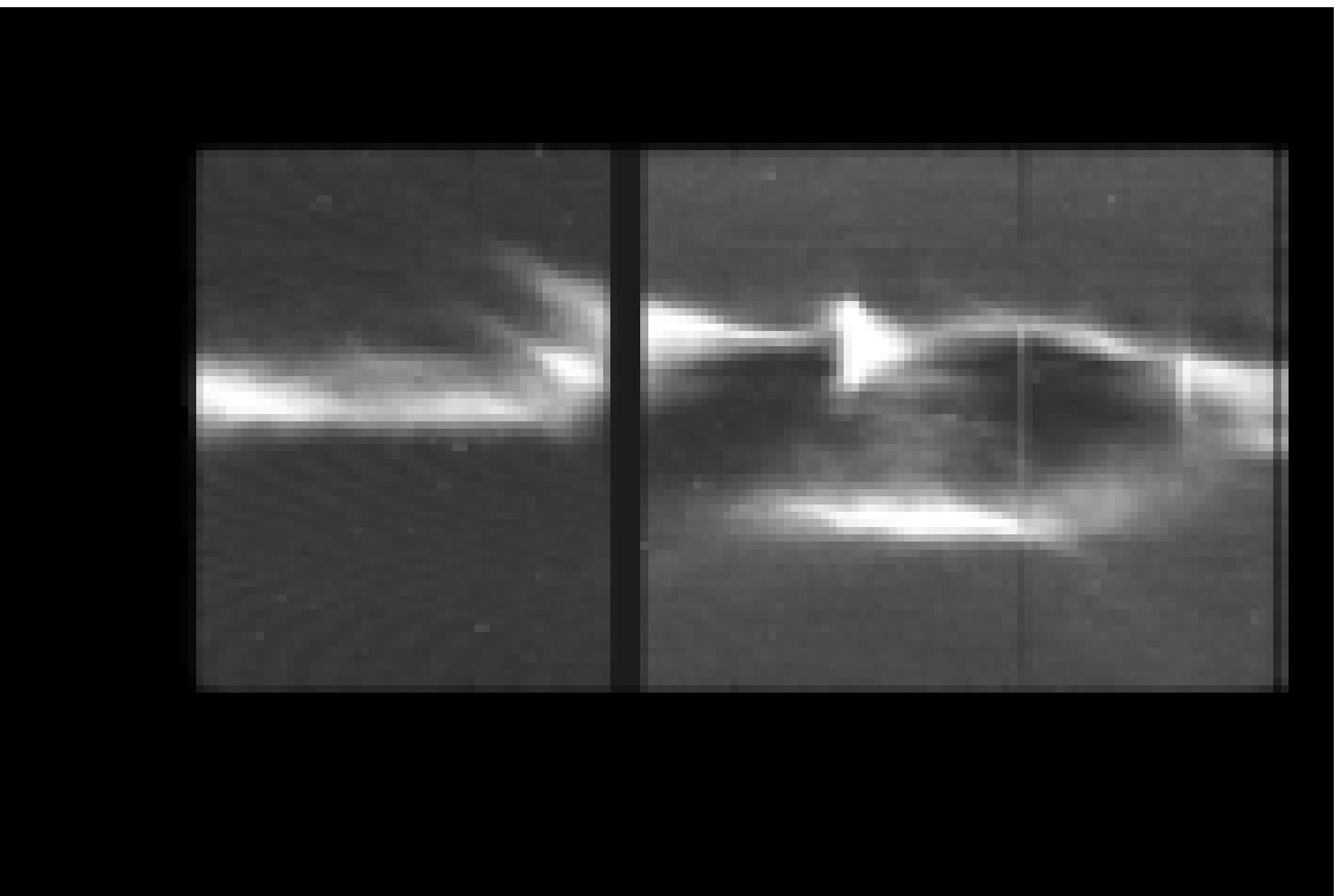}
\includegraphics[width=0.49\textwidth]{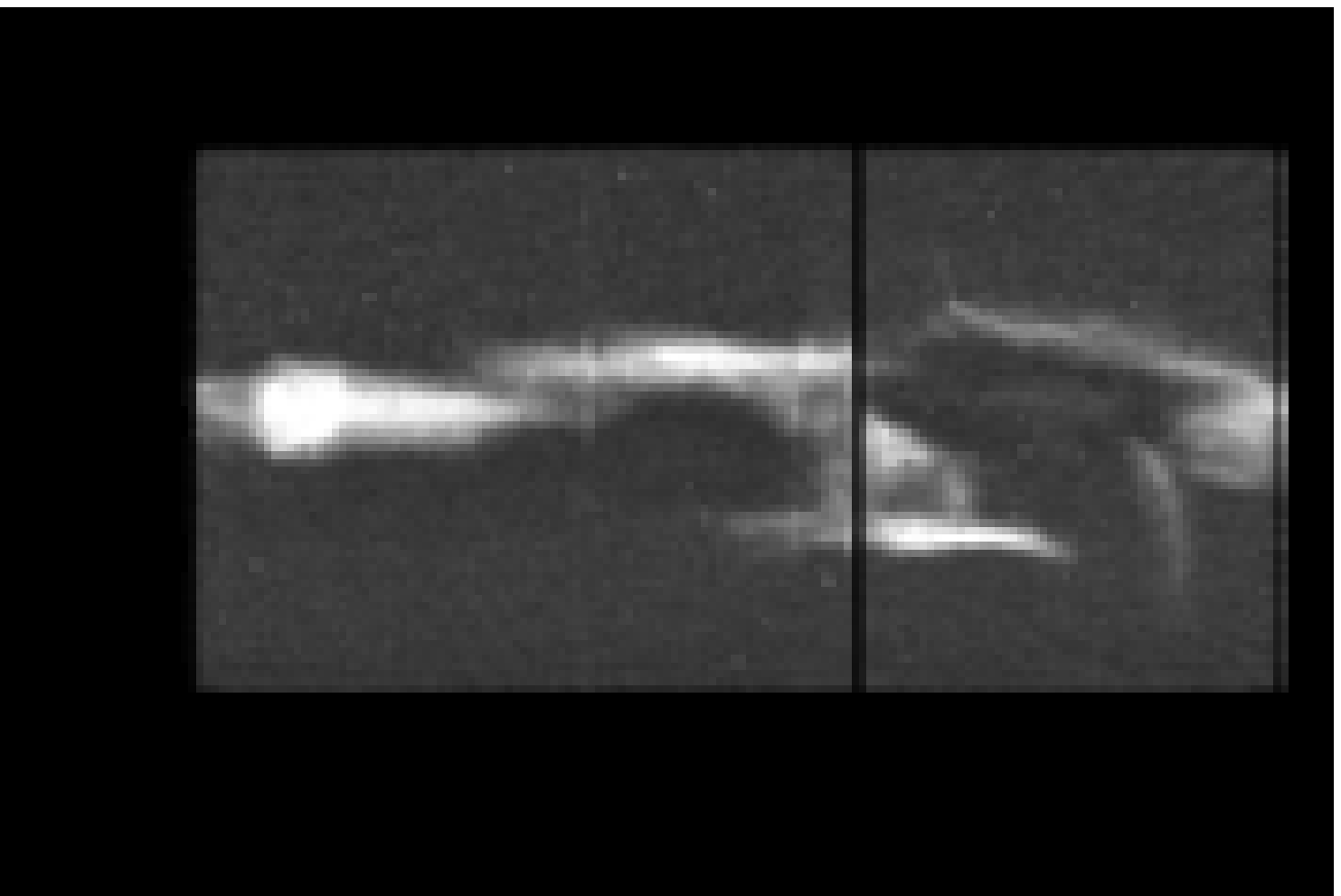}
\includegraphics[width=0.49\textwidth]{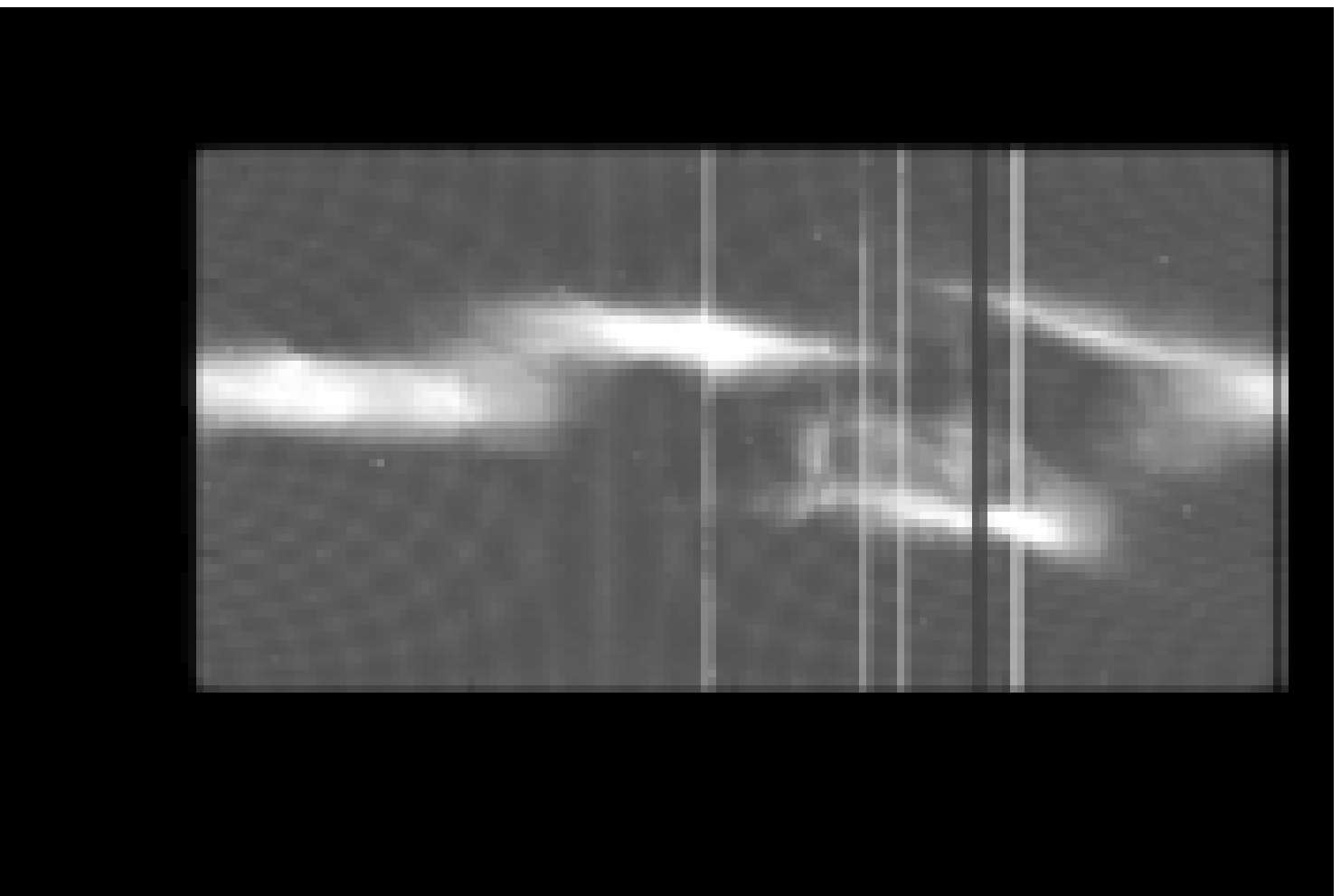}
\includegraphics[width=0.49\textwidth]{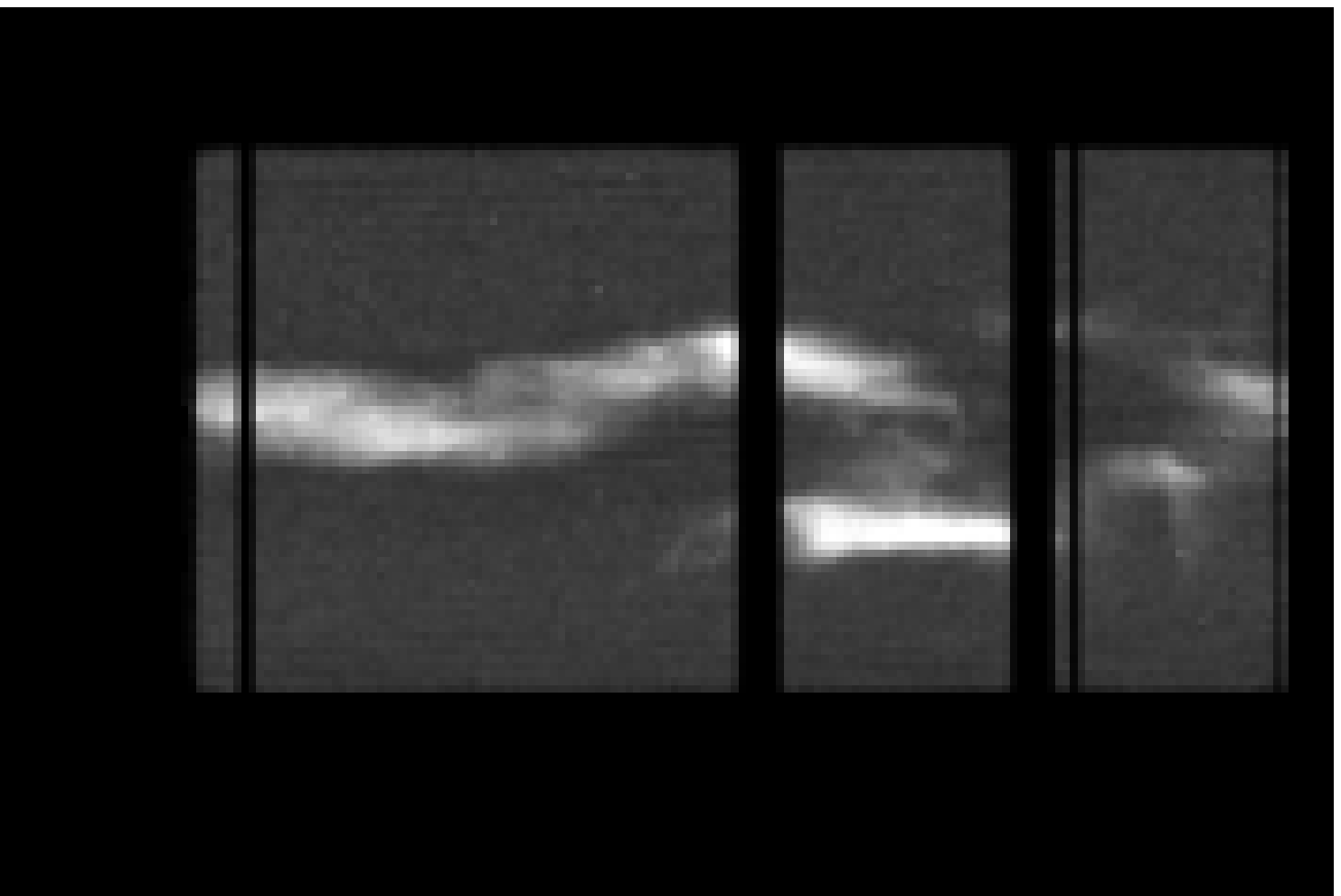}
\includegraphics[width=0.49\textwidth]{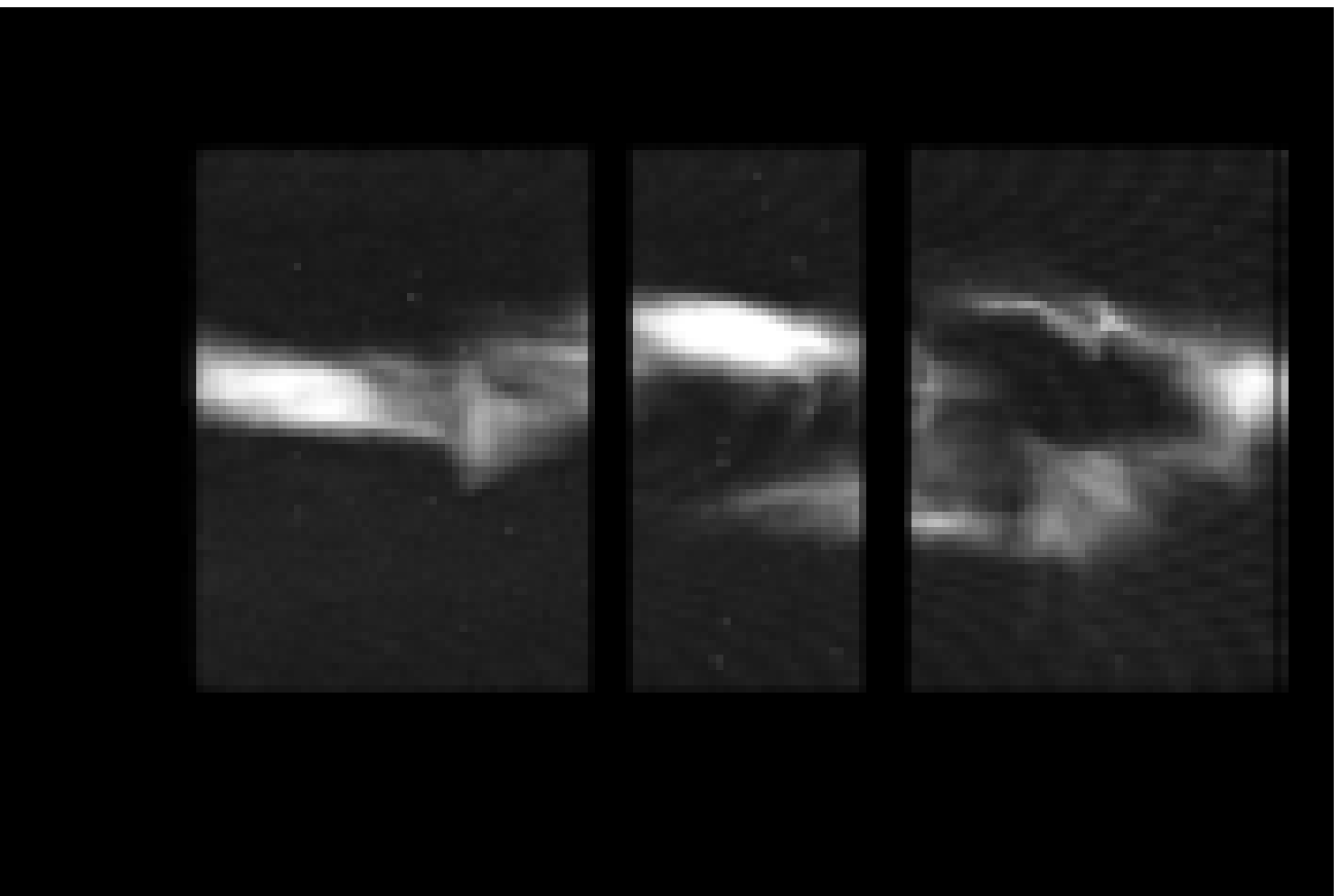}
\end{center}
\caption{STEREO/SECCHI/COR1 synoptic maps for CR 2067 (top pictures), CR 2068 (middle pictures) and CR 2069 (bottom pictures) from the ahead (left pictures) and behind (right pictures) spacecraft.  These maps show the streamer brightness at $2.6 R_S$ off the east limb, where $R_s$ is the solar radius.  The equivalent maps for the West limb appear very similar.}
\label{fig:stereoeastlimb}
\end{figure}


\begin{figure} 
\begin{center}
\includegraphics[width=0.49\textwidth]{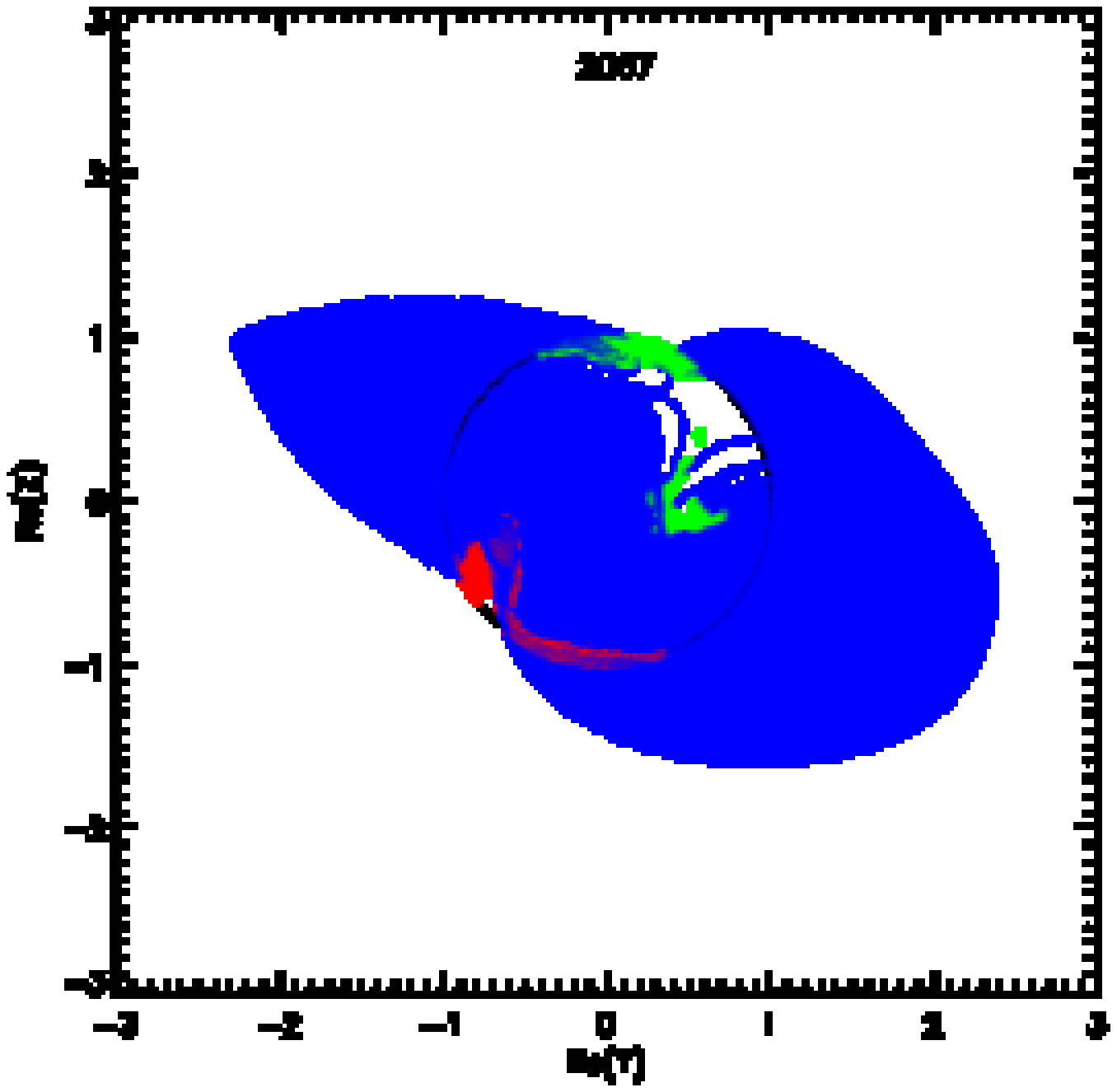}
\includegraphics[width=0.49\textwidth]{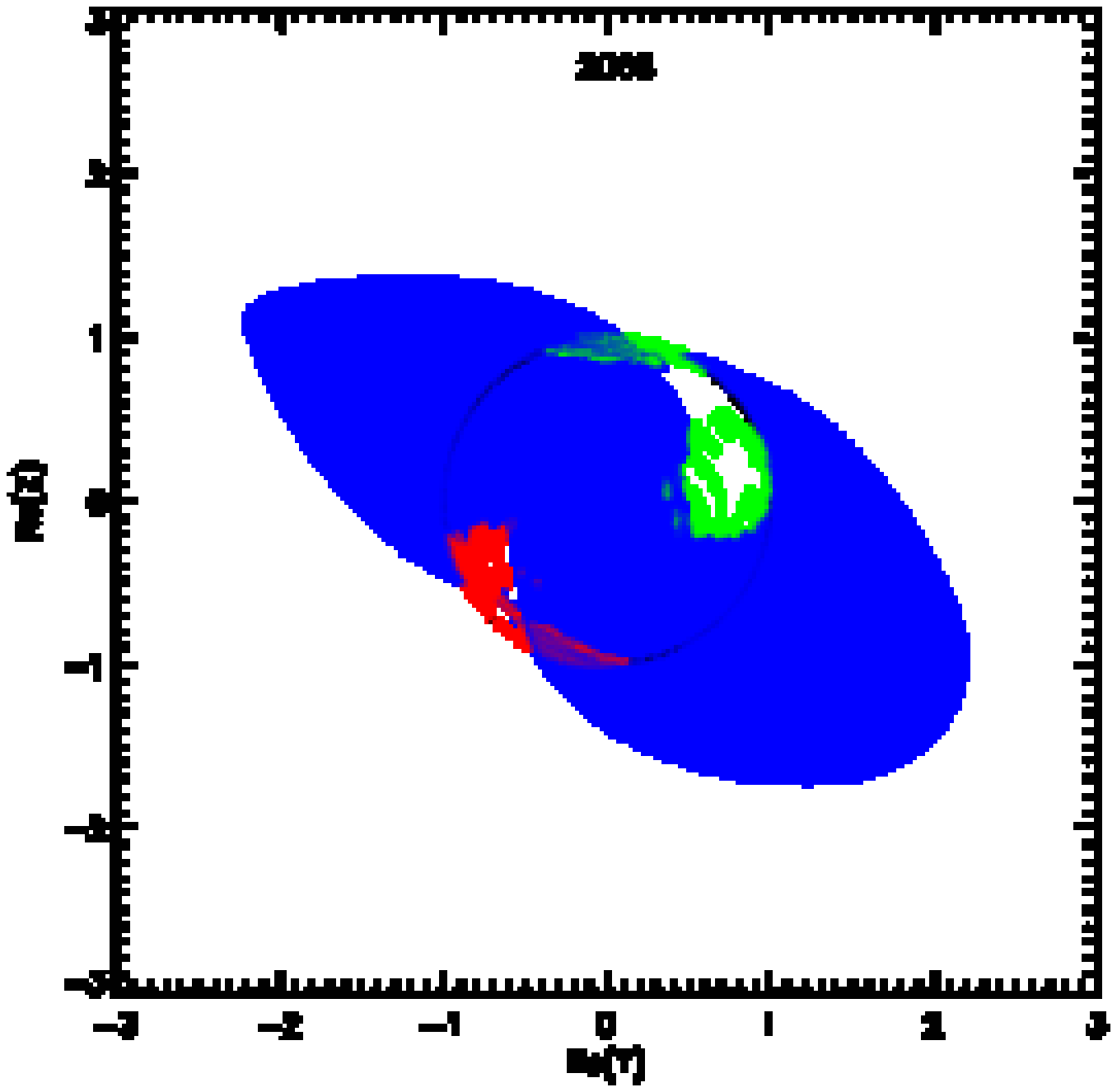}
\includegraphics[width=0.49\textwidth]{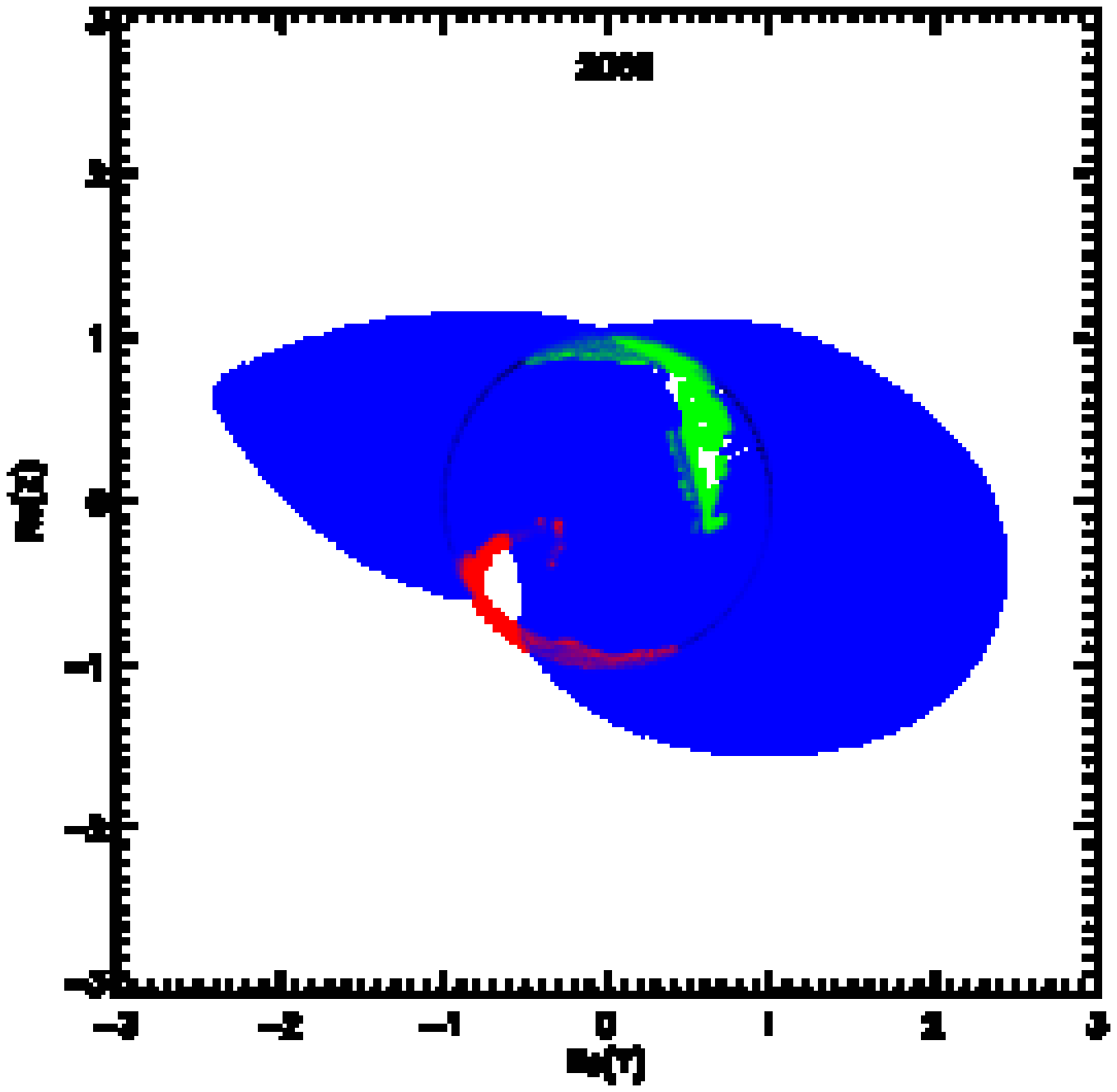}
\end{center}
\caption{The PFSS models of Figure~\ref{fig:synmodels} extrapolated from the GONG synoptic magnetograms of Figure~\ref{fig:synmags} for Carrington rotations 2067 (top left), 2068 (WHI, top right) and 2069 (bottom).  Positive and negative coronal holes are colored red and green.  Streamer-belt fields are plotted in blue.}
\label{fig:dat240models}
\end{figure}

\begin{figure} 
\begin{center}
\includegraphics[width=0.99\textwidth]{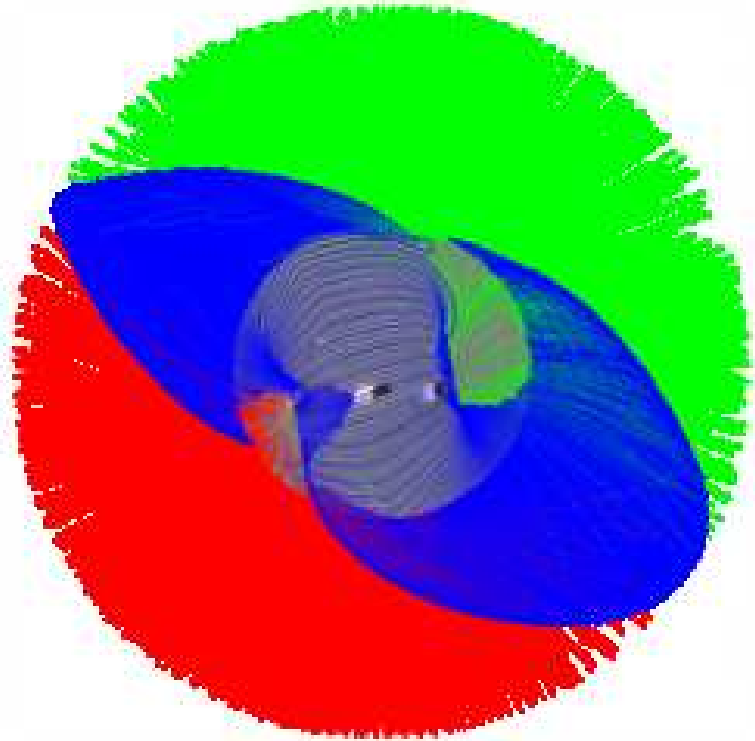}
\end{center}
\caption{``Hairy ball'' plot of the PFSS model of Figure~\ref{fig:synmodels} extrapolated from the GONG synoptic magnetogram for CR 2068.  The magnetogram is plotted in greyscale.  The streamer-belt fields are plotted in blue and the open positive and negative fields in red and green.}
\label{fig:hairyball}
\end{figure}

\begin{figure} 
\begin{center}
\includegraphics[width=0.99\textwidth]{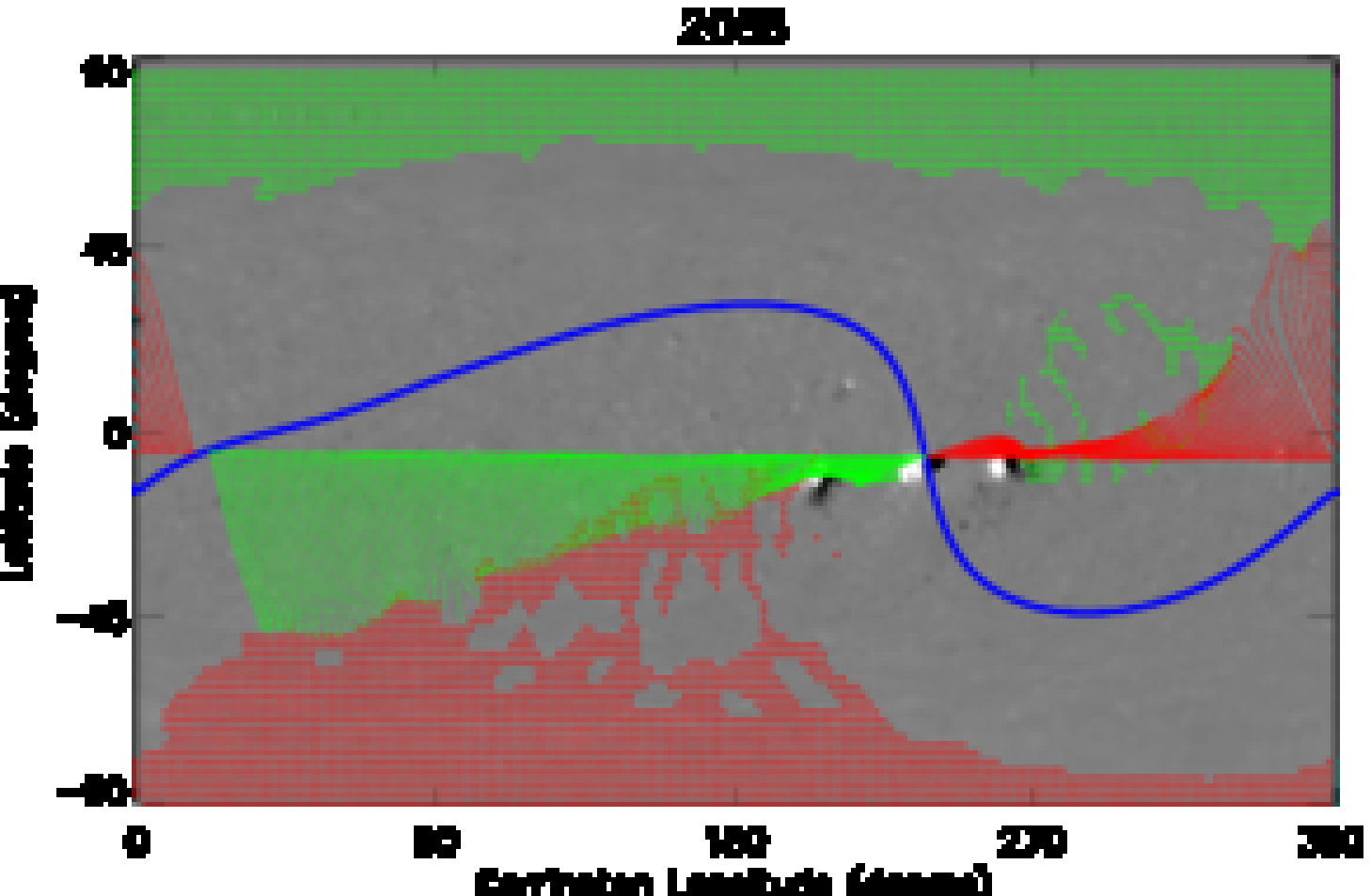}
\end{center}
\caption{Plot of the positive (red) and negative (green) ecliptic-plane fields of the PFSS model for CR 2068.  Also plotted are the streamer-belt neutral line in blue, the positive and negative coronal holes represented by red and green dots and the GONG magnetogram in greyscale.}
\label{fig:ecliptic}
\end{figure}

Figure~\ref{fig:synmags} shows the GONG synoptic maps for Carrington rotations 2067-9.  These show that rotation 2068 had three well-defined active regions, NOAA Active Regions 10987, 10988 and 10989 at approximately equal latitude, about $-10^{\circ}$, and separated by about $30^{\circ}$ of longitude from each other; 10989 at about $205^{\circ}$, 10988 at $235^{\circ}$ and 10987 at $265^{\circ}$.  All are consistent with Hale's polarity law in having leading negative polarity but their tilt angles appear to differ.  The maps also suggest that AR10989 was shorter-lived than the other two regions.  It is dispersed in the 2069 map while another less intense region (not numbered by NOAA) appears at the same Carrington longitude but North of the equator and with opposite leading polarity.  Meanwhile regions 10987 and 10988 reappear in rotation 2069, both showing evidence of wear and tear from turbulent diffusion.  The map of rotation 2067, on the other hand, shows little coherent structure but only diffuse field around longitude $235^{\circ}$, latitude $-10^{\circ}$ where 10988 will appear.

Figure~\ref{fig:synmodels} show features of the PFSS model for rotations 2067-9 based on the synoptic map of Figure~\ref{fig:synmags}.  These plots show the main neutral line which is formed by the apexes of the set of tallest closed field lines projected radially downwards onto the photosphere.  The coronal streamer belt is expected to be located immediately above this neutral line, separating open field regions of opposite polarity.  The thin black lines in the top left plot denote locations of pseudo-streamers, streamer structures sometimes found above multipolar loop systems, separating regions of open field of the same polarity.  The principal morphological difference between the two types of streamers in coronagraph images is that the helmet streamer cusps are located a solar radius or two above the solar surface whereas only the long stalks of pseudo-streamers are visible above such a height.  In the PFSS model the closed portion of the streamer must extend to the outer boundary because a current sheet cannot appear in the domain of a regular potential field.  On the other hand, the cusp of a pseudo-streamer can appear at any height in the domain.  These are found by following \inlinecite{WangSheeleyRich2007}.  We define a point on the source surface to be a pseudo-streamer if at least two of its neighbors have foot-point separation$> (\pi /12) R_s$, where $R_s$ is the solar radius, and are of the same polarity.  Also shown in the model plots are the predicted coronal holes, regions formed by the set of foot points of field lines open to the outer boundary of the model.  The STEREO/EUVI 171~\AA\  maps in Figure~\ref{fig:stereosyn171} show the coronal hole locations as dark patches while the STEREO/COR1 synoptic maps in Figure~\ref{fig:stereoeastlimb} summarize the  brightness distribution of the East limb over the rotation.  The modeled coronal hole and streamer locations match the observations reasonably well.

The three aligned active regions produce a warp in the neutral line during the WHI.  They do this by creating an East-West magnetic moment on a spatial scale large enough to influence the global field structure.  During much of 2008 there was activity at approximately these Carrington longitudes, corresponding to a major filament system at the same longitudes observed throughout 2008 by Patrick McIntosh - see \inlinecite{Webbetal2010} for examples.  It is this asymmetric activity pattern combined with relatively weak polar fields \cite{PetriePatrikeeva2009} that results in a global coronal field configuration that is far from dipolar.  Comparison of the spherical harmonic components of the potential field shows why the field is so tilted and non-dipolar.  The equatorial dipole component is approximately half as strong as the polar dipole, producing a combined dipole tilted at about 60$^{\circ}$ to the rotation axis.  Even at the source surface the full dipole component is only three times stronger than the quadrupole and octupole components of the field.  Figure~\ref{fig:dat240models} shows the models for rotations 2067-9 in spherical coordinates, viewed from longitude $240^{\circ}$.  This viewpoint shows the warp and tilt of the streamer belt.  The strength and alignment of the active regions during the WHI produces a slightly larger warp and tilt in the model for 2068 compared to 2067 and 2069.  However, the displacement of the streamer belt from the equator is approximately the same for all three rotations according to Figure~\ref{fig:stereoeastlimb}.

Figure~\ref{fig:hairyball} displays in spherical coordinates the photospheric surface field for rotation 2068 in greyscale with the streamer-belt field in blue and the positive and negative open field in red and green.  Figure~\ref{fig:ecliptic} shows the magnetogram for 2068 in greyscale with features of the coronal field structure over-plotted.  The neutral line is plotted in blue and the positive and negative coronal holes are represented by regions of red and green dots.  The red and green lines represent the positive and negative field trajectories that connect the base of the corona to where the ecliptic plane intersects the outer boundary of the model.  These fields are therefore the predicted trajectories of solar wind plasma that will eventually reach Earth.  Unusually for solar minimum, much of this ecliptic flux originates from midlatitude coronal holes.  This is another indication of how far from a dipolar state the global coronal field was during this rotation.

\section{Structure of the active regions}
\label{s:activeregions}

\begin{figure} 
\begin{center}
\rotatebox{270}{\includegraphics[width=0.7\textwidth]{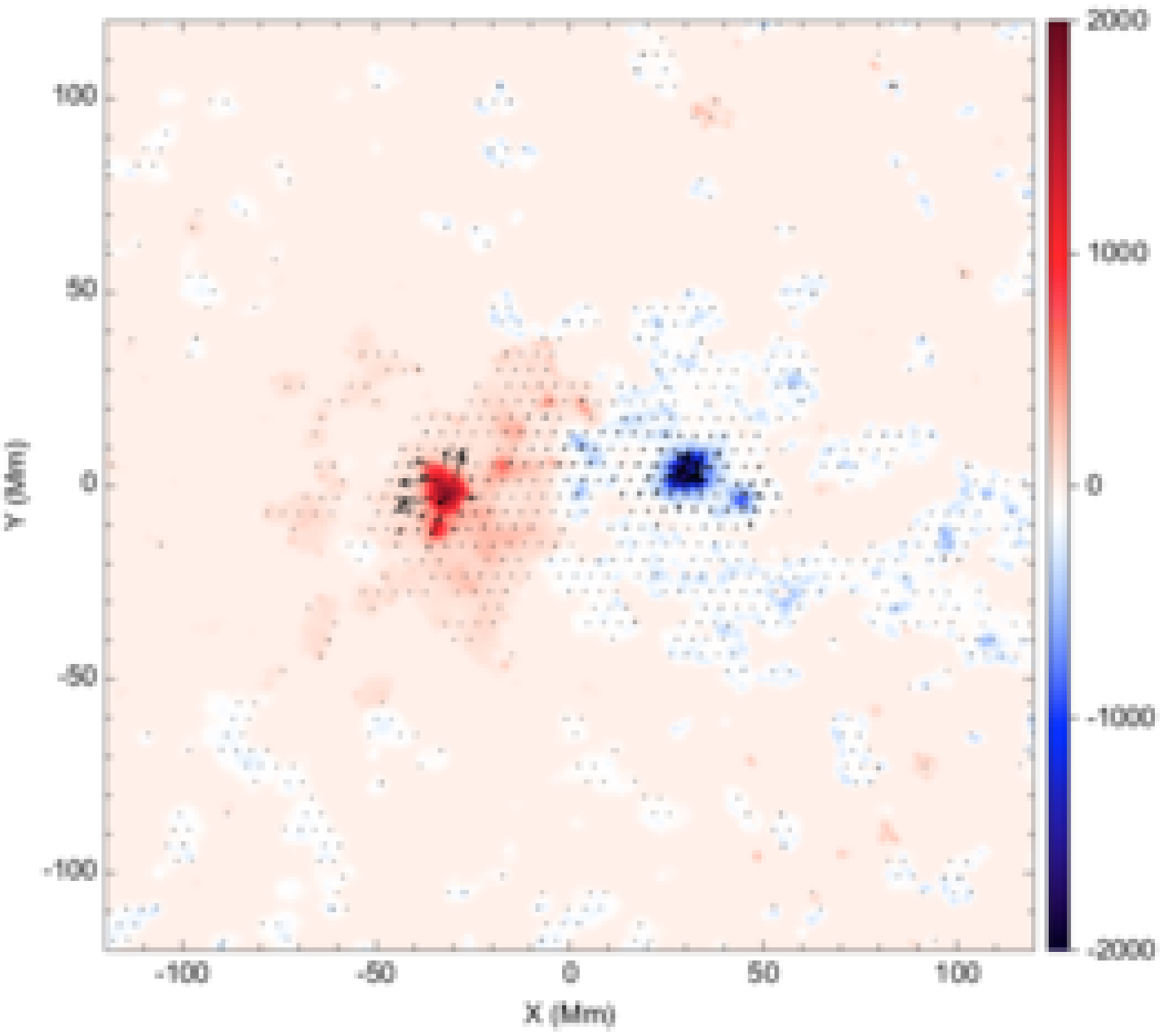}}
\rotatebox{270}{\includegraphics[width=0.7\textwidth]{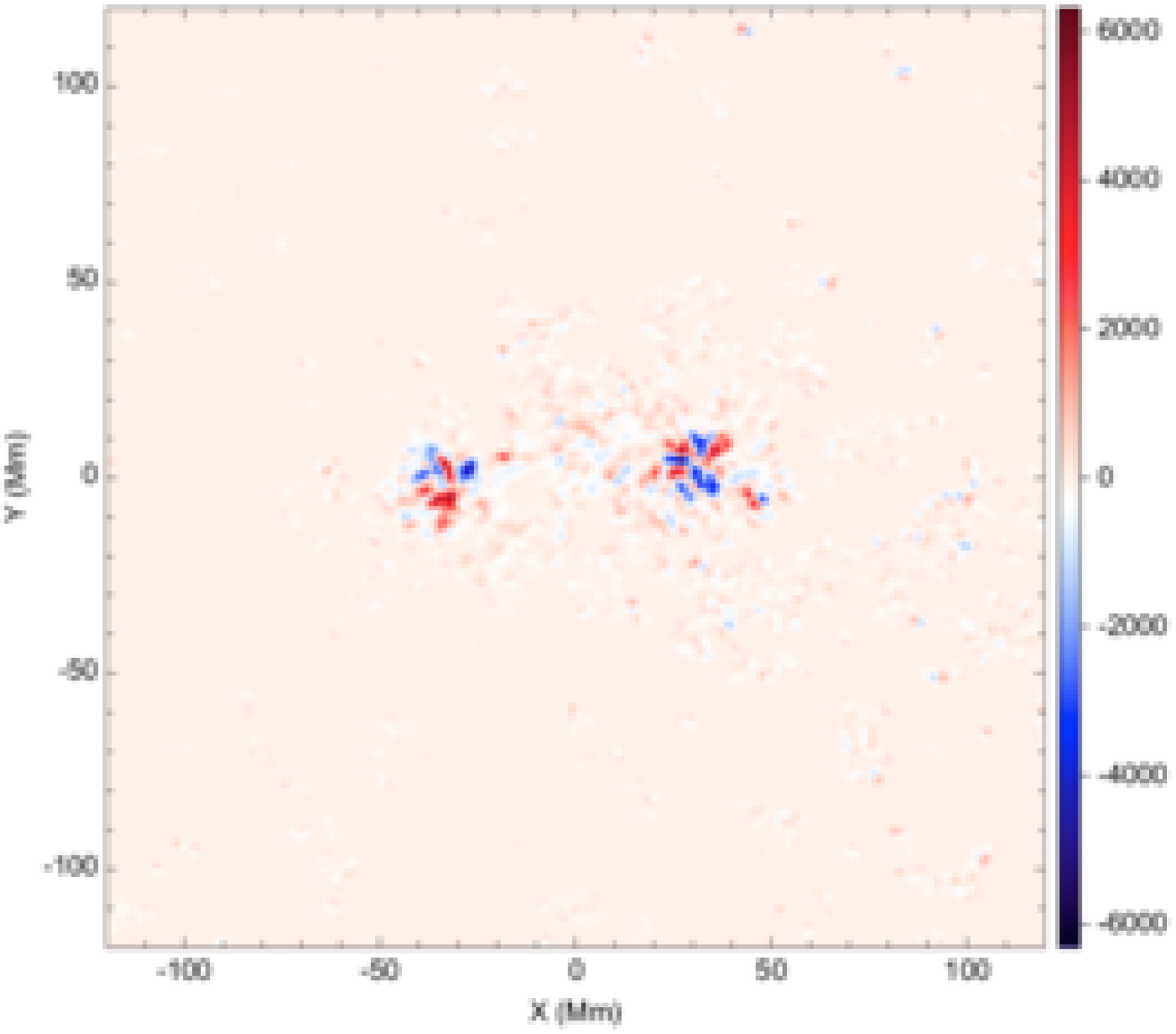}}
\end{center}
\caption{SOLIS vector magnetogram (top) and corresponding vertical electric current density (bottom) of AR~10987 on 2008 March 27 at 1535~UT.  In the plot of the vector magnetogram red and blue denote positive and negative vertical magnetic field, with the color bar showing the values in Gauss, and the black arrows show the transverse field vectors.  In the electric current plot red and blue denote positive and negative current with the color bar showing the values in statampere/cm$^2$.}
\label{fig:vec10987}
\end{figure}

\begin{figure} 
\begin{center}
\rotatebox{270}{\includegraphics[width=0.7\textwidth]{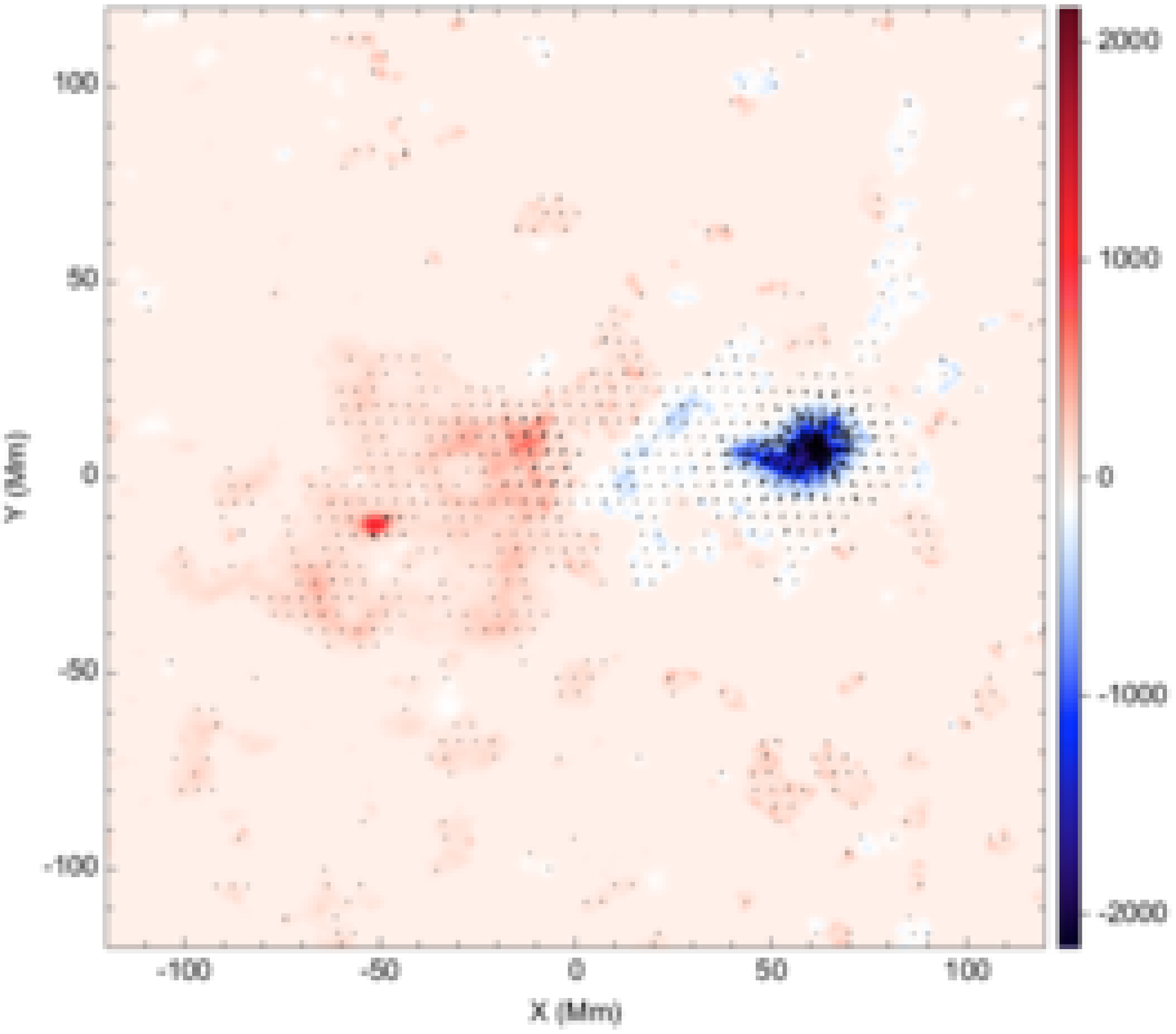}}
\rotatebox{270}{\includegraphics[width=0.7\textwidth]{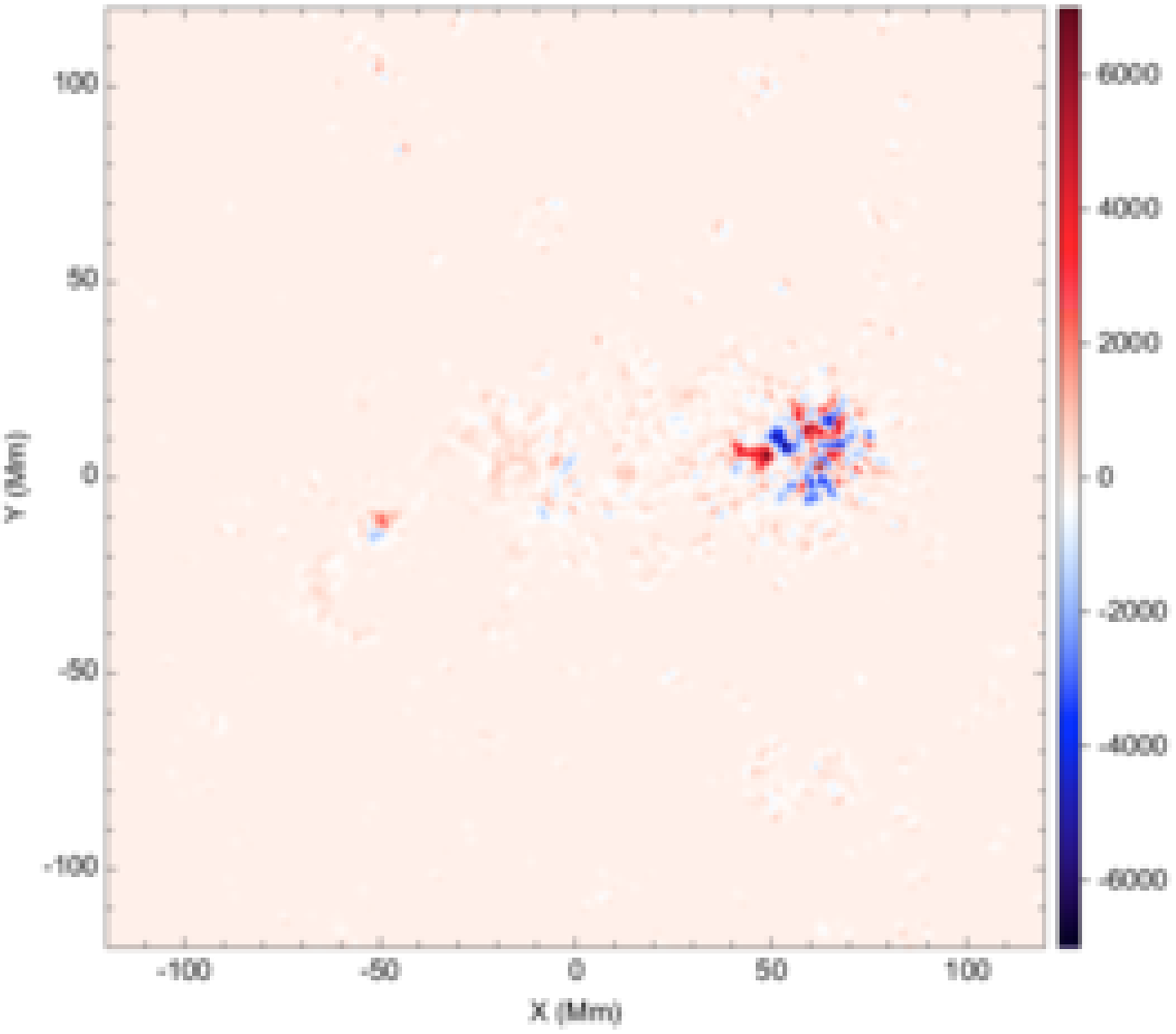}}
\end{center}
\caption{SOLIS vector magnetogram (top) and corresponding vertical electric current density (bottom) of AR~10988 on 2008 March 29 at 1528~UT.  In the plot of the vector magnetogram red and blue denote positive and negative vertical magnetic field, with the color bar showing the values in Gauss, and the black arrows show the transverse field vectors.  In the electric current plot red and blue denote positive and negative current with the color bar showing the values in statampere/cm$^2$.}
\label{fig:vec10988}
\end{figure}

\begin{figure} 
\begin{center}
\rotatebox{270}{\includegraphics[width=0.7\textwidth]{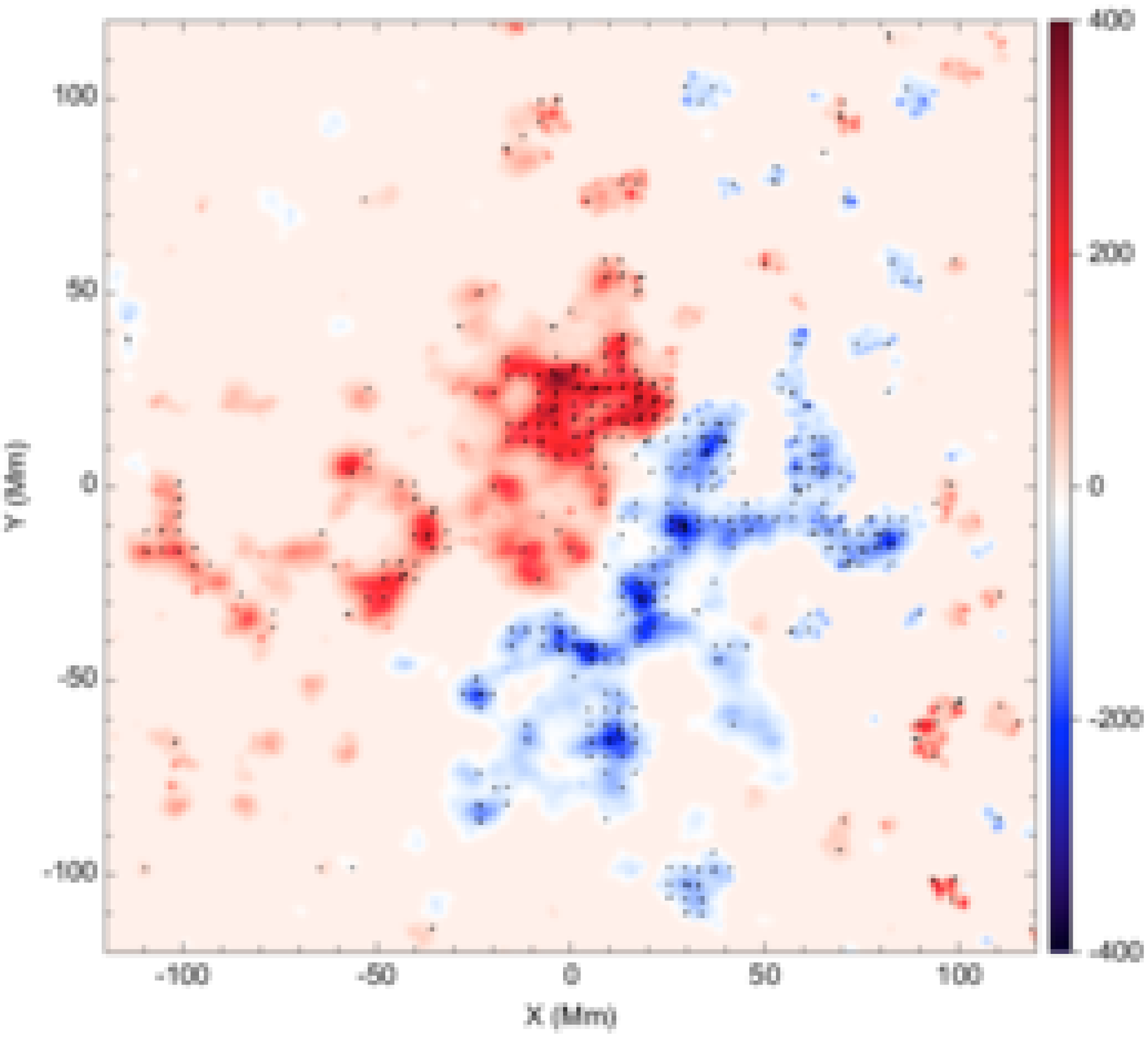}}
\rotatebox{270}{\includegraphics[width=0.7\textwidth]{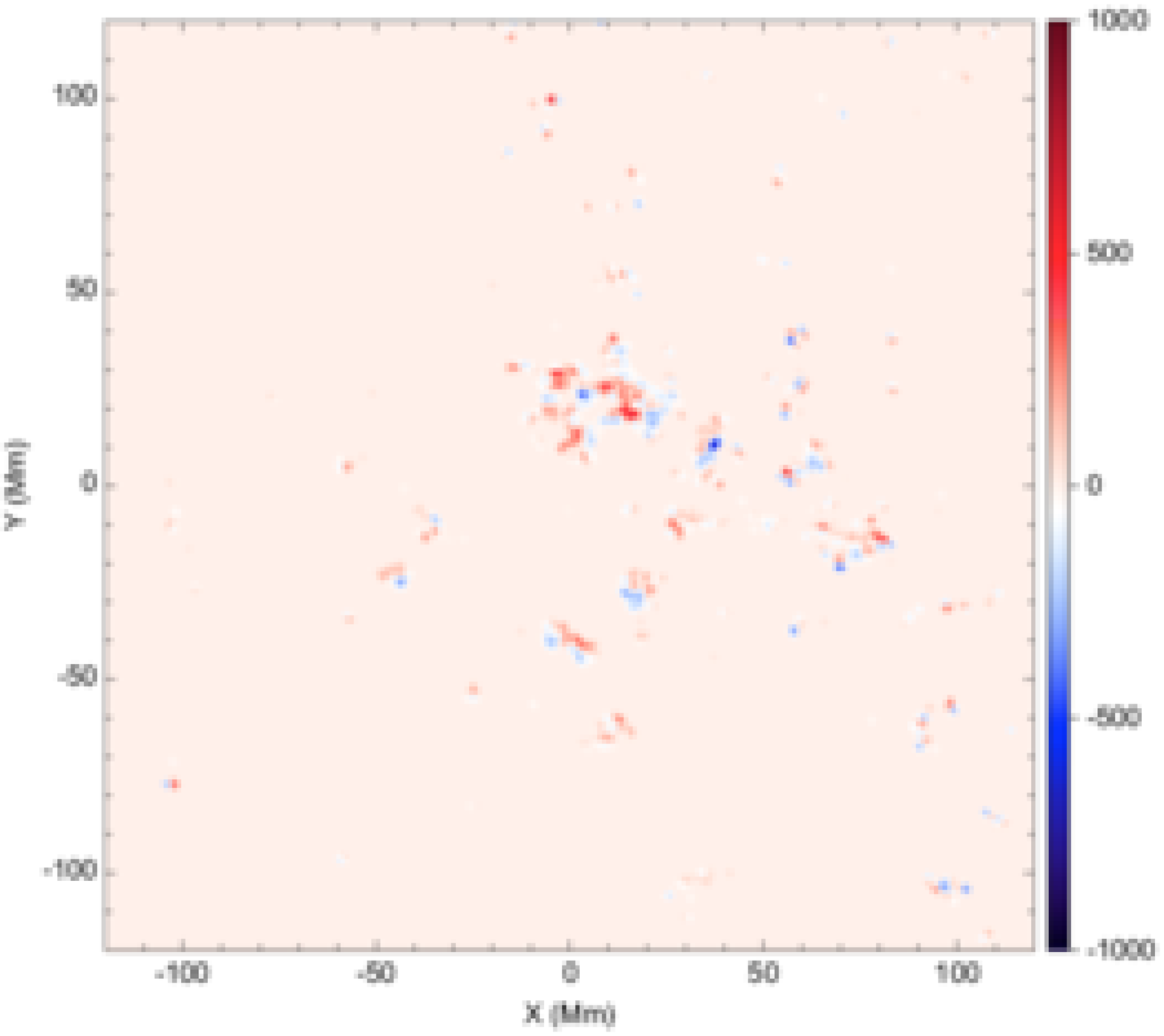}}
\end{center}
\caption{SOLIS vector magnetogram (top) and corresponding vertical electric current density (bottom) of AR~10989 on 2008 March 31 at 1742~UT.  In the plot of the vector magnetogram red and blue denote positive and negative vertical magnetic field, with the color bar showing the values in Gauss, and the black arrows show the transverse field vectors.  In the electric current plot red and blue denote positive and negative current with the color bar showing the values in statampere/cm$^2$.}
\label{fig:vec10989}
\end{figure}


\begin{figure} 
\begin{center}
\resizebox{0.75\textwidth}{!}{\includegraphics*[100,90][700,700]{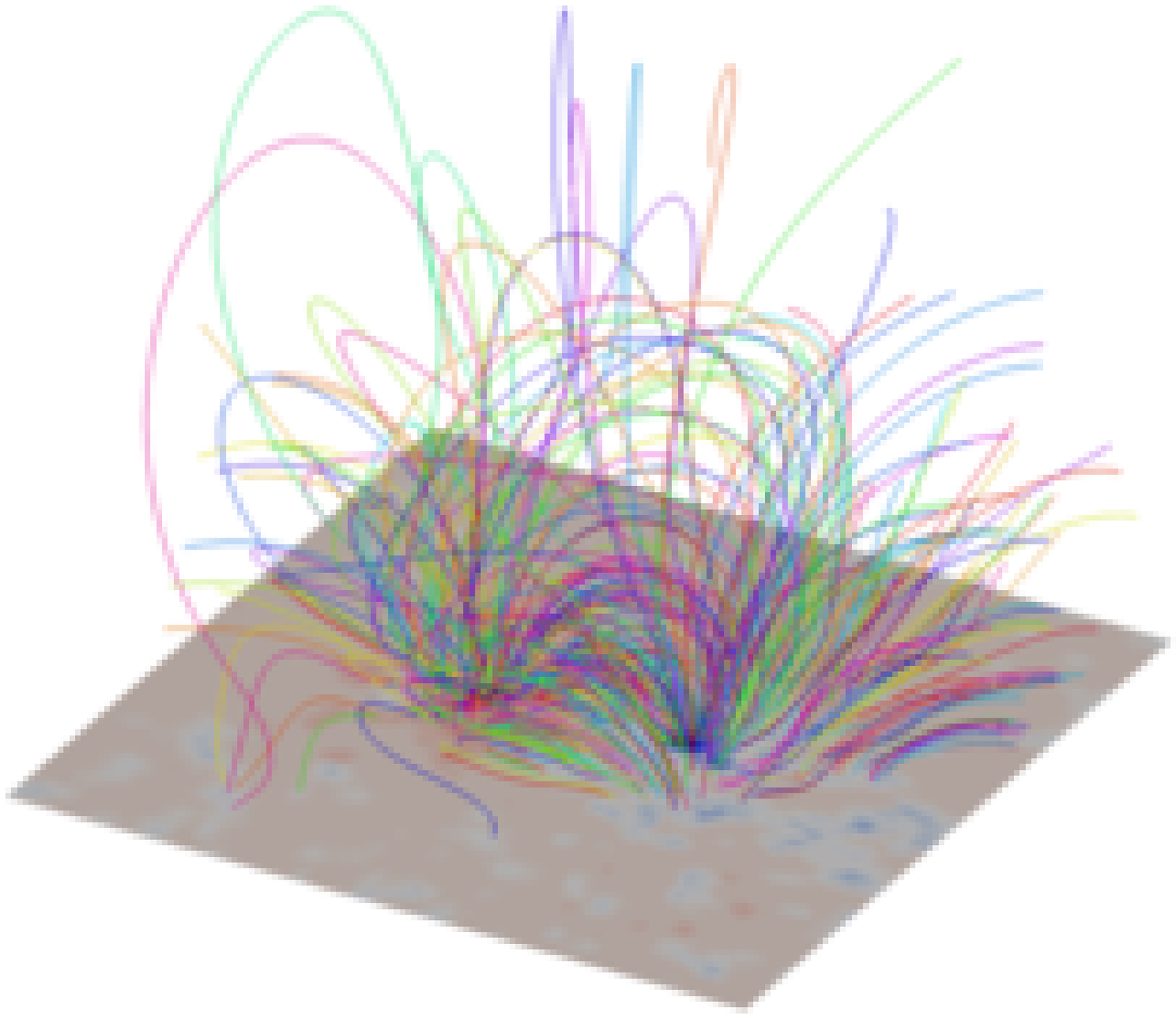}}
\resizebox{0.75\textwidth}{!}{\includegraphics*[150,150][650,650]{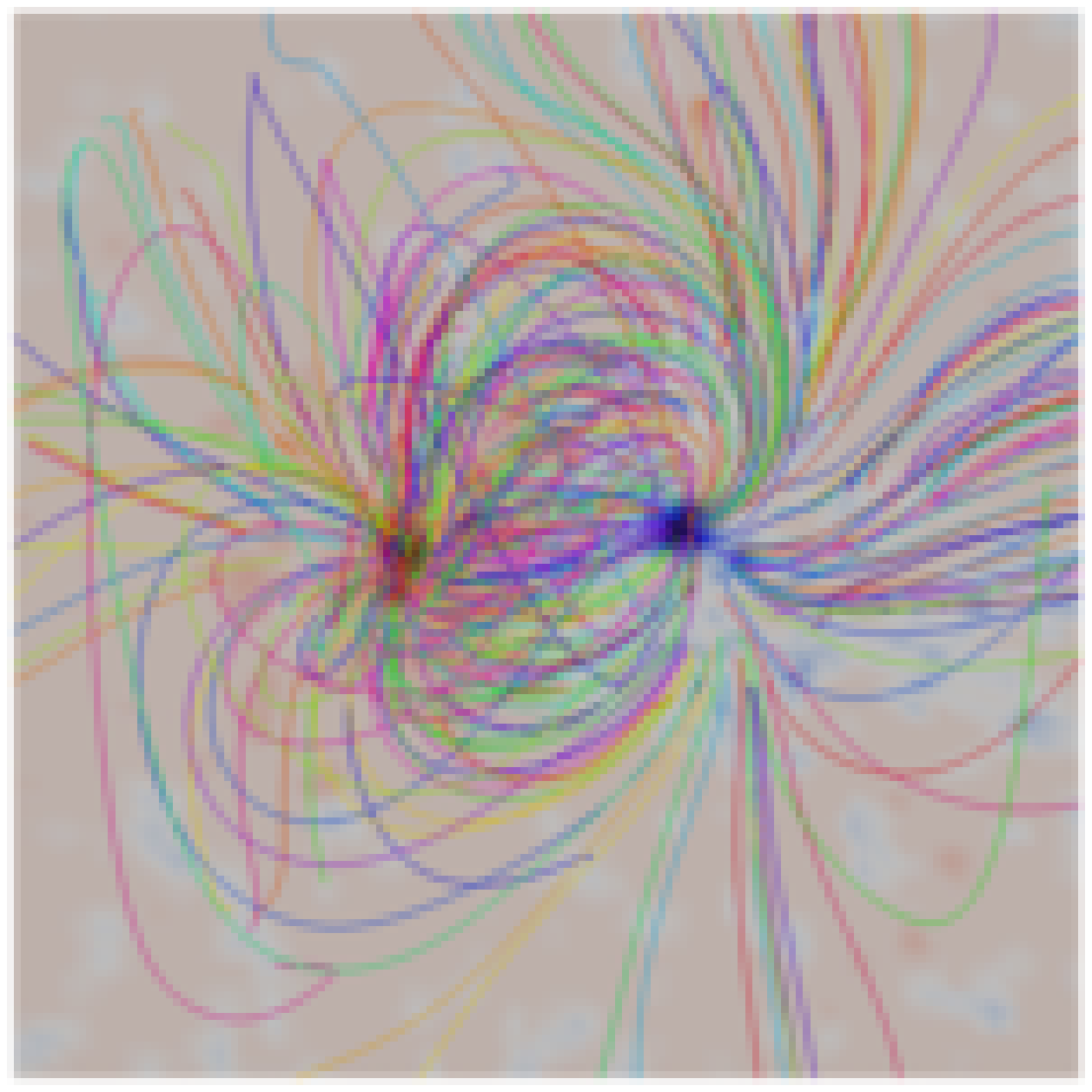}}
\end{center}
\caption{Oblique and top projections of selected representative 3D magnetic field lines of the NLFFF model for AR~10987 on March 27.  On the lower boundary, the vertical magnetic flux distribution is indicated by red (positive field) and blue (negative field) coloring.  The colors of the field lines are not significant.}
\label{fig:flines10987}
\end{figure}

\begin{figure} 
\begin{center}
\resizebox{0.75\textwidth}{!}{\includegraphics*[100,90][700,700]{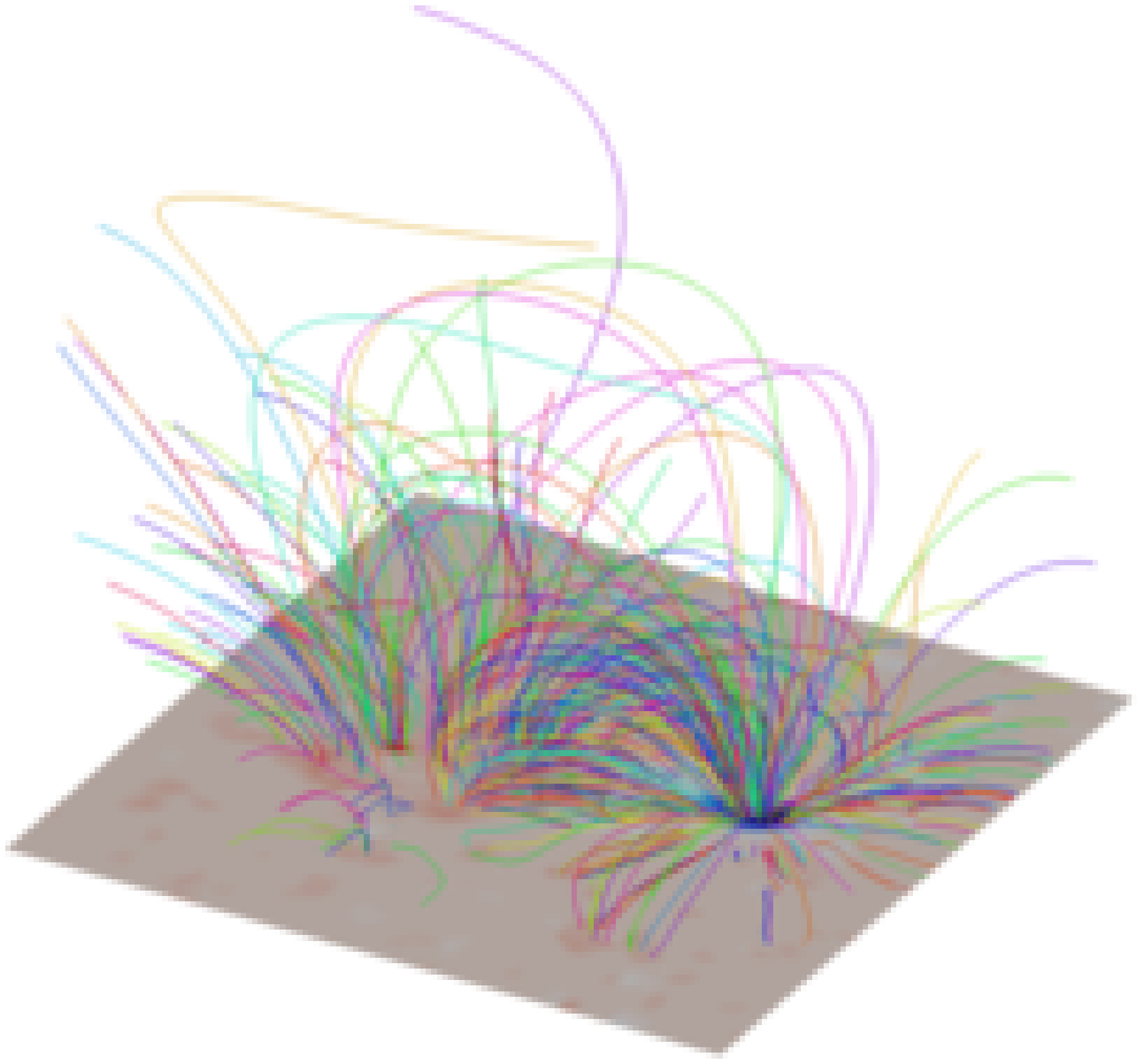}}
\resizebox{0.75\textwidth}{!}{\includegraphics*[150,150][650,650]{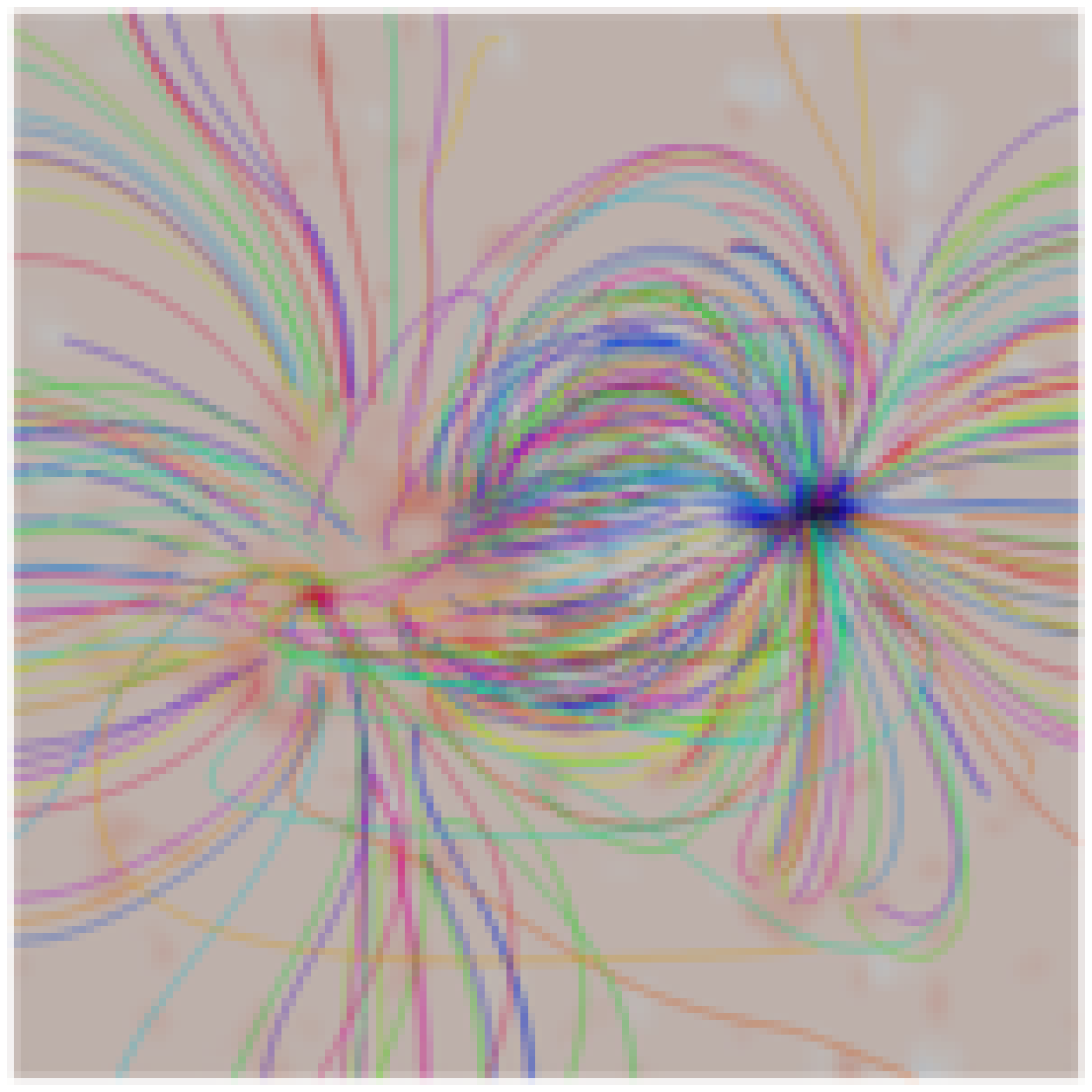}}
\end{center}
\caption{Oblique and top projections of selected 3D magnetic field lines of the NLFFF model for AR~10988 on March 29.  On the lower boundary, the vertical magnetic flux distribution is indicated by red (positive field) and blue (negative field) coloring.  The colors of the field lines are not significant.}
\label{fig:flines10988}
\end{figure}

\begin{figure} 
\begin{center}
\resizebox{0.75\textwidth}{!}{\includegraphics*[100,90][700,700]{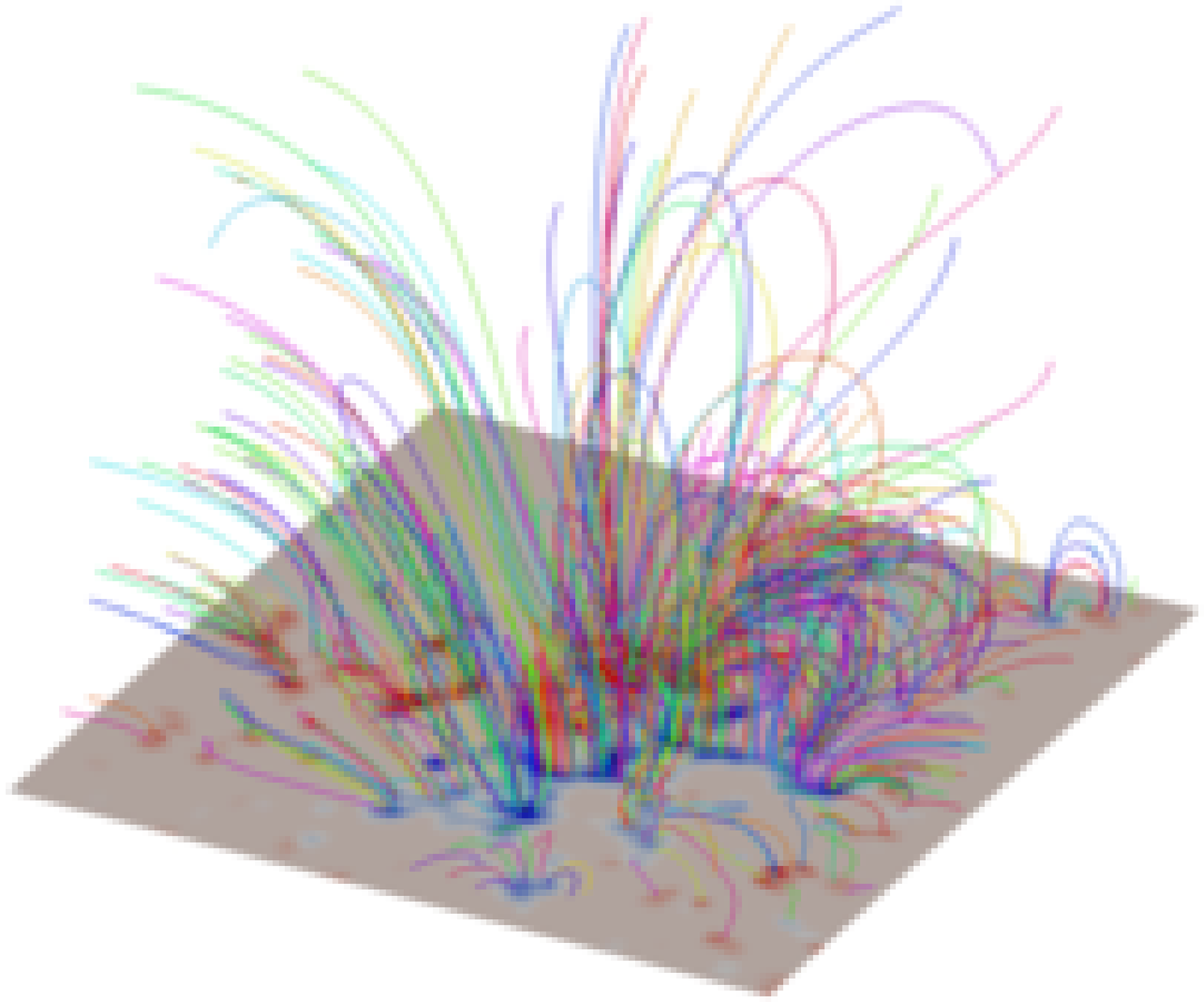}}
\resizebox{0.75\textwidth}{!}{\includegraphics*[150,150][650,650]{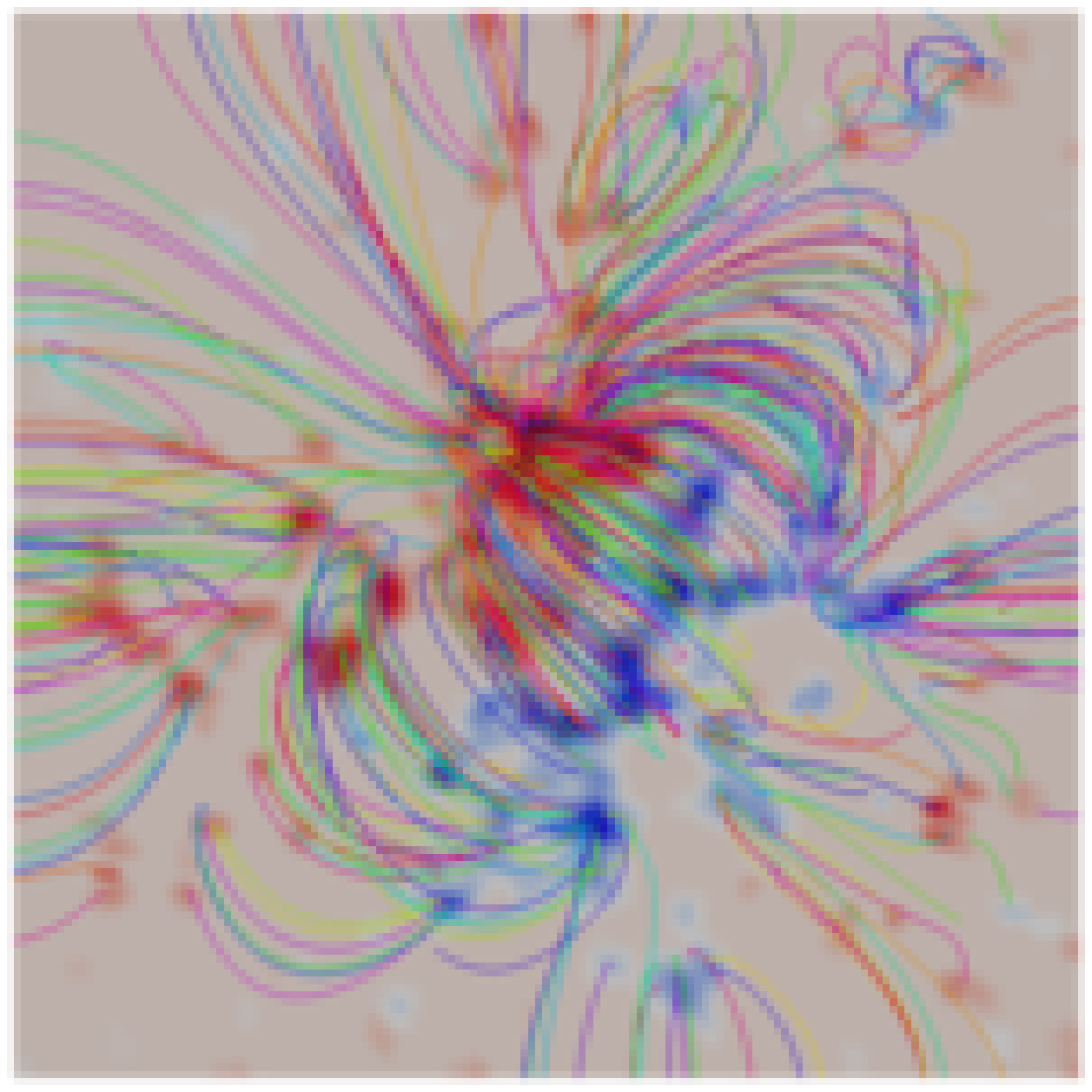}}
\end{center}
\caption{Oblique and top projections of selected representative 3D magnetic field lines of the NLFFF model for AR~10989 on March 31.  On the lower boundary, the vertical magnetic flux distribution is indicated by red (positive field) and blue (negative field) coloring.  The colors of the field lines are not significant.}
\label{fig:flines10989}
\end{figure}

\begin{figure} 
\begin{center}
\resizebox{0.75\textwidth}{!}{\includegraphics*[100,90][700,700]{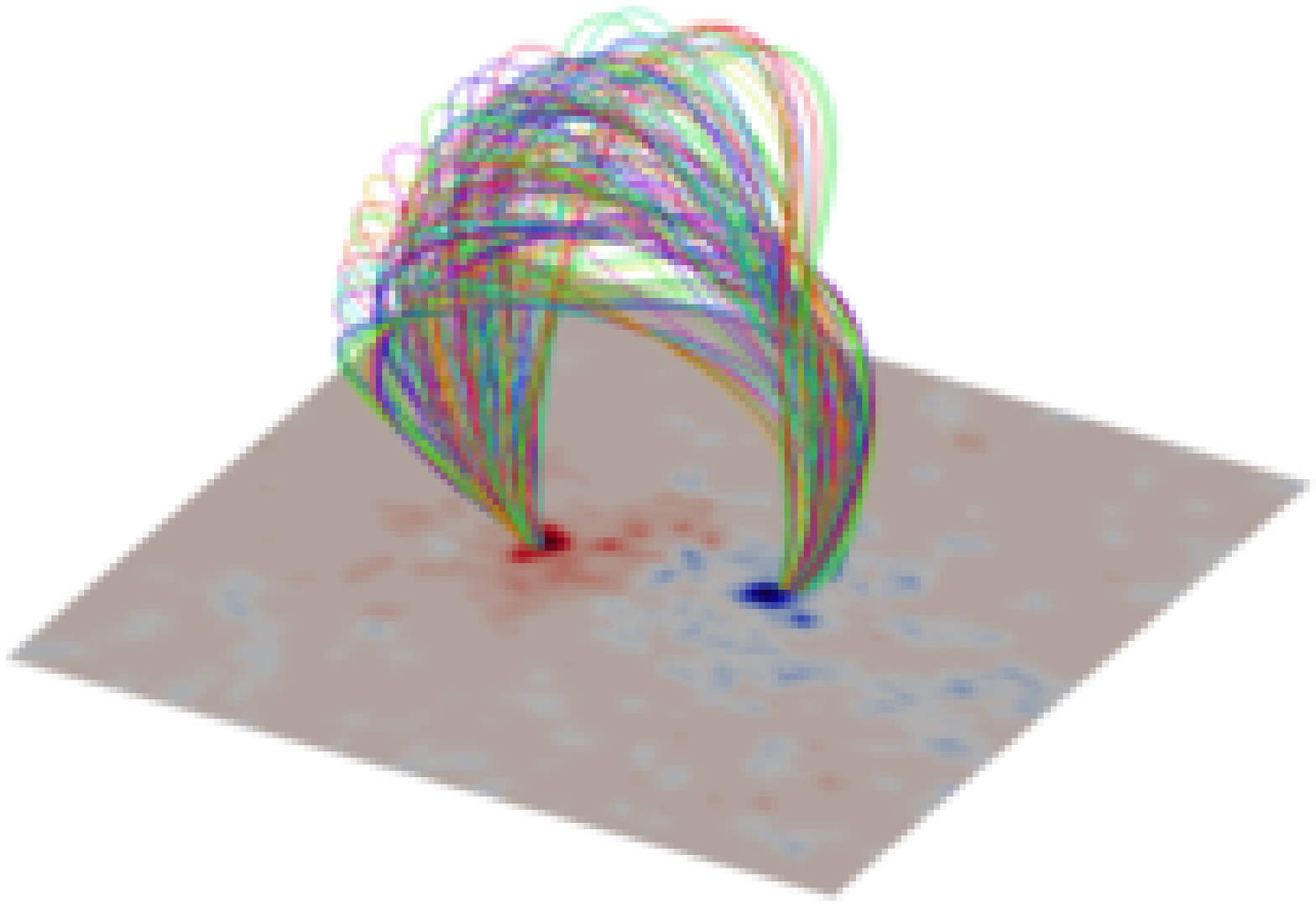}}
\resizebox{0.75\textwidth}{!}{\includegraphics*[150,150][650,650]{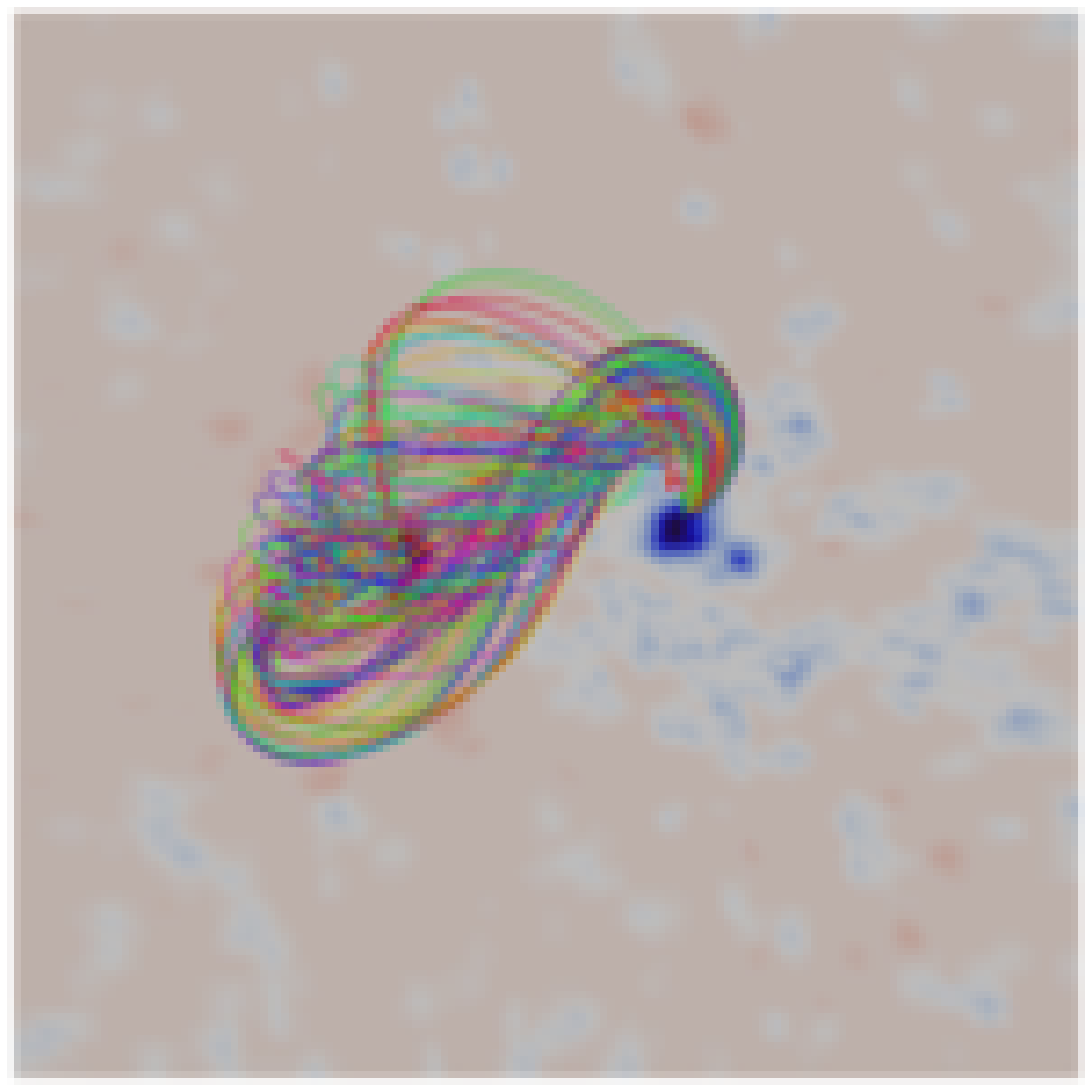}}
\end{center}
\caption{Oblique and top projections of selected  3D magnetic field lines for AR~10987 on March 27.  On the lower boundary, the vertical magnetic flux distribution is indicated by red (positive field) and blue (negative field) coloring.  The colors of the field lines are not significant.  For this plot the field lines represent the flux through an electric current concentration in the positive magnetic flux concentration.  The resulting field lines together resemble a sigmoidal or S-shaped structure.}
\label{fig:sigmoid10987}
\end{figure}

The three main active regions of the WHI have been compared and contrasted in terms of their CME productivity and flare activity.    The CMEs and other transient phenomena during the WHI have been surveyed by \inlinecite{WebbGT2010}, \inlinecite{Cremadesetal2010} and \inlinecite{Sterling2010}.  These surveys were carried out independently and used different data sources and analysis techniques.  \inlinecite{WebbGT2010} combined these surveys into a comprehensive catalog and \inlinecite{Webbetal2010} extended this catalog to CR~2067 and CR~2069.  About two thirds of the CMEs from CRs~2067-9 occurred during CR~2068 and most of those between March 21 and April 11.  According to GOES X-ray flux measurements significantly more flaring activity occurred during CR~2068 than during CR~2067 and CR~2069, mainly between March 24 and April 6 \cite{Webbetal2010}.  This flaring activity was low to moderate.  Even during the relatively active period of CR~2068 there were only three flares with peak fluxes above the X-ray C level ($10^{-6}$~W/m$^2$): an M1.7 flare on March 25, by far the most energetic flare during the WHI, and two C1.2 flares on April 3.  Most of the CMEs were associated with the three ARs~10987-9, in particular with AR~10989.  As Figure~10 in \inlinecite{Webbetal2010} shows, significantly more CMEs were traced to locations close to (mainly to the South of) AR~10989 than to AR~10987 and AR~10988.  The two largest CMEs during the WHI occurred on March~25 and April~9.  The March~25 event, associated with the M1.7 flare mentioned above, occurred at the East limb and was associated with AR~10989.  (The April~9 event was associated with a prominence eruption.)  AR~10989 was the most CME-productive of the three regions, producing an M2 flare on March 25th and various ejecta, while AR~10987 was less active and AR~10988 least active of all.   \inlinecite{WelschMcTiernan2010} augmented the standard GOES flare list with Sam Freeland's ``Latest Events'' database\footnote{http://www.lmsal.com/solarsoft/latest\_events }  and the RHESSI
flare catalog and found that the activity levels among the three active regions were more evenly balanced when the ``Latest Events'' database was used to associate more of the GOES flares to source active regions, and when RHESSI events were manually associated with the active regions.

According to several predictors of flare activity based on the photospheric magnetic field (total unsigned magnetic flux, magnetic flux near polarity inversion lines, amount of cancelled flux, proxy Poynting flux, relative magnetic helicity flux), \inlinecite{WelschMcTiernan2010} found that AR~10988 should have been the most active region of the three and AR~10989 the least.  Of the magnetic parameters they studied, the characteristic helicity, the dimensionless quantity defined as the cumulative relative magnetic helicity flux normalized by the square of the mean magnetic flux over the time the active region was observed, was the only one for which AR~10989 had a higher value than the other two regions.    Magnetic helicity is not easily dissipated in the corona \cite{Berger1984} and is thought to accumulate there until bodily removed by CMEs \cite{Low2001}.

We are fortunate that the SOLIS VSM was collecting full-disk vector-field measurements daily during the disk passage of the three active regions.  We therefore have a time series of measurements including an observation close to disk-center for each region.  While the three regions appear quite similar in the synoptic magnetogram for rotation 2068 shown in Figure~\ref{fig:synmags}, closer examination reveals major differences between them.  The top pictures in Figures~\ref{fig:vec10987}-\ref{fig:vec10989} show SOLIS vector magnetograms for the three active regions.  Each region is represented by the magnetogram closest to central meridian.  AR~10987 is a bipolar region with approximately equally strong concentrations of positive and negative flux.  AR~10988 has a strong negative leading polarity, the most intense field in the triplet of regions, while the trailing positive flux is significantly weaker and more diffuse.  Finally, AR~10989 is the weakest and most diffuse region.  It has no flux concentration comparable to those of the other two regions.  These patterns are also found in the GONG continuum intensity images (not shown).  AR~10987 has a sunspot of each polarity and they are of approximately equal size.  AR~10988 has a large leading-polarity sunspot, the largest among the three regions, but no comparable trailing sunspot.  AR~10989 has no significant sunspot.  The vertical unsigned and net magnetic fluxes, $F_{tot}$ and $F_{net}$, associated with the magnetograms in Figures~\ref{fig:vec10987}-\ref{fig:vec10989} are recorded in Table~\ref{NLFFFtable}.  ARs~10987 and 10988 have substantial flux imbalances, over 30\% and just under 20\%.  These imbalances are of opposite signs, hinting at some magnetic connection between the two regions.  Meanwhile, AR~10989 is more nearly flux-balanced with an imbalance of just under 10\%.

The vertical electric current maps in the bottom pictures of Figures~\ref{fig:vec10987}-\ref{fig:vec10989} also show current concentrations of approximately equal size in AR~10987, a strong leading current concentration and weak, diffuse following current in AR~10988 and no significant current structure in AR~10989.  These maps also show that the vertical electric current is well balanced in each case.  Therefore the electrical currents must flow in both directions along the field trajectories, from positive to negative polarity and vice versa, in approximately equal quantities.  Connection due to flux imbalance thus corresponds to non current carrying field lines between the three active regions.  


\begin{table}
\scriptsize
\caption{Properties of the field measurements and NLFFF models of ARs~10987-9 in Figures~\ref{fig:flines10987}-\ref{fig:flines10989}.  The parameter $CW\!sin$ is the current-weighted average of the sine
of the angle between the current density and the magnetic field.}
\begin{tabular}{lccc}
\hline\hline
 & AR~10987 & AR~10988 & AR~10989 \\
 \hline
Total magnetic flux $F_{tot}$~(Mx)  & $2.3\times 10^{22}$ & $2.2\times 10^{22}$ & $1.1\times 10^{22}$\\
Net magnetic flux $F_{net}$~(Mx) & $-8.0\times 10^{21}$ & $4.1\times 10^{21}$ & $9.7\times 10^{20}$\\
Potential field magnetic energy $W_{\pi}$~(erg) & $2.9\times 10^{32}$ & $3.8\times 10^{32}$ & $6.0\times 10^{31}$\\
NLFFF magnetic energy $W[{\bf B}]$~(erg) & $3.1\times 10^{32}$ & $4.0\times 10^{32}$ & $6.1\times 10^{31}$\\
Normalized magnetic energy $W[{\bf B}]/W_{\pi}$ & 1.07 & 1.05 & 1.02\\
Relative magnetic helicity $\Delta H$~(G$^2$cm$^4$) & $2.9\times 10^{42}$ & $2.0\times 10^{42}$ & $2.2\times 10^{41}$\\
Characteristic helicity $\Delta H/F_{tot}^2$ & $5.5\times 10^{-3}$ & $4.1\times 10^{-3}$ & $1.7\times 10^{-3}$\\
$CW\!sin$ (see caption) & 0.05 & 0.07 & 0.1\\
\hline
\end{tabular}
\label{NLFFFtable}
\end{table}

Figures~\ref{fig:flines10987}-\ref{fig:flines10989} shows plots of representative field lines from nonlinear force-free models of the three regions.  These plots contain the same field of view as the  plots in Figures~\ref{fig:vec10987}-\ref{fig:vec10989}.  The model for AR~10987 includes some twisted lines connecting the two sunspots while the models for AR~10988 and AR~10989 appear less twisted.  Like AR~10987, AR~10988 has a simple bipolar appearance with long loops connecting fairly localized polarities.  AR~10989, on the other hand, has more dispersed magnetic polarities and many short loops distributed over a large area.  Table~\ref{NLFFFtable} shows selected physical parameter values for the magnetograms and models.  The magnetic energy for a field $\bf B$ is defined by the volume integral $\int (|{\bf B}|^2)/8\pi dV$.  $W_{\pi}$  and W [{\bf B}] are the energies associated with the unique potential field ${\bf B}_{\pi}$ determined from the boundary data for the vertical field component and the NLFF field $\bf B$, and
the difference between them is the free magnetic energy, i.e., the maximum energy
that can be released during a flare and/or a CME.  ARs~10987 and 10988 have approximately equal unsigned magnetic flux, and significantly more than AR~10989.  On the other hand, the model for AR~10988 has the most magnetic energy, significantly more than the model for AR~10987, and the model for AR~10989 has the least.  We estimate from the PFSS model for CR~2068 that the global coronal field (out to 2.5 solar radii) contains about $1.8\times 10^{33}$~ergs of magnetic energy, about 40\% of which is accounted for by the NLFFF models for the active regions.  AR~10989 also has significantly less free magnetic energy than the other two regions which have approximately equal free magnetic energy.  AR~10987 does, however, contain proportionally the largest free magnetic energy.  AR~10987 also contains significantly more relative magnetic helicity than AR~10988 and AR~10989 contains very little relative magnetic helicity.  The relative magnetic helicity can easily be computed using
XTRAPOL, which solves for the vector potential $\bf A$, via the volume integral
$\Delta H = \int ({\bf A} + {\bf A}_{\pi} ) \cdot({\bf B} - {\bf B}_{\pi} )\ dV $ \cite{FinnAntonsen1985}.  The characteristic helicity \cite{WelschMcTiernan2010}, the relative magnetic helicity normalized by the square of the mean magnetic flux, shows this difference between the regions more starkly.  The field of the AR~10987 model does indeed appear to be more twisted than the fields of the other two models.  Figure~\ref{fig:sigmoid10987} shows a subset of field lines whose foot-points are located in a strong current concentration in the negative sunspot.  These twisted, S-shaped lines highlight the twist in this active region's field structure.  According to the NLFFF models the other two regions did not include such twisted structure.  Nevertheless, according to the models, none of these regions contains enough free magnetic energy to power a major flare.

\inlinecite{Wheatlandetal2000} introduced a parameter measuring the force-freeness of a numerical model, also included in Table~\ref{NLFFFtable}.  The quantity $CW\!sin$ is the current-weighted average of the sine
of the angle between the current density and the magnetic field. This parameter
is equal to zero for an exact force-free magnetic field.  For the three models for ARs~10987-9 plotted in  Figures~\ref{fig:flines10987}-\ref{fig:flines10989} the $CWsin$ parameter took reasonable values.

\begin{figure} 
\begin{center}
\includegraphics[width=0.45\textwidth]{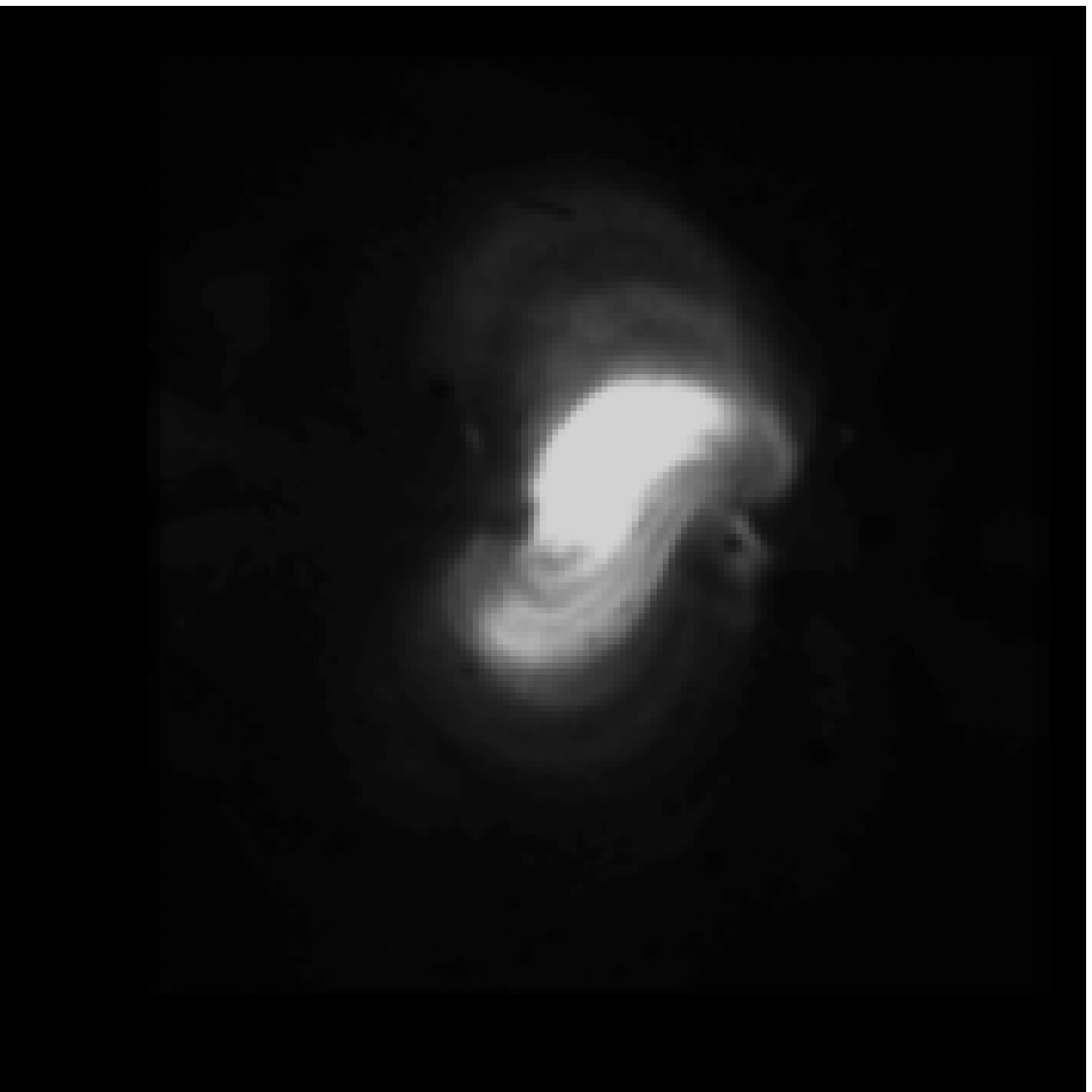}
\includegraphics[width=0.45\textwidth]{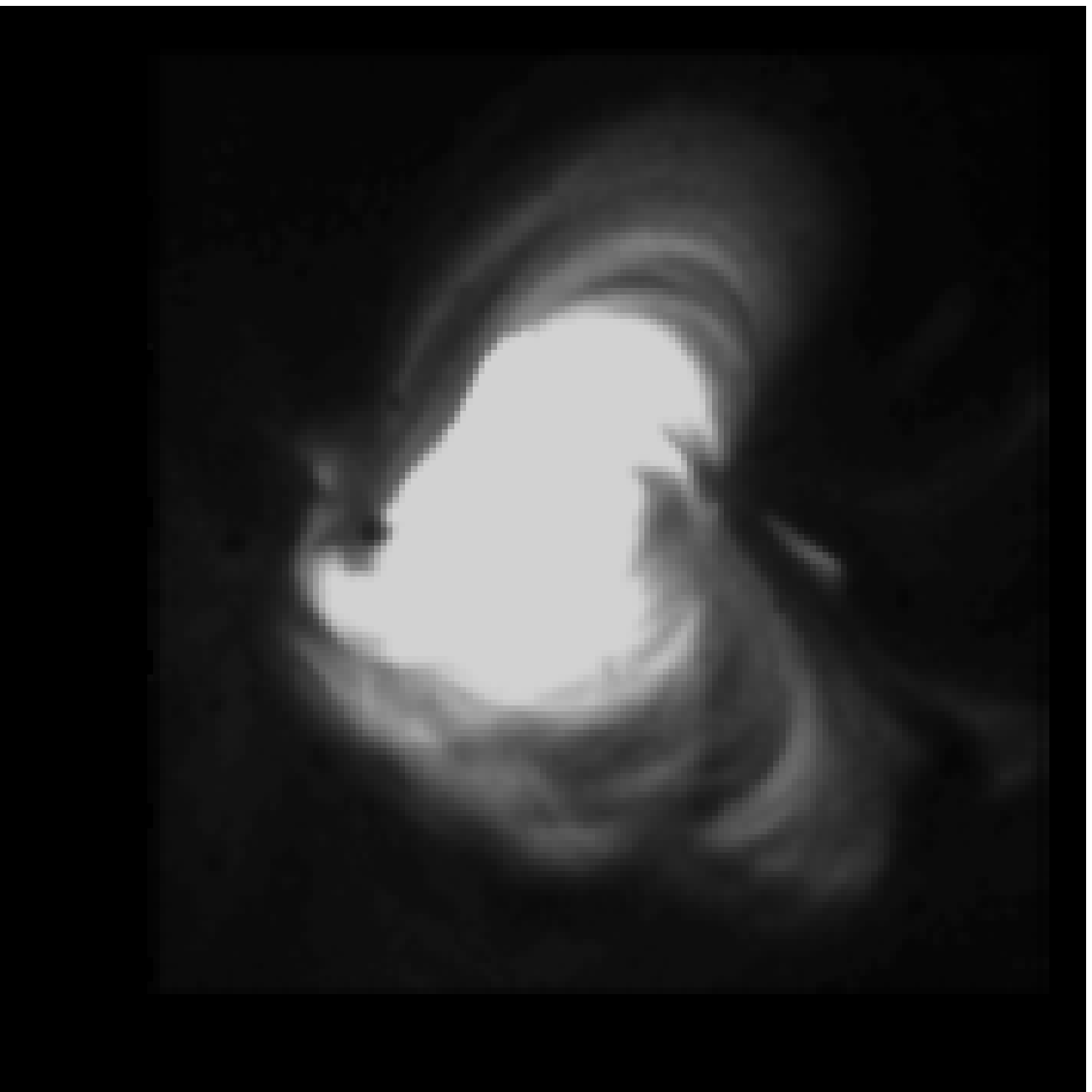}
\includegraphics[width=0.45\textwidth]{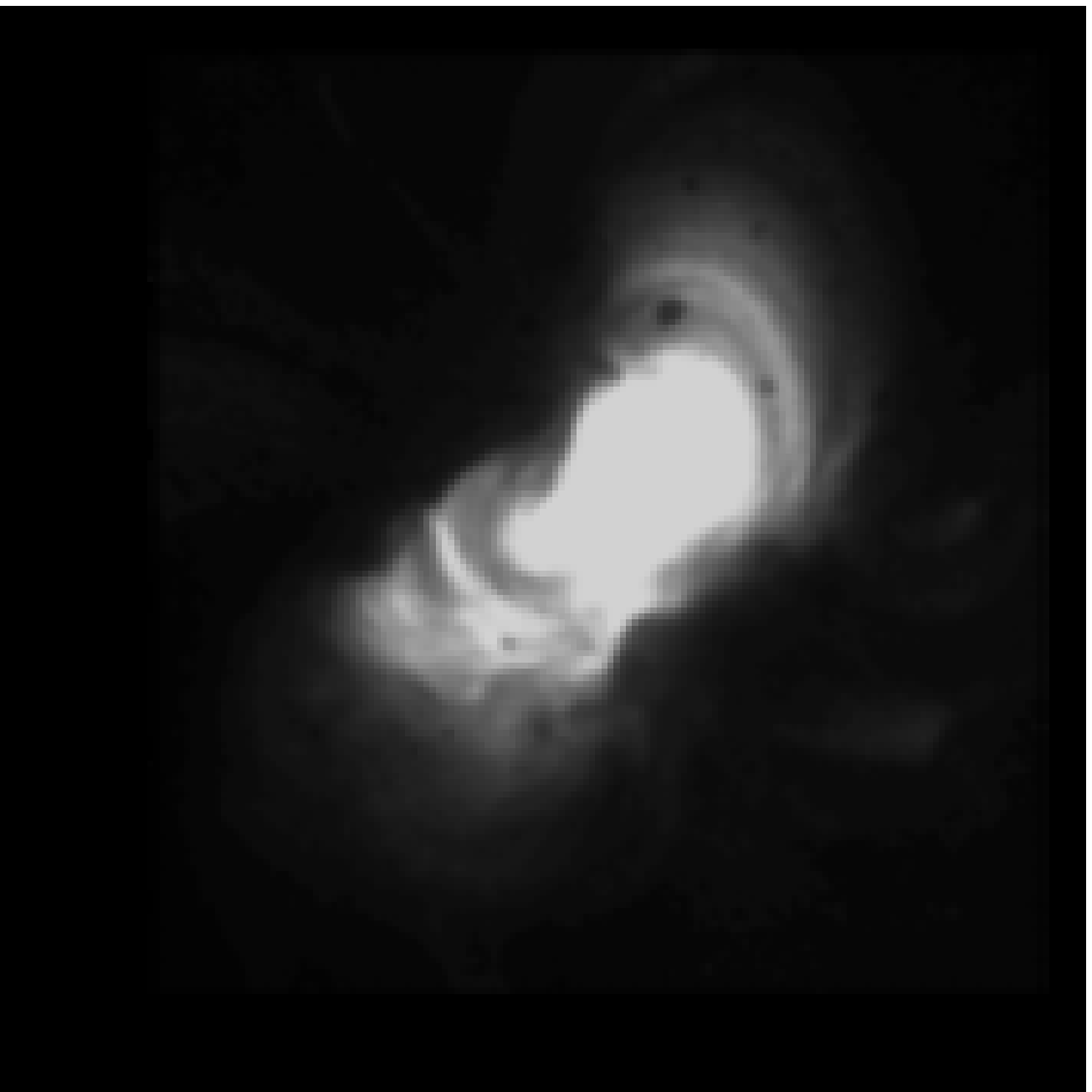}
\end{center}
\caption{Hinode XRT images of AR~10987 on 2008 March 27 at 0617~UT (top left), AR~10988 on 2008 March 29 at 1759~UT (top right) and AR~10989 on 2008 March 31 at 1802~UT (bottom).}
\label{fig:hinodears}
\end{figure}



\begin{figure} 
\begin{center}
\includegraphics[width=0.45\textwidth]{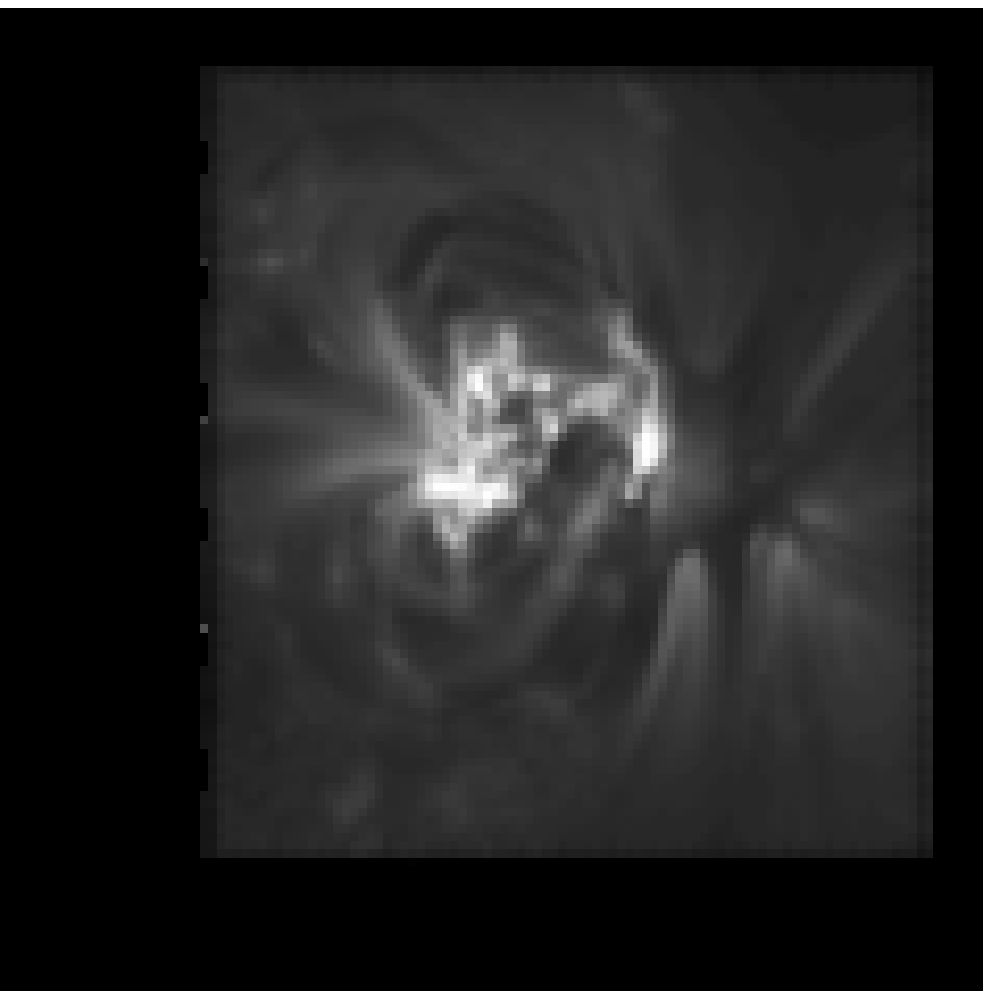}
\includegraphics[width=0.45\textwidth]{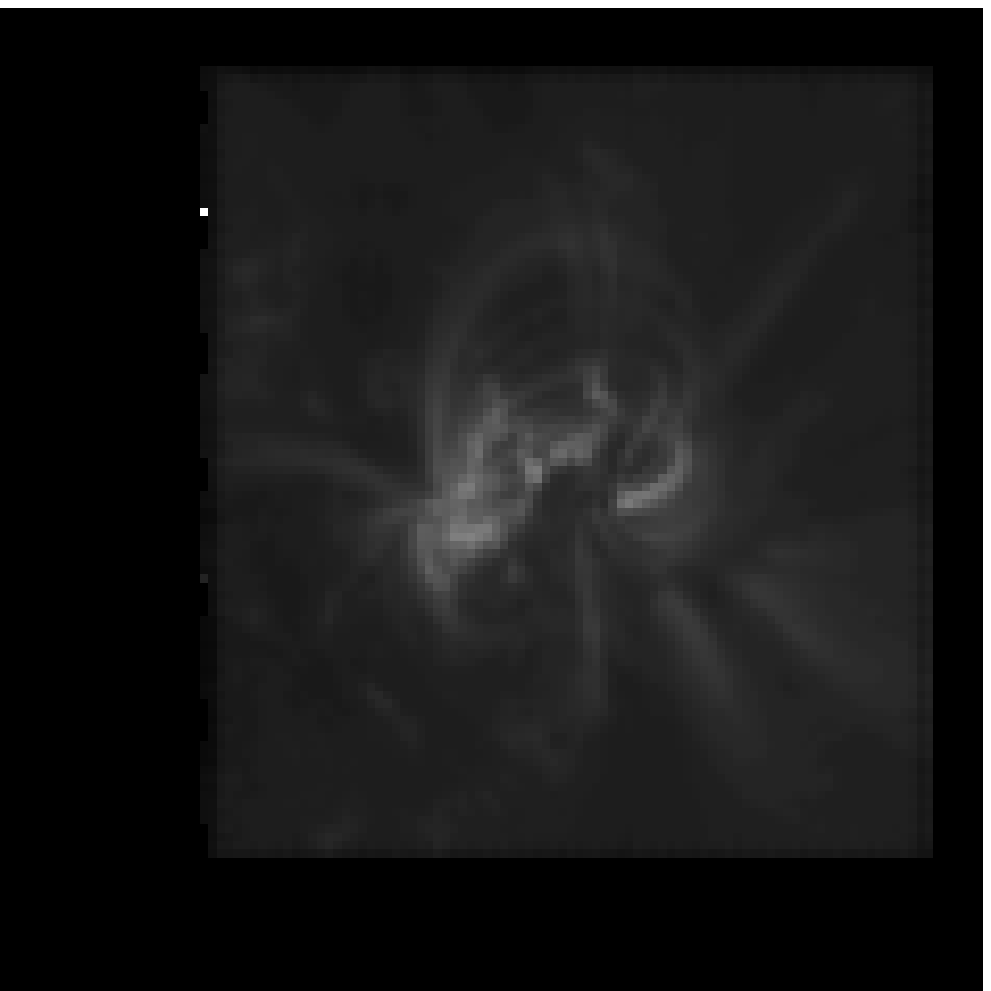}
\includegraphics[width=0.45\textwidth]{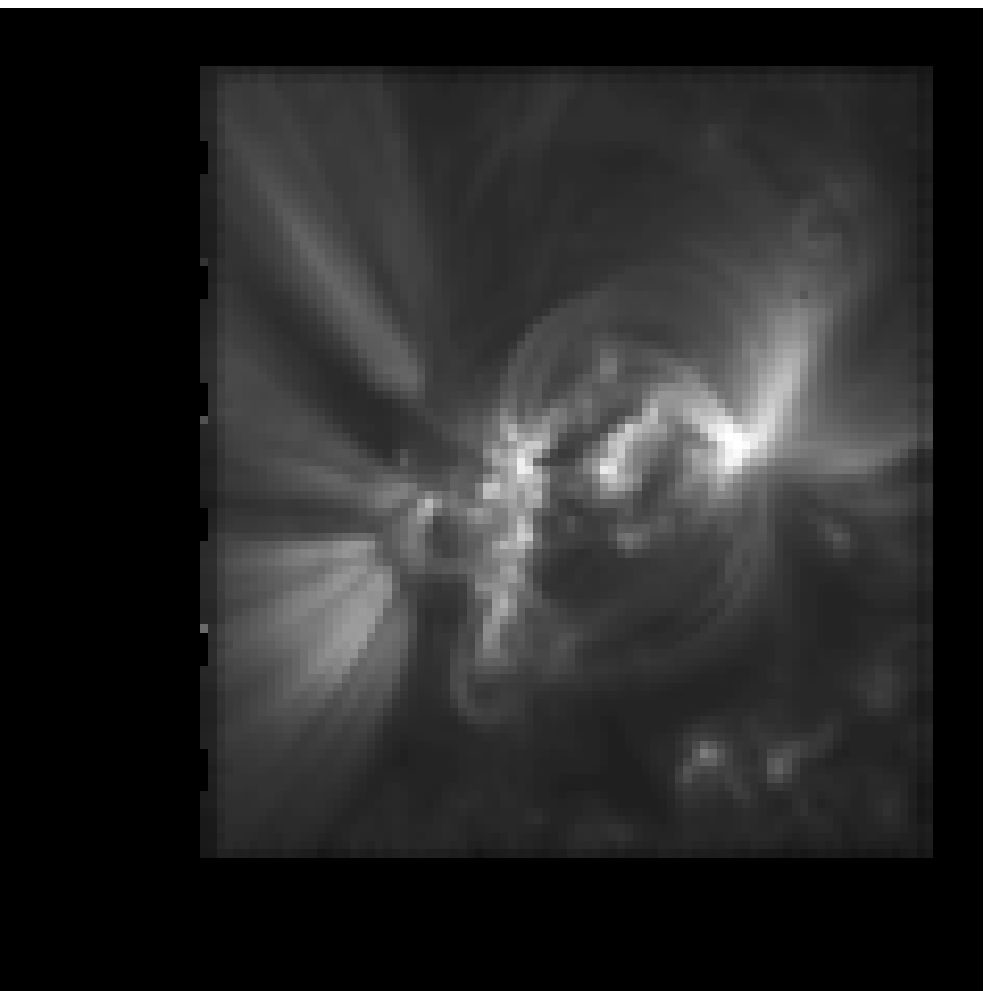}
\includegraphics[width=0.45\textwidth]{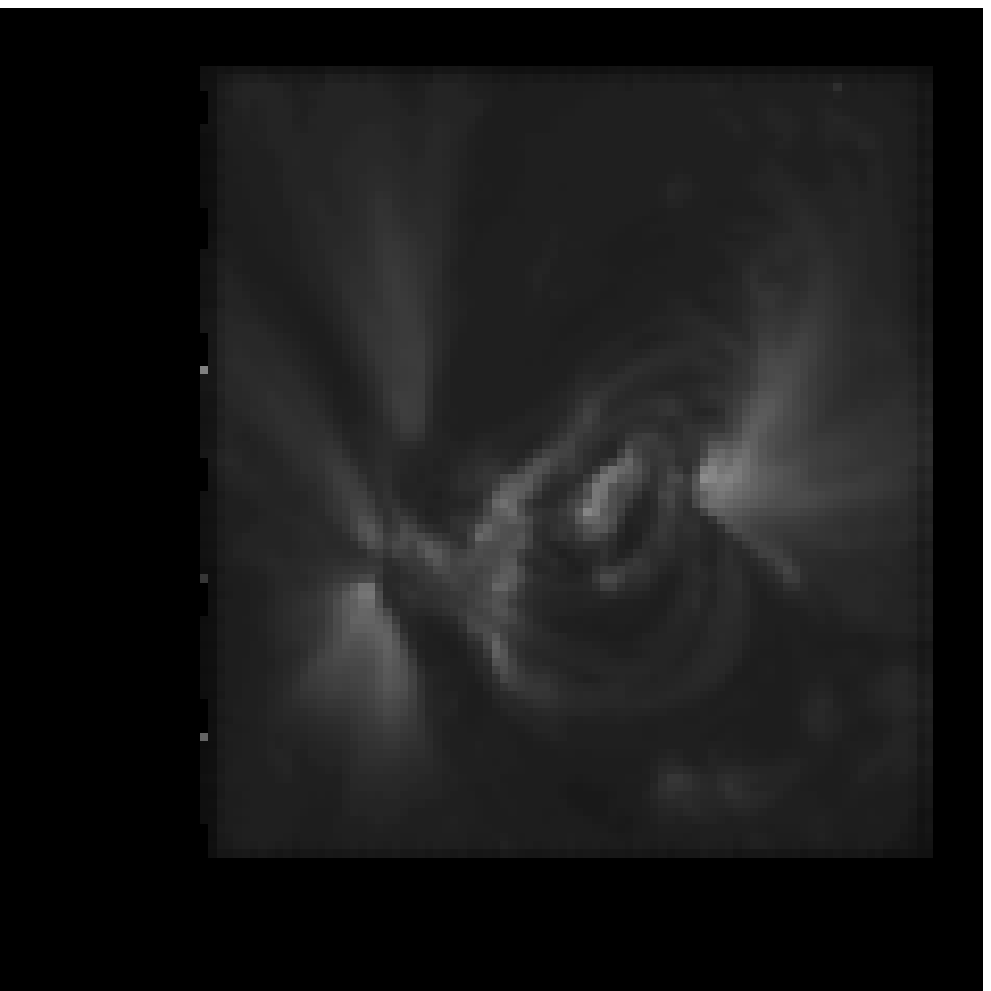}
\includegraphics[width=0.45\textwidth]{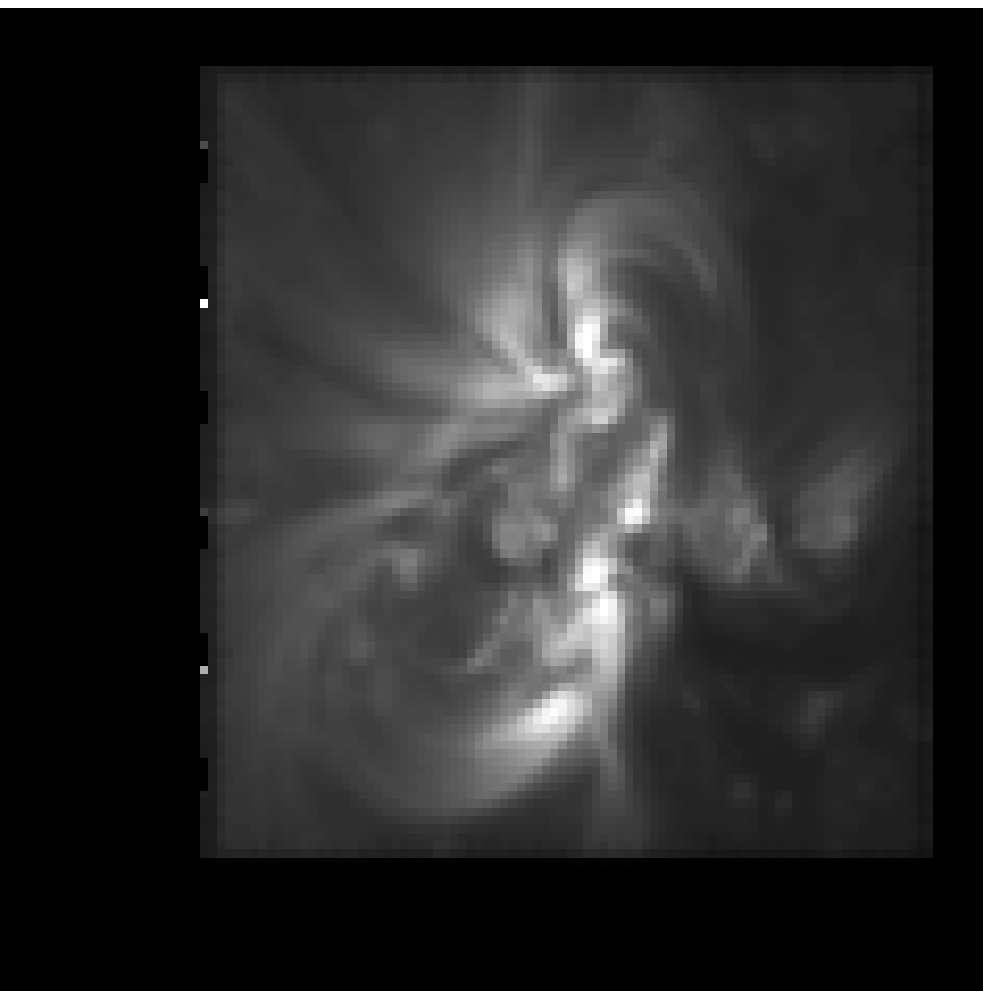}
\includegraphics[width=0.45\textwidth]{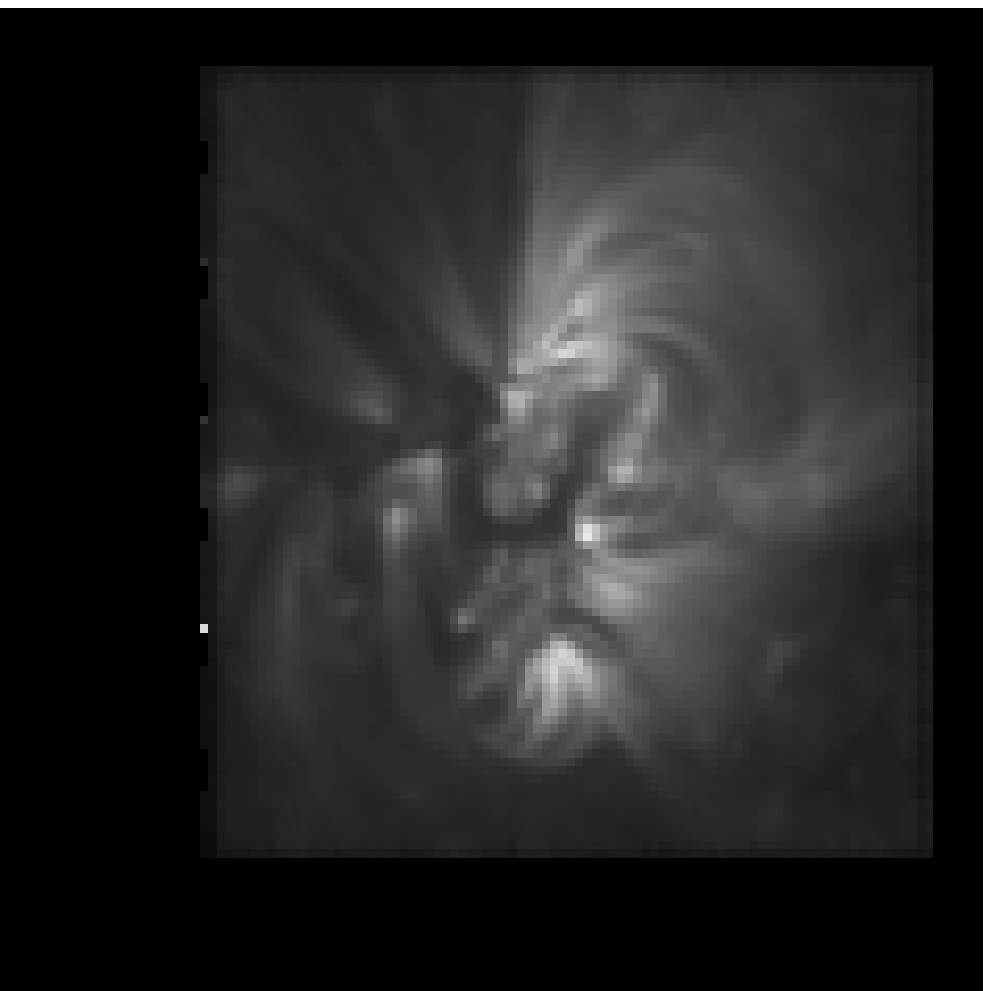}
\end{center}
\caption{STEREO/SECCHI/EUVI 171~\AA\   images for AR~10987 on 2008 March 27 at 1536~UT (top pictures), AR~10988 on 2008 March 29 at 1528 and 1526~UT (middle pictures) and AR~10989 on 2008 March 31 at 1741~UT (bottom pictures) from the ahead (left pictures) and behind (right pictures) spacecraft.}
\label{fig:stereoar171}
\end{figure}

\begin{figure} 
\begin{center}
\includegraphics[width=0.45\textwidth]{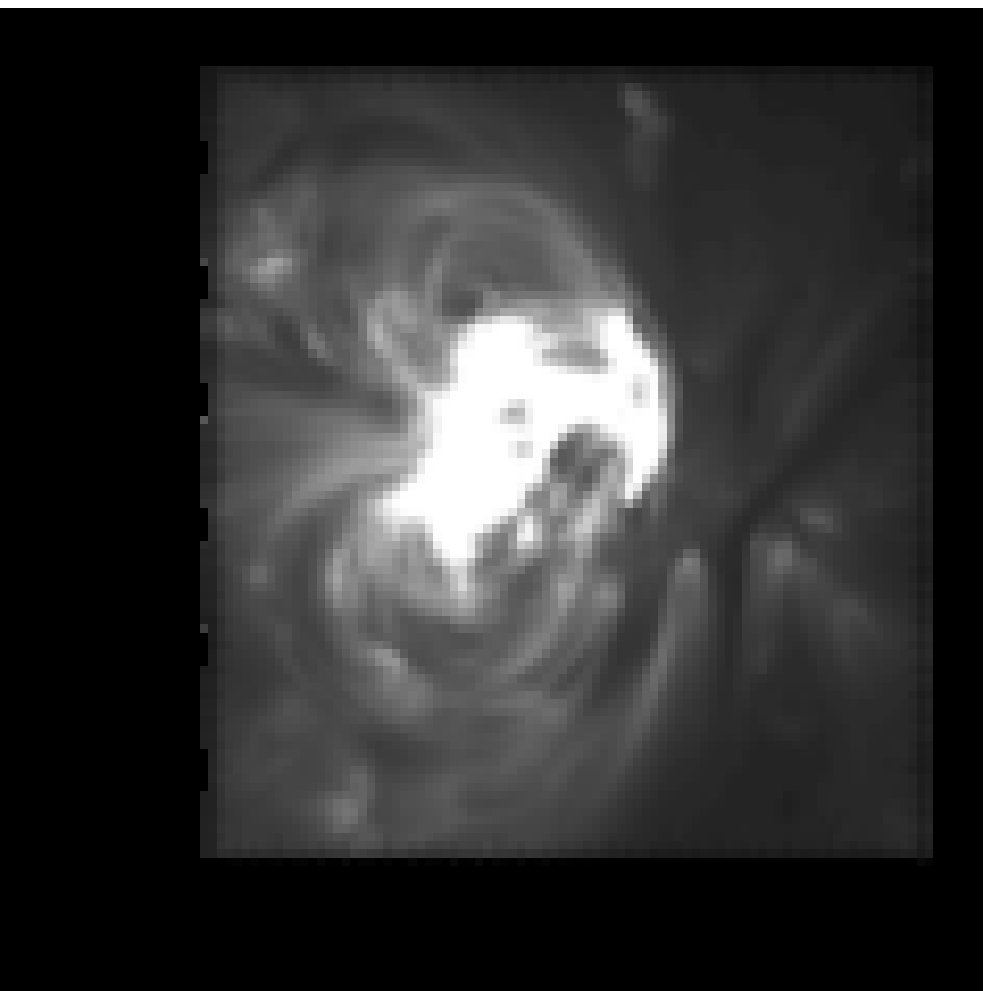}
\includegraphics[width=0.45\textwidth]{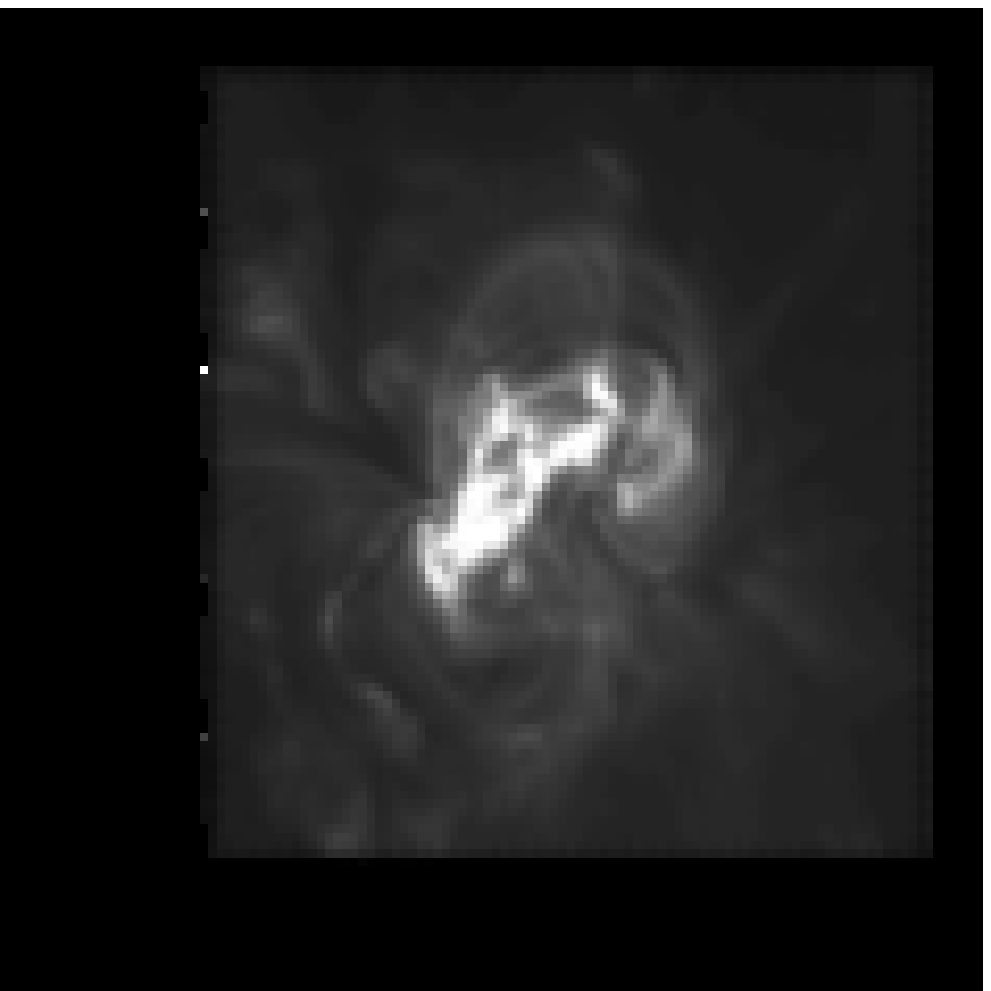}
\includegraphics[width=0.45\textwidth]{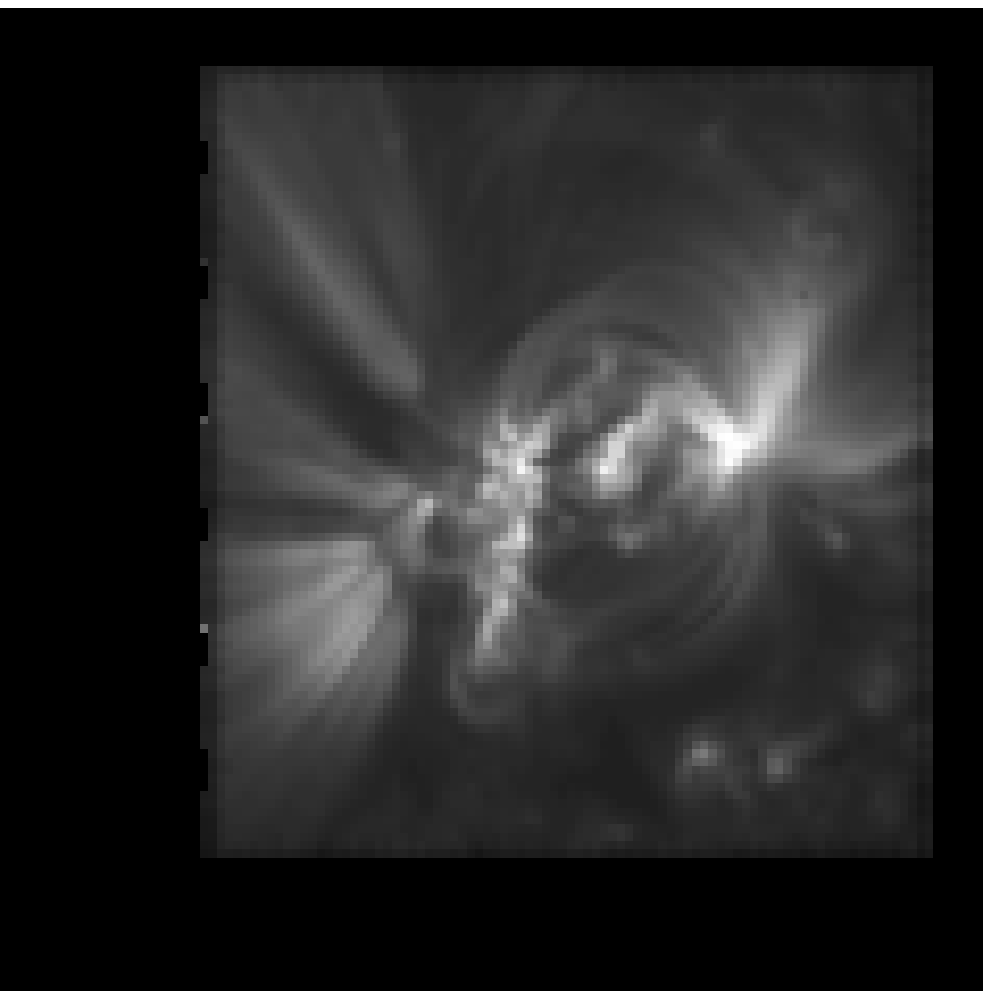}
\includegraphics[width=0.45\textwidth]{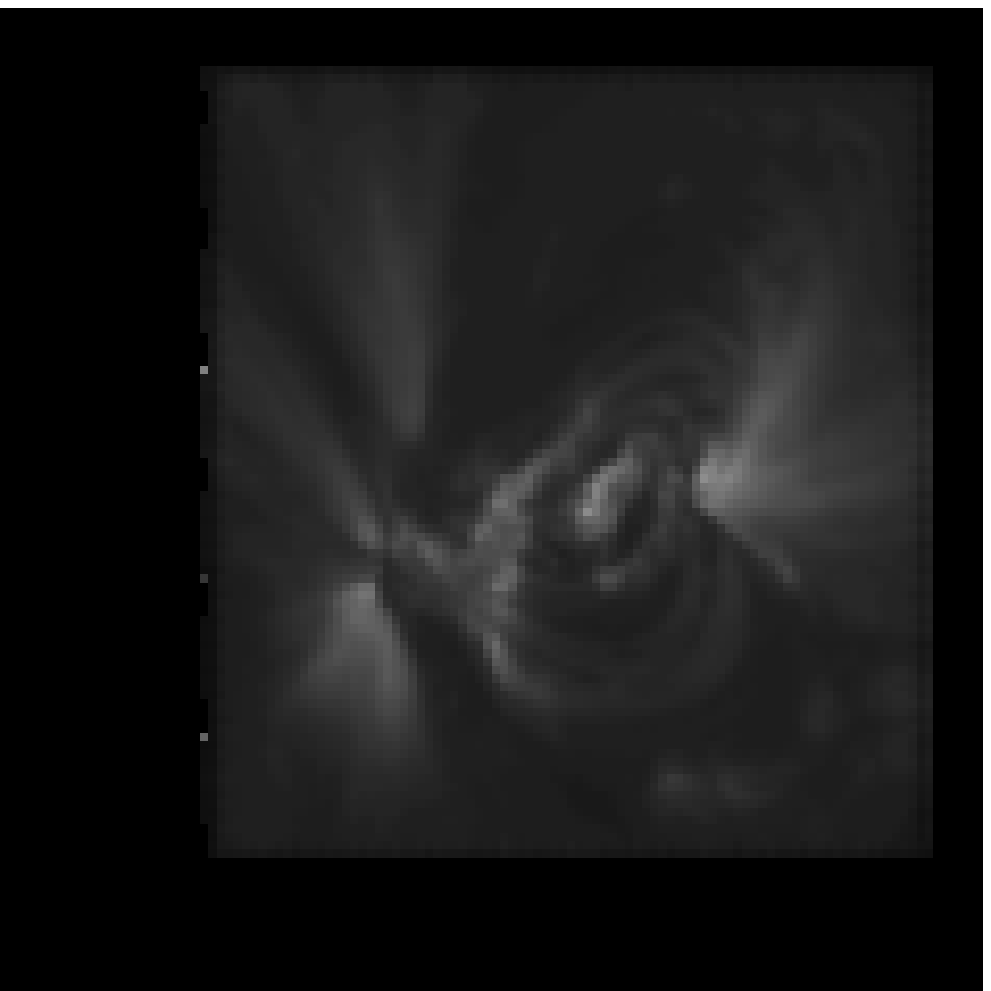}
\includegraphics[width=0.45\textwidth]{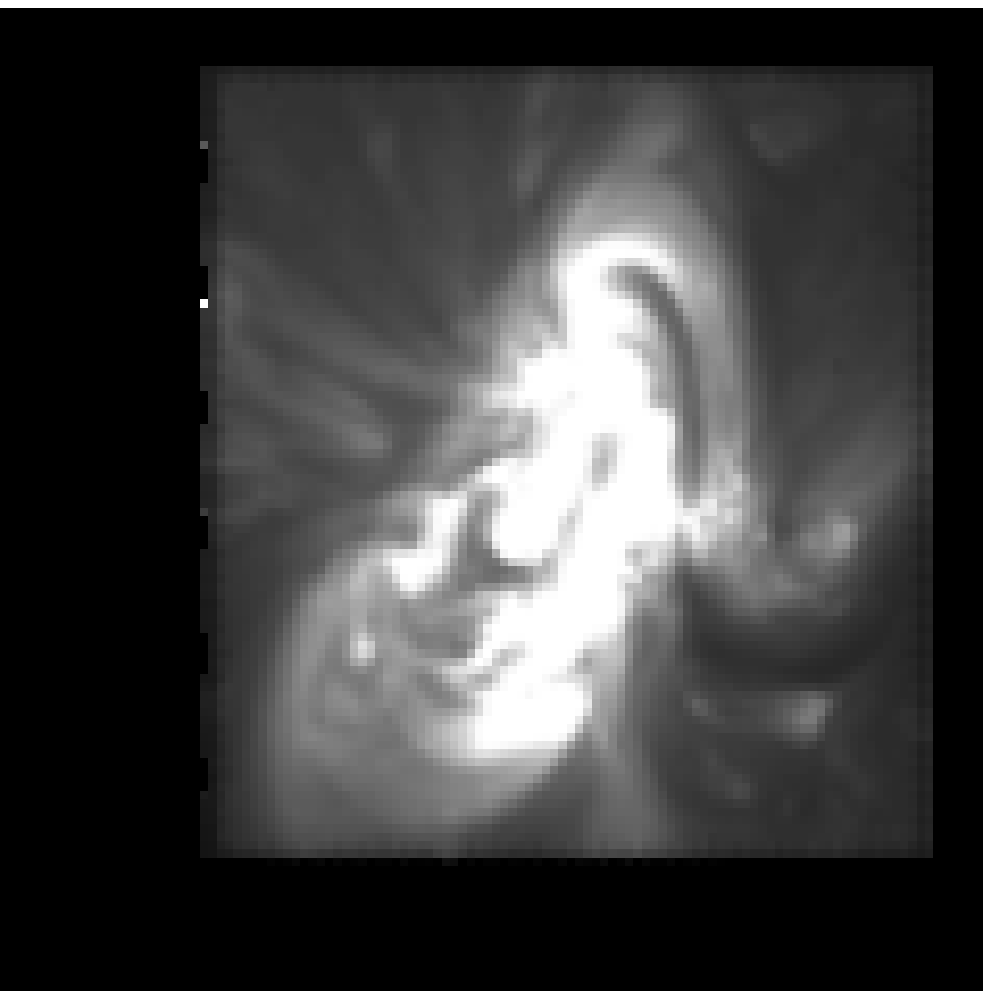}
\includegraphics[width=0.45\textwidth]{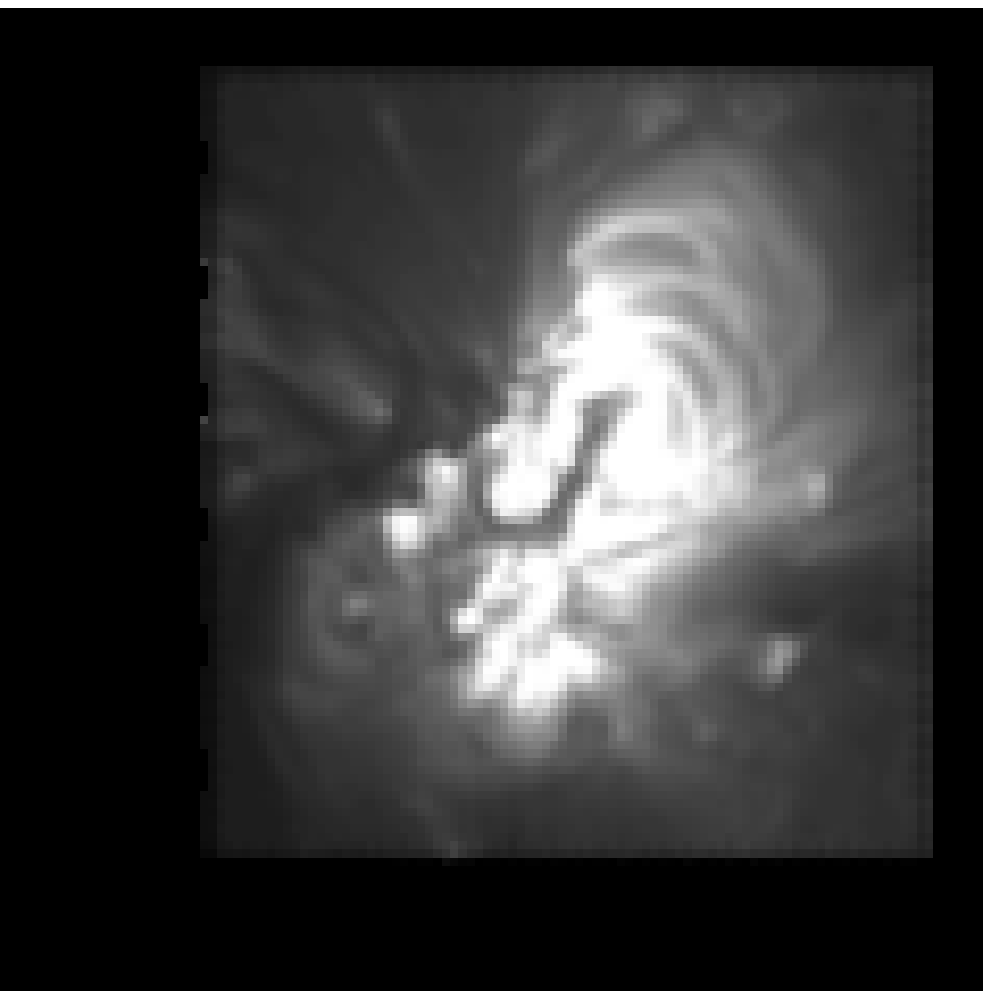}
\end{center}
\caption{STEREO/SECCHI/EUVI 195~\AA\   images for AR~10987 on 2008 March 27 at 1535~UT (top pictures), AR~10988 on 2008 March 29 at 1521~UT (middle pictures) and AR~10989 on 2008 March 31 at 1745~UT (bottom pictures) from the ahead (left pictures) and behind (right pictures) spacecraft.}
\label{fig:stereoar195}
\end{figure}

\begin{figure} 
\begin{center}
\includegraphics[width=0.45\textwidth]{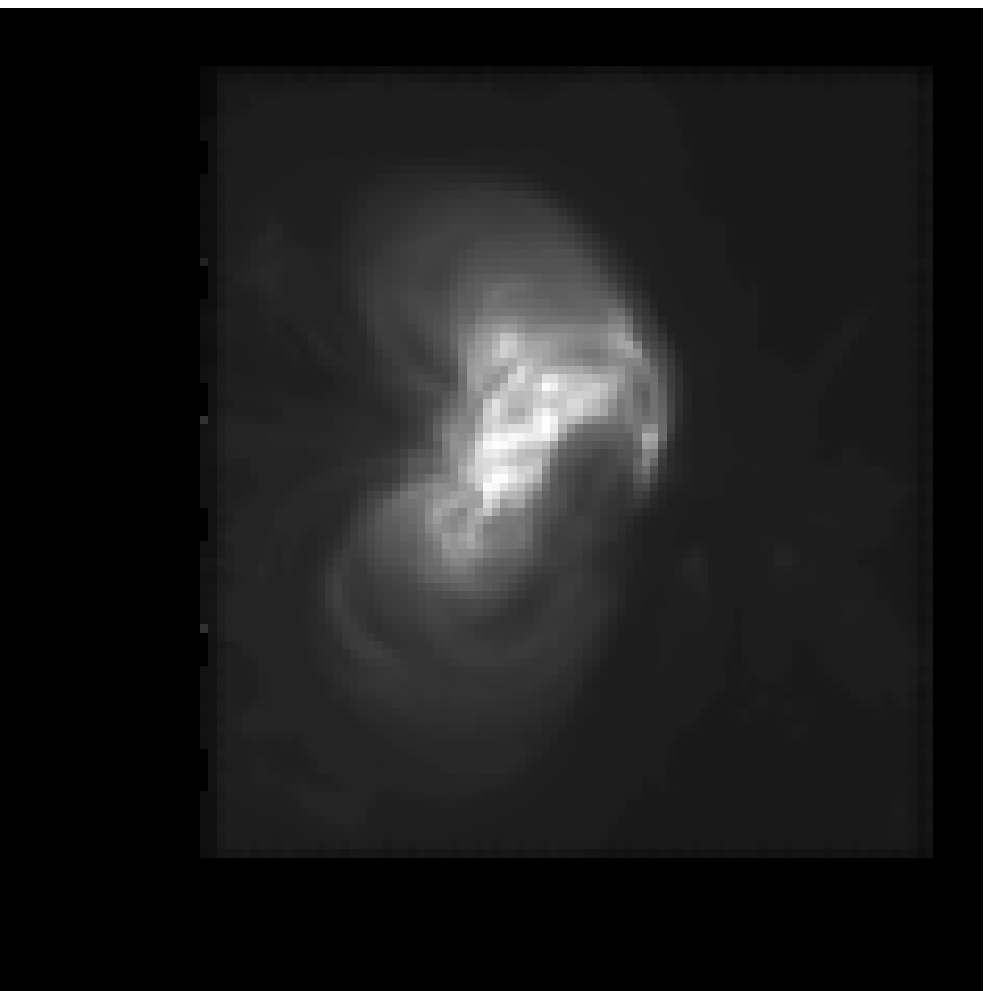}
\includegraphics[width=0.45\textwidth]{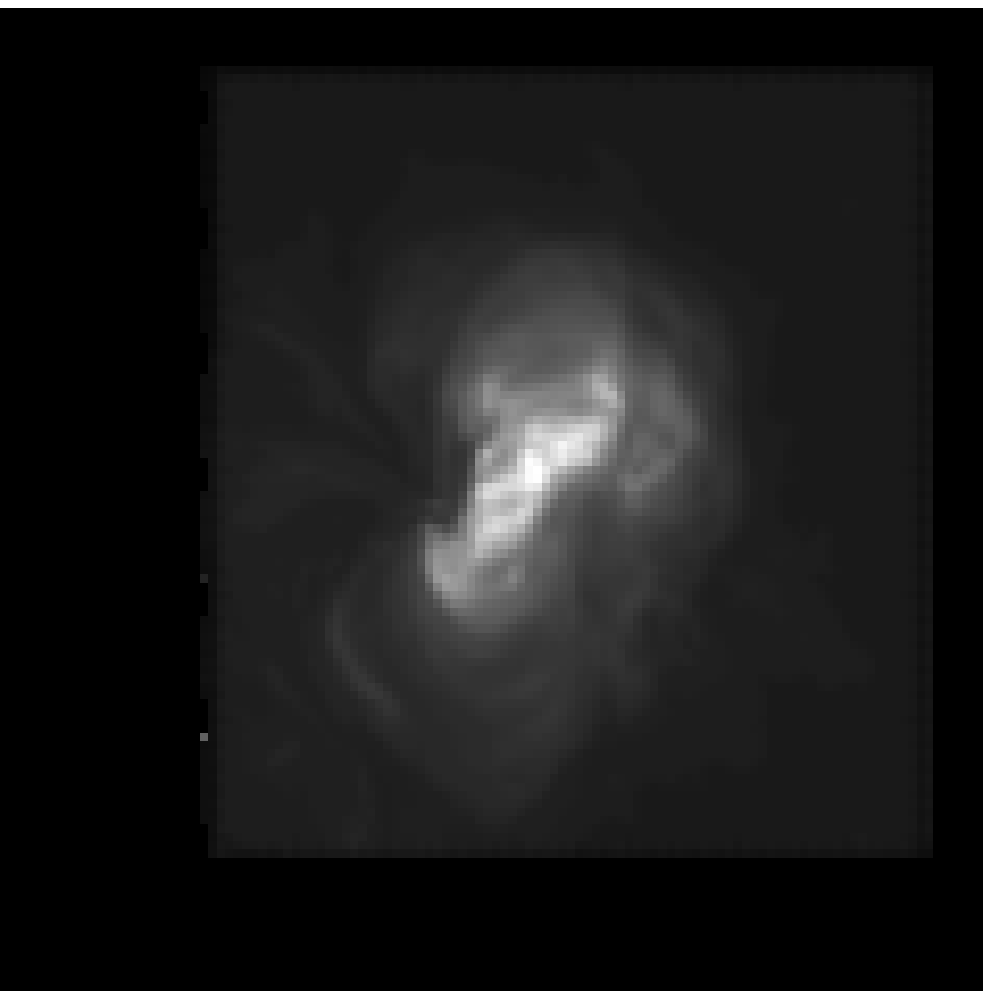}
\includegraphics[width=0.45\textwidth]{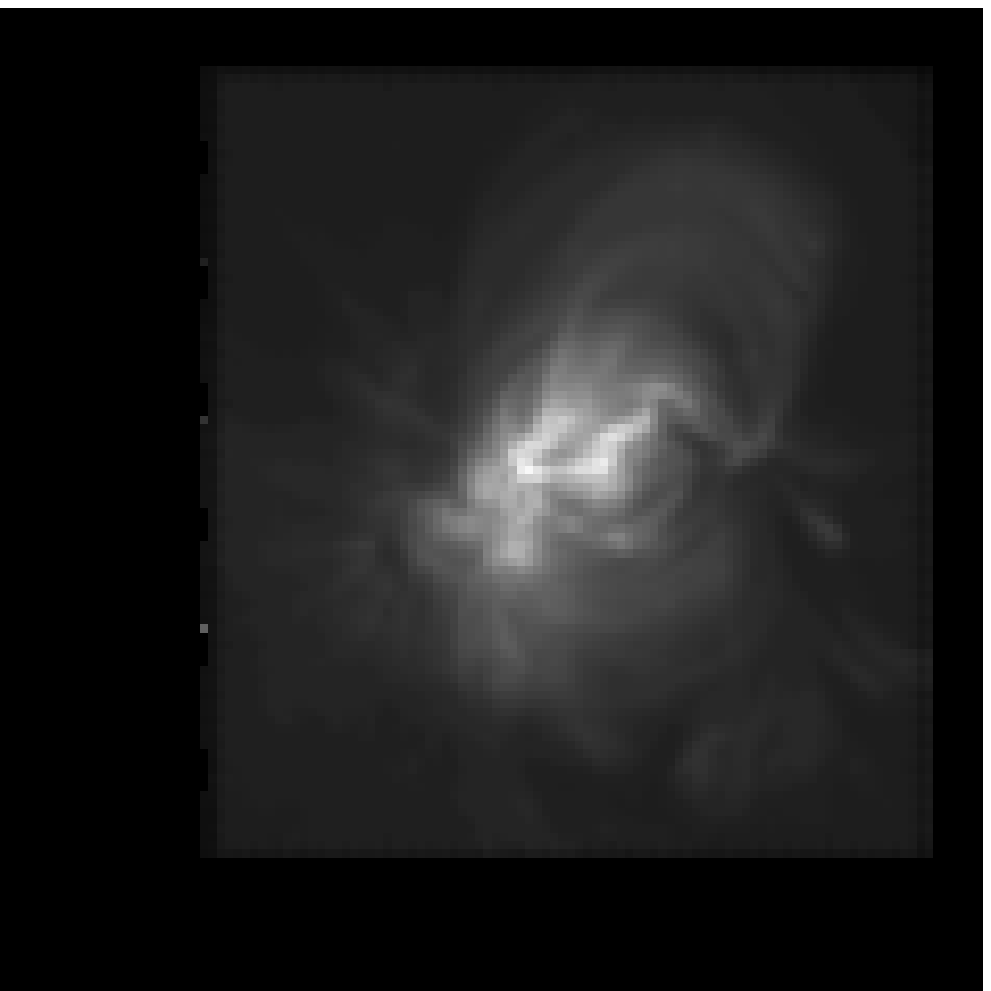}
\includegraphics[width=0.45\textwidth]{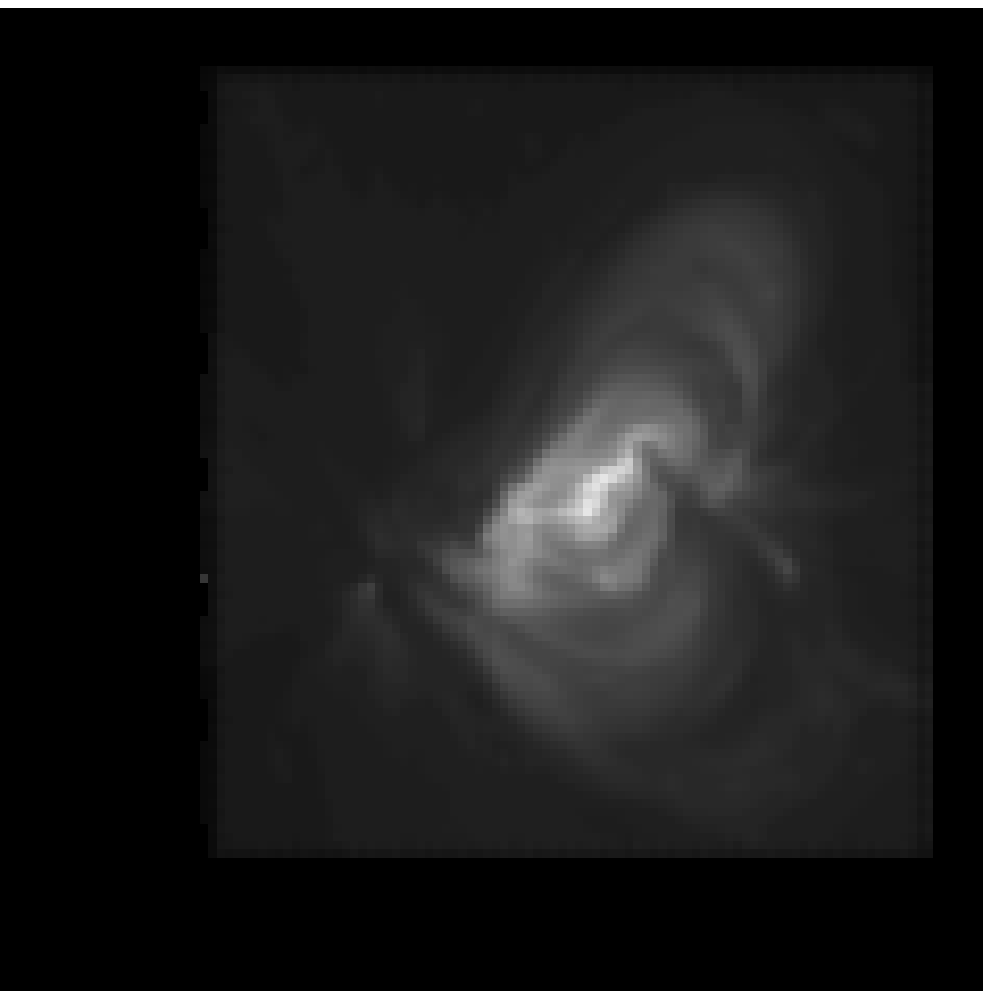}
\includegraphics[width=0.45\textwidth]{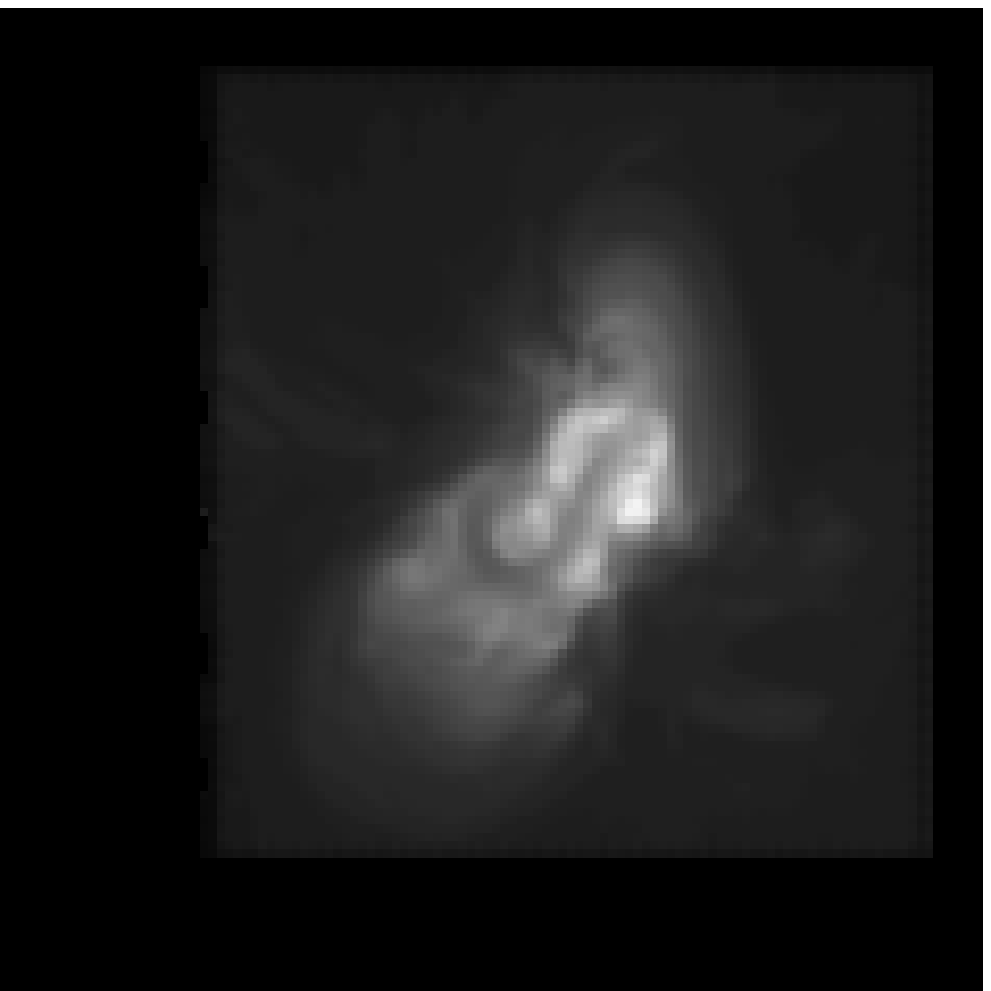}
\includegraphics[width=0.45\textwidth]{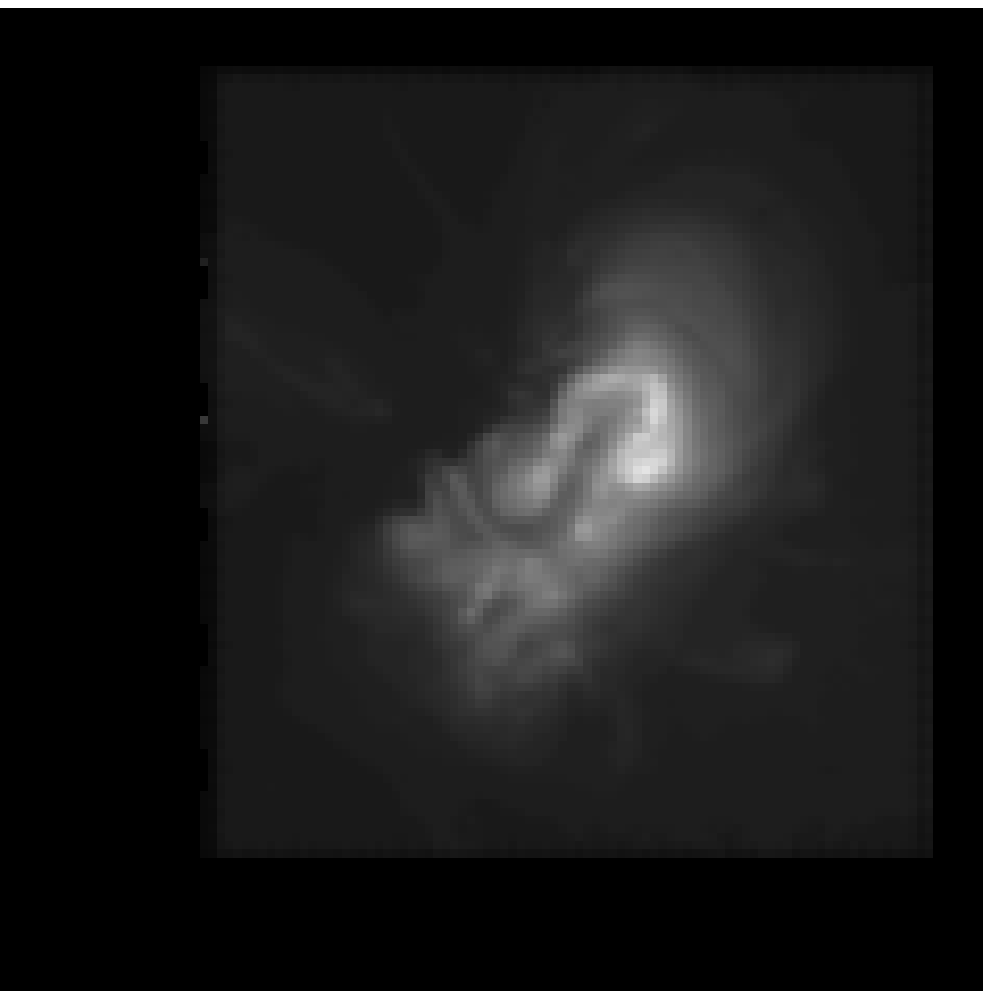}
\end{center}
\caption{STEREO/SECCHI/EUVI 284~\AA\   images for AR~10987 on 2008 March 27 at 1546~UT (top pictures), AR~10988 on 2008 March 29 at 1526~UT (middle pictures) and AR~10989 on 2008 March 31 at 1746~UT (bottom pictures) from the ahead (left pictures) and behind (right pictures) spacecraft.}
\label{fig:stereoar284}
\end{figure}


The Hinode XRT soft X-ray image of AR~10987 on March 27 shown in Figure~\ref{fig:hinodears} features an S-shaped structure commonly referred to as a sigmoid.  The magnetic field model plots in Figure~\ref{fig:flines10987} show evidence of a twisted field joining the two sunspots, with twist of the same handedness as the structure in the Hinode image.  However, the plots of the modeled field also shows much structure that does not appear in the Hinode image.  If one plots only field lines with foot points located within an intense concentration of vertical current, shown in Figure~\ref{fig:sigmoid10987}, an S-shaped structure stands out, bearing a striking resemblance to the S-shaped structure in the Hinode image.  The X-ray image appears brightest where the twist appears greatest and where the overlying field lines meet the twisted structure at a large angle.

The Hinode XRT image of AR~10988 on March 29 in Figure~\ref{fig:hinodears} shows plasma structure fanning out from the leading sunspot to the more diffuse following flux.  In soft X-rays AR~10987 appears more compact than AR~10988 and AR~10989.  In the plots of the modeled field for AR~10987 in Figure~\ref{fig:flines10988}, the asymmetric structure of the short field lines reproduces features of the Hinode image.  For example, the lines ending at the South edge of the leading sunspot originate from locations due South of the spot, corresponding to a distinct lobe at the South-West of the plasma structure seen in the Hinode image.  Meanwhile, structures ending at the North, East and South-East edges of the spot curve in from the East in a manner common to the Hinode image and the model.

Despite the lack of significant current structure in AR~10989, the Hinode XRT image of AR~10989 on March 31 shows contiguous bright structure.  Its brightness distribution in soft X-rays is more compact than that of AR~10988 but less so than that of AR~10987.  The set of short closed field lines connecting intense field concentrations in Figure~\ref{fig:flines10989} together resembles the bright X-ray structure.

The XRT is sensitive to a broad range of temperatures, peaking in sensitivity between about 6 - 8~MK.  For further information we also plot in Figures~\ref{fig:stereoar171}-\ref{fig:stereoar284} images from STEREO/SECCHI/EUVI with three narrow-band filters, 171~\AA\  (corresponding to about 1~MK), 195~\AA\  (about 1.5-1.6~MK) and 284~\AA\  (about 2.5~MK).

\begin{figure} 
\begin{center}
\rotatebox{270}{\includegraphics[width=0.7\textwidth]{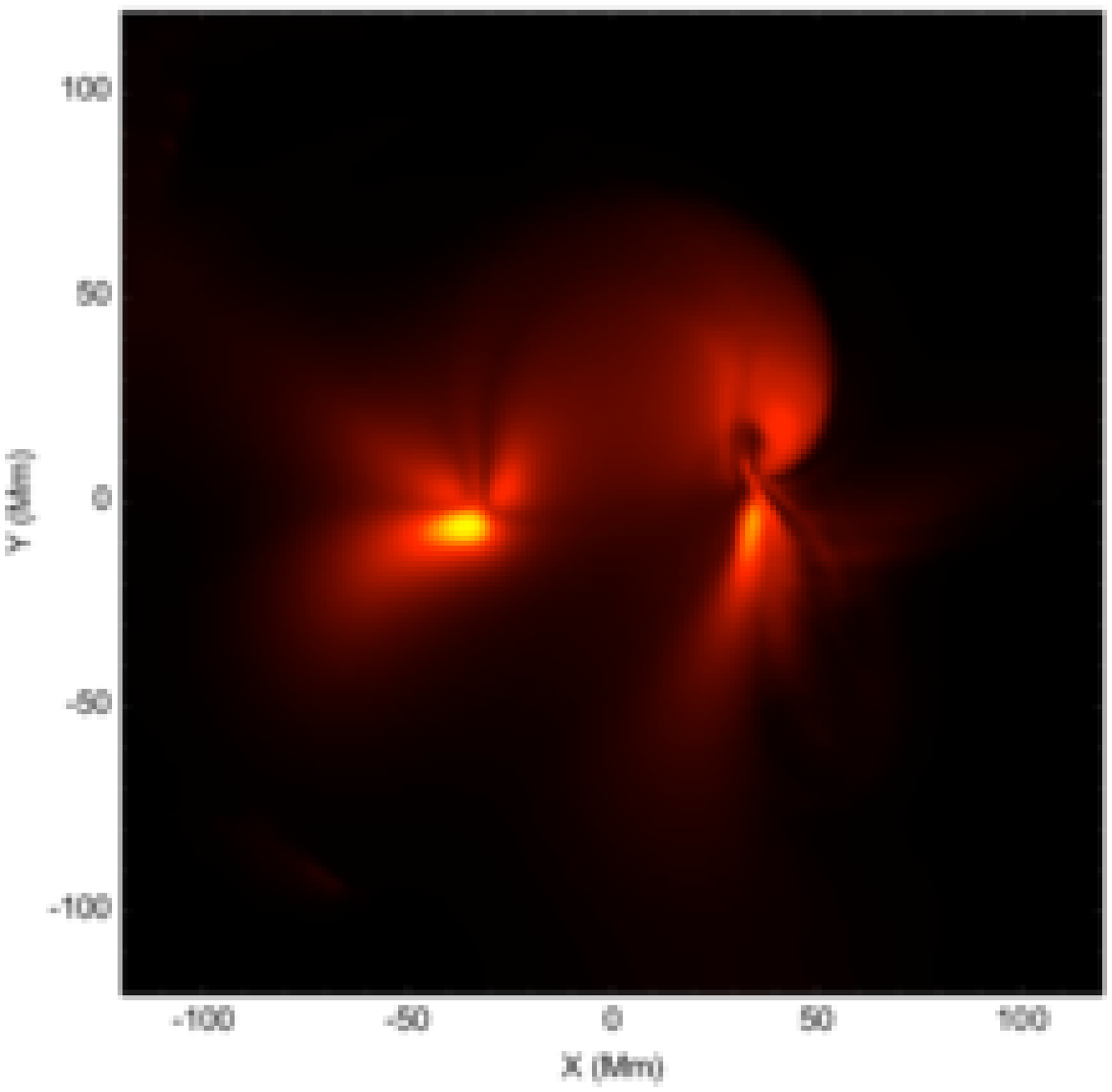}}
\rotatebox{270}{\includegraphics[width=0.7\textwidth]{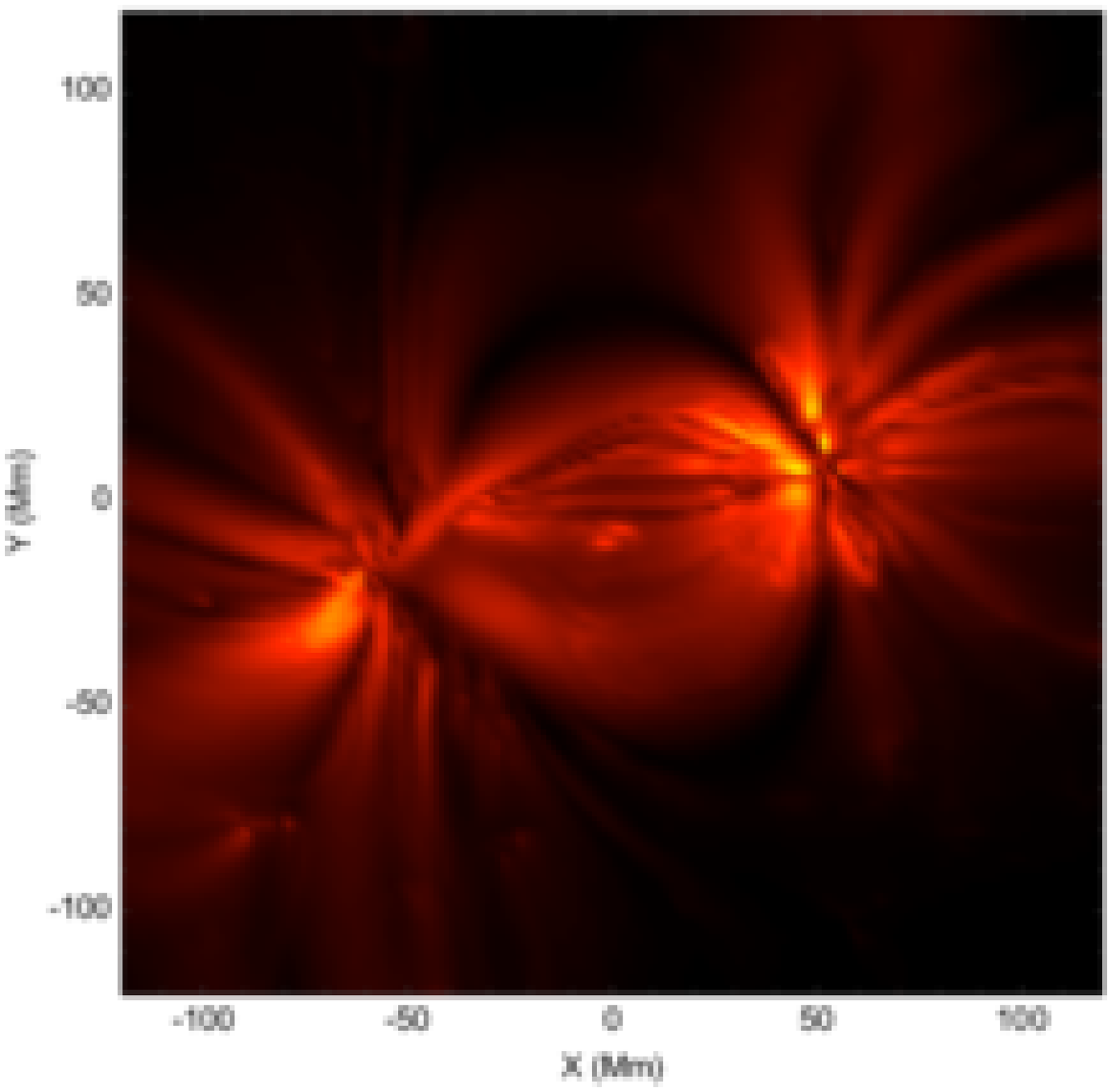}}
\end{center}
\caption{The vertically integrated modeled coronal current densities for ARs~10987 (top) and 10988 (bottom) on March 27.}
\label{fig:coronalcurrents}
\end{figure}

Figure~\ref{fig:stereoar171} shows in the top pictures the STEREO/SECCHI/EUVI 171~\AA\  ahead and behind images of AR~10987 on March 27 and Figure~\ref{fig:coronalcurrents} shows the vertical integral of the electric current squared ${\bf J}\cdot{\bf J}$.  There is no sign of twisted S-shaped structure in any of these plots.  It appears that the sigmoidal structure in the Hinode XRT image is not well populated with plasma emitting at EUV wavelengths.  The middle pictures of Figure~\ref{fig:stereoar171} shows the STEREO behind and ahead images of AR~10988 on March 29.  There is some resemblance between the modeled field plots in Figure~\ref{fig:flines10988} and STEREO images of AR~10988.  While there are clearly significant misalignment angles between the modeled field trajectories and the plasma loops there are striking qualitative similarities between them.   This might suggest that most of the AR~10987 field is populated with X-ray plasma whereas AR~10988 contains plasma at a greater range of EUV and X-ray temperatures.  Finally, the bottom pictures of Figure~\ref{fig:stereoar171}, the STEREO EUV 171~\AA\  images of AR~10989, do not reveal so much organized loop structure in AR~10989 as in AR~10987 and AR~10988.

Figure~\ref{fig:stereoar195} shows the corresponding STEREO/SECCHI/EUVI 195~\AA\  images.  The main difference is that ARs 10987 and 10989 appear brighter in this temperature range.  More of the peripheral structure of these regions is visible at 195~\AA\  than at 171~\AA\ , and both are brighter than AR~10988.  The 195~\AA\  images of AR~10988 appear very similar to the 171~\AA\  images, retaining the resemblance to the field structure shown in Figure~\ref{fig:flines10988}.  There is a hint of sigmoidal structure in the behind image of AR~10987 but it is not a clear sigmoidal signature.

The near-simultaneous STEREO/SECCHI/EUVI 284~\AA\  images are plotted in Figure~\ref{fig:stereoar284}.  These images differ significantly from those in Figures~~\ref{fig:stereoar171} and ~\ref{fig:stereoar195}.  Here the structure is more diffuse and only the structure at the cores of the active regions are generally visible.  In fact these images are more like the soft X-ray images from Hinode shown in Figure~\ref{fig:hinodears}.  At 284~\AA\  AR~10987 has a pronounced sigmoidal structure very similar to appearance in soft X-rays.  The asymmetric loop system of AR~10988 is clearly delineated in the 284~\AA\  behind plot, the middle right plot of Figure~\ref{fig:stereoar284}, and in the Hinode image, the top right plot of \ref{fig:hinodears}.

According to the Hinode XRT and STEREO EUVI images, then, AR~10987 has a sigmoidal structure that is populated with hot plasma and this structure corresponds to a set of `S'-shaped current-carrying field lines in the NLFFF model.  The surrounding fields in the model shows little resemblance to the structure in the cooler EUVI images.  AR~10988 has an asymmetric loop system that is visible at all temperatures and structures leaving the field of view to the East and West visible at the cooler temperatures.  These structures correspond reasonably well to features in the NLFFF model.  AR~10989 is much less coherently structured than the other two regions but is visible at all temperatures, although its peripheral structures are only visible at the cooler temperatures.  Its basic shape resembles the set of the most intense field structures of the model but there is little correspondence between individual plasma loops and modeled field trajectories.  The other two regions had significantly more intense fields than this one.

Figure~\ref{fig:coronalcurrents} shows the vertical integral of the modeled coronal current density squared ${\bf J}\cdot{\bf J}$ for ARs~10987 and 10988 both on March 27.  The vertically integrated
current density squared (or ohmic heating for a uniform resistor) does not forward-model actual soft X-ray or EUV emission
but it is often used as a proxy for a qualitative comparison.  Comparing the AR~10987 plots to Figures~\ref{fig:flines10987}, \ref{fig:sigmoid10987} and \ref{fig:hinodears} the integrated current structure shows no sign of the twisted, sigmoidal structure of the field trajectories and XRT brightness pattern.  Indeed, none of the models for AR~10987 show sigmoidal structure in their integrated vertical current profiles.  The twisted current-carrying fields seen in Figure~\ref{fig:sigmoid10987} perhaps extend to high into the atmosphere and are perhaps too weak to compete with the lower-lying strong field that dominates the integrated current image.  The plot for AR~10988, however, does resemble the structure of the field in Figures~\ref{fig:flines10987} and the EUV emission patterns in the middle pictures of Figures~\ref{fig:stereoar171} and \ref{fig:stereoar195}.  On the other days during its disk passage the current structure of AR~10988 looks similar but some of the details are missing.

\begin{figure}
\begin{center}
\includegraphics[width=\textwidth]{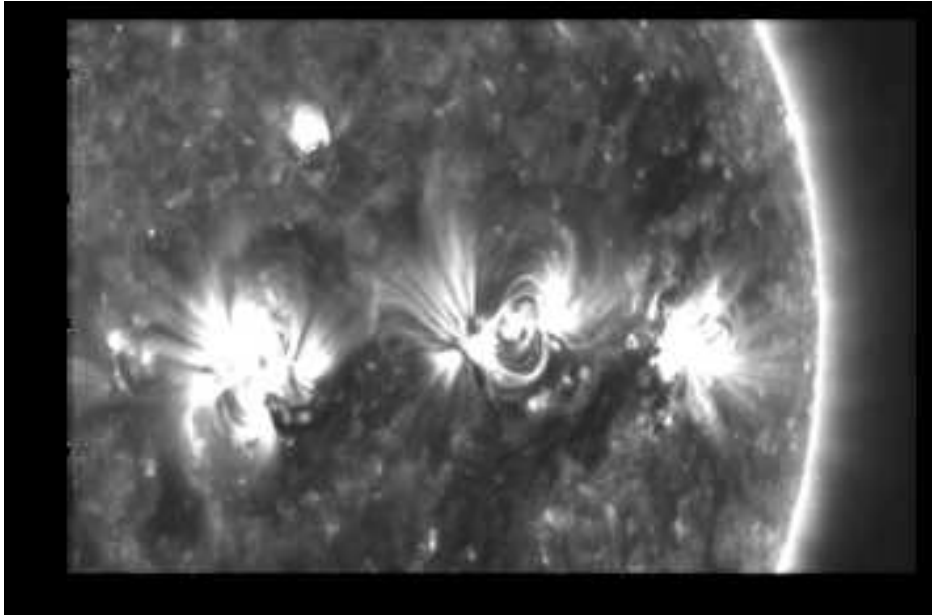}
\end{center}
\caption{STEREO/SECCHI/EUVI 171~\AA\   image of the three active regions 10987-9 on 2008 March 29 at 0409~UT from the behind spacecraft.}
\label{fig:stereoconnected}
\end{figure}

\begin{figure}
\begin{center}
\resizebox{0.75\textwidth}{!}{\includegraphics*[190,80][370,200]{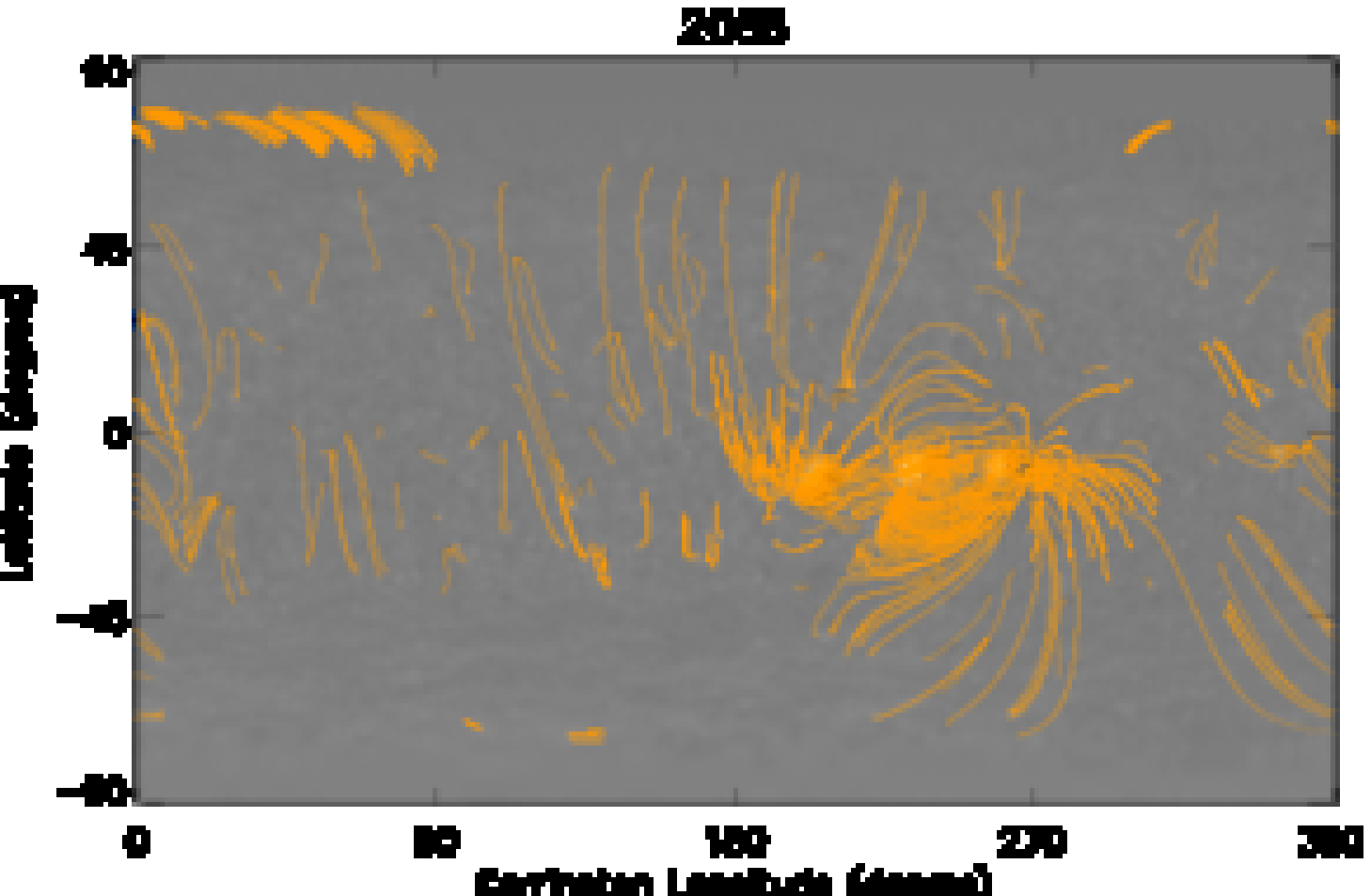}}
\resizebox{0.75\textwidth}{!}{\includegraphics*[190,80][370,200]{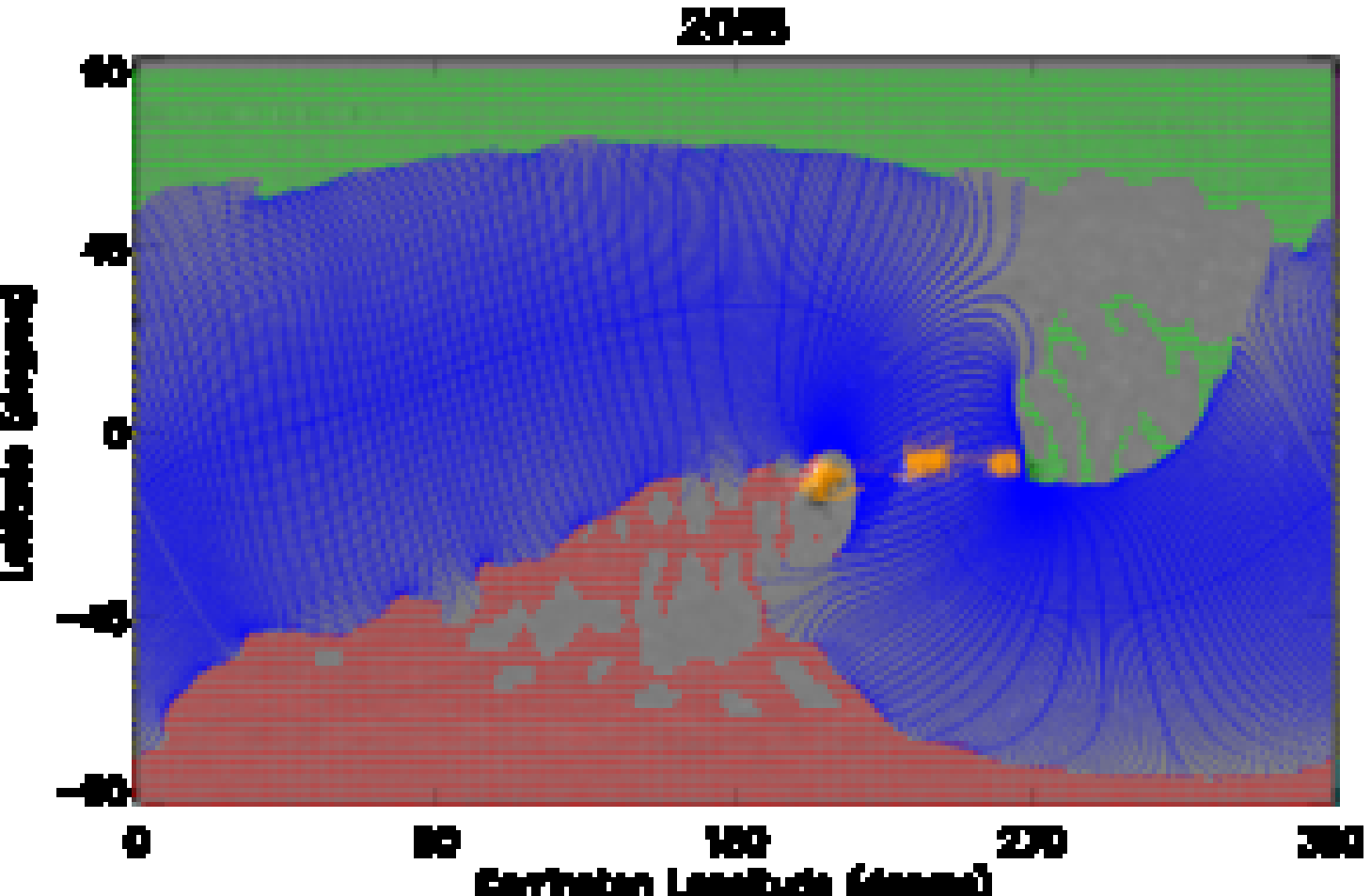}}
\end{center}
\caption{Two pictures of the PFSS model for CR 2068.  The two pictures show the synoptic magnetogram in greyscale.  The top picture shows field lines with foot-point fields of strength $15~G$ or higher in yellow.  The bottom picture shows field lines with foot-point fields of strength $45~G$ or higher in yellow, streamer-belt field lines in blue and positive and negative coronal holes in red and green dots.}
\label{fig:modelconnected}
\end{figure}

The STEREO/EUVI 171~\AA\ image in Figure~\ref{fig:stereoconnected} shows evidence that the three active regions are magnetically connected.  In particular, AR~10989, which is approximately flux-balanced, has structures extending outwards both to the East and to the West.

As the top panel of Figure~\ref{fig:modelconnected} shows, the PFSS model for the global coronal field of rotation 2068 also indicates that all three active regions are magnetically connected.  In the model, magnetic flux directed towards the West side of AR~10989 originates from the East side of AR~10988 and, the flux of AR~10989 being nearly balanced,  positive flux from the East side of AR~10989 connects to quiet-Sun fields to the East.  The connectivities of AR~10987 and AR~10988 are more unusual.  Because these regions have significant positive and negative flux imbalances, respectively, there is less connection between their neighboring positive and negative polarities than between their more separated negative and positive fields.  Flux connecting these fields must loop over both active regions.  The second plot shows the same model but also includes coronal holes and streamer-belt field lines.  Comparing the two plots, the tall, looping field lines connecting AR~10987 and AR~10988 are contained within the streamer belt while a significant quantity of AR~10987's flux, directed towards the Western part of the region, is open field. All but this Western part of AR~10987 is contained within the streamer belt.  Almost all of AR~10988 is also within the streamer belt but the part outside, the Eastern edge, is mostly connected to AR~10989 to the East.  AR~10989 is unique in lying entirely outside albeit close to the streamer belt.  An active region, consisting of closed loop structures, located mostly outside the streamer belt among ambient open field may be expected to have more fields in contact with topologically different domains and therefore be more flare-productive than a region embedded within closed fields under the streamer belt. This may help to explain why, although this region has the least free magnetic energy and helicity of the three regions, it is also the most flare-productive \cite{Webbetal2010}.

\begin{figure}
\begin{center}
\includegraphics[width=\textwidth]{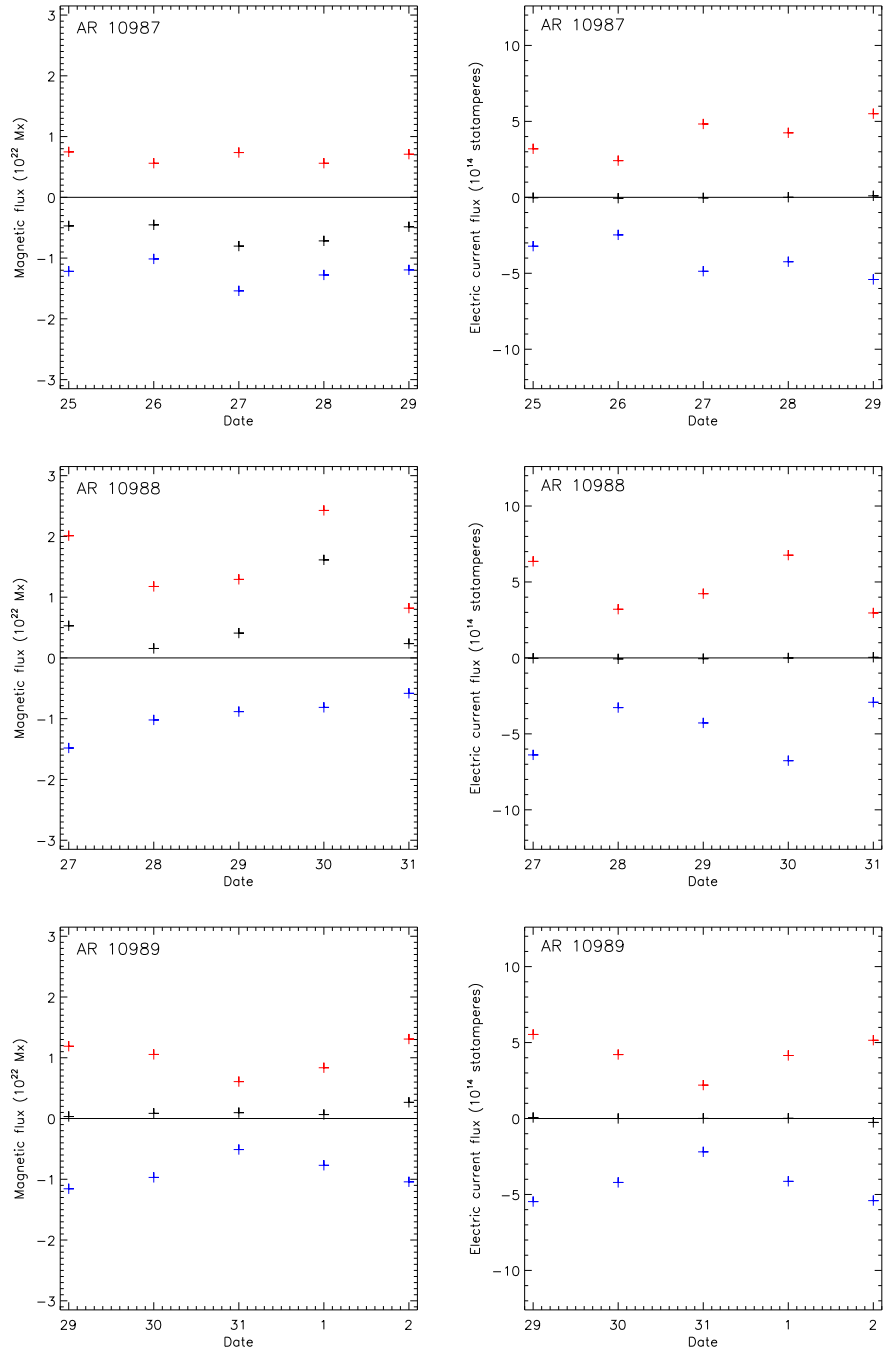}
\end{center}
\caption{Temporal evolution over five-day intervals of the magnetic (left plots) and electric current (right plots) fluxes for AR~10987 (top plots) 10988 (middle plots) and 10989 (bottom plots).  Red/blue/black symbols represent positive/negative/net fluxes.}
\label{fig:obstimeseries}
\end{figure}

\begin{figure}
\begin{center}
\includegraphics[width=\textwidth]{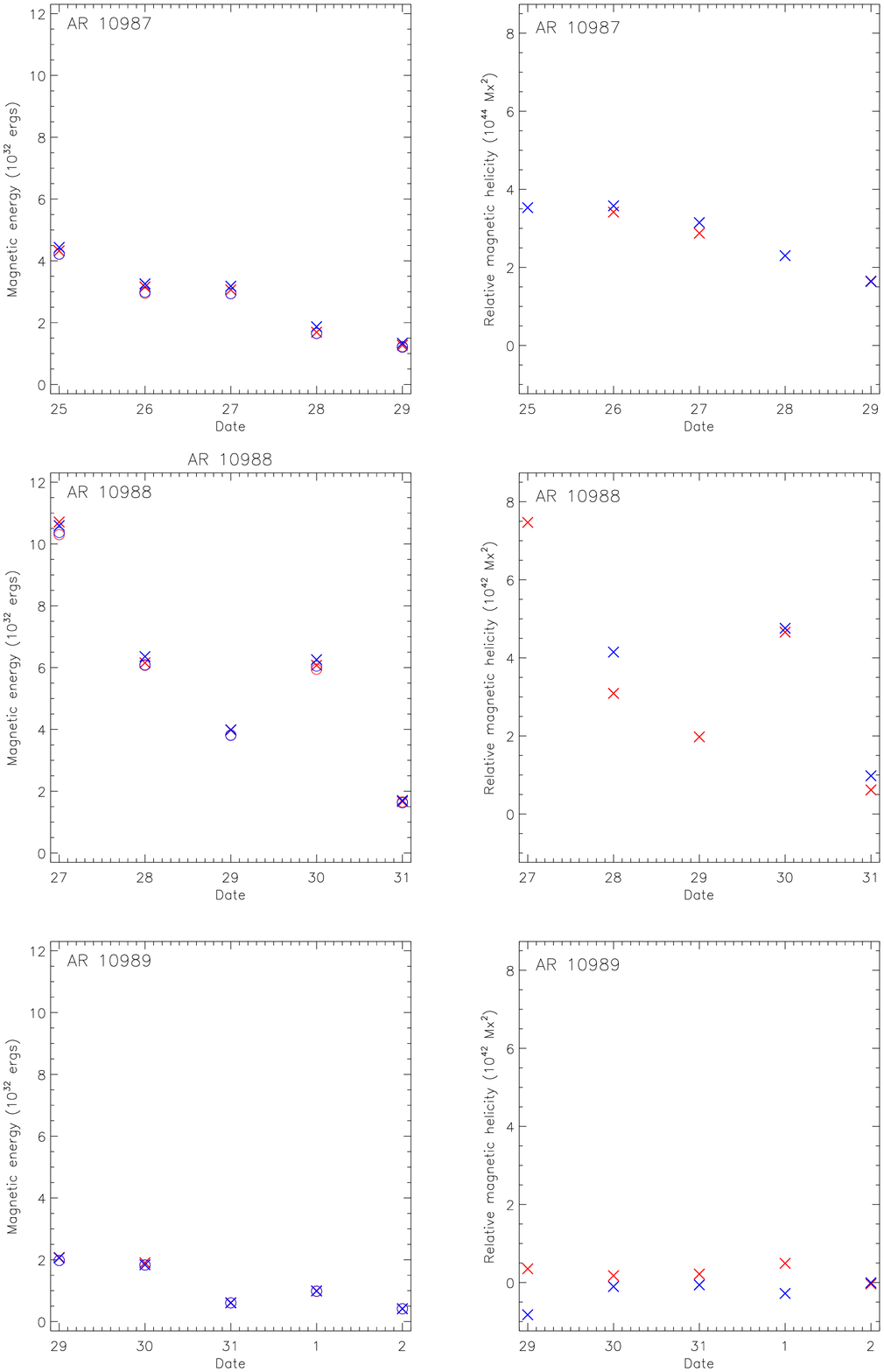}
\end{center}
\caption{Temporal evolution over five-day intervals of the modeled magnetic energies (left plots) and relative magnetic helicities (right plots) for AR~10987 (top plots) 10988 (middle plots) and 10989 (bottom plots).  Crosses represent NLFFF models and circles represent the corresponding potential-field (current-free) models.  The potential-field models have zero relative magnetic helicity.  Red/blue symbols represent models integrated from positive/negative vertical magnetic fields.}
\label{fig:modtimeseries}
\end{figure}

Figure~\ref{fig:obstimeseries} shows the net, positive and negative magnetic fluxes of the three active regions over five daily observations.  Almost all measurements show a significant positive flux imbalance in AR~10987, a significant negative flux imbalance in AR~10988 and approximately balanced flux in AR~10989.  The measurements for AR~10987 seem to be approximately steady in time whereas AR~10988 and AR~10989 show significant variation.  While AR~10988's negative flux declines steadily in the plots, its positive flux has a large peak on March 30th, resulting in a particularly large flux imbalance on that day.  Inspection of the images reveals anomalously strong plage field measurements on that day that seem unlikely to be accurate.  The positive and negative fluxes of AR~10989 increase in size with distance from disk-center.  This apparent dependence on position on the solar disk is most likely an artifact due to noise.  The further the region is displaced from disk-center, the more the signal for $B_z$ derives from transverse as opposed to longitudinal measurements.  Transverse field measurements are intrinsically noisier than longitudinal field measurements.  It seems that the measurements for AR~10989 are more affected by this noise than are the measurements for the other two regions because AR~10989's fields are relatively weak.

Also plotted are the net, positive and negative electric current fluxes of the regions.  The current fluxes are very well balanced in these measurements.  AR~10987 has a gradual overall increase in current.  Moreover, the current is well balanced at each sunspot, suggesting that local twisting motions are adding twist to the field.  AR~10988's graphs are more ambiguous.  The electric current flux of AR~10989 shows the same dependence on position on the solar disk as its magnetic flux.  

Figure~\ref{fig:modtimeseries} shows plots of the free magnetic energies and helicities of the models of the three regions based on the same five-day daily observations.  The plots include models based on both positive- and negative-field boundary conditions.  As we mentioned in Section~\ref{s:pfssnlfff} the reconstruction method can use boundary conditions for $\alpha$ either where the vertical field is positive or where it is negative.  While innovative methods have been developed to take into account both polarities by \inlinecite{WheatlandRegnier2009} and \inlinecite{AmariAly2010}, our active regions are nearly potential so that the positive- and negative-polarity solutions tend to match quite well.  When they do not match, there is generally a simple explanation.  Hence for simplicity we retain the simpler approach of integrating from each polarity separately.  In Figure~\ref{fig:modtimeseries} the left plots of magnetic energy also include potential-field models for comparison.  It is clear that none of these models is far from a potential state.  AR~10988 has the most photospheric flux of the three regions and also has the most magnetic energy.  All three regions show decreasing trends in magnetic energy, with outlying data points for AR~10988 on March 30 and for AR~10989 on March 31 interrupting these trends.  All three regions appear to have decreasing free magnetic energy, perhaps due to cumulative effects of turbulent diffusion.  It is notable that the model parameters for AR~10989 do not show the same dependence on position on the solar disk as the magnetic and electric current fluxes do for that region.  AR~10987 has proportionally significantly more free magnetic energy than the other regions, reflected in its twisted, sigmoidal structure.  However, the free energy contained in even this region is much less than would be needed to power a major flare.

The graphs of magnetic helicity are more complicated because this quantity is very sensitive to problems in the data and model.  In extreme cases the model can even give the wrong sign for the helicity. For example, the positive-polarity model for AR~10988 on March 27 had a problem because the large positive flux balance of the region caused a significant amount of current-carrying flux to leave the numerical domain, instead of closing within the region.  Because magnetic helicity depends on the twist and interlinking of fields, such a problem can significantly affect the relative helicity of the modeled field.  Similar problems occur with the negative-polarity model for AR~10987 on March 25th and 28th and for the positive-polarity model for AR~10988 on March 29th.  In such cases it is clear that the anomalous models should be disregarded.  Judging from those models that are free of such problems, whose helicities are plotted in the right half of Figure~\ref{fig:modtimeseries}, ARs~10987 and 10988 most likely have positive magnetic helicity while AR~10989 does not have significant helicity.  These results are  consistent with the statistical result \cite{pevtsovetal1995,pevtsovetal2001} that active regions in the northern/southern hemisphere tend to have negative/positive helicity.

 \inlinecite{WelschMcTiernan2010} calculated from time series of MDI longitudinal magnetograms that AR~10988 contains the most magnetic flux and AR~10989 the least.  They find that, while the three regions are fed similar quantities of relative helicity flux, the characteristic magnetic helicity flux is greatest for AR~10989 because of its relatively small magnetic flux.  This is the only evidence they derived from the MDI magnetograms that AR~10989 should have been the most active of the regions.  As \inlinecite{WelschMcTiernan2010} have noted, we find little evidence of significant relative magnetic helicity in the NLFFF models for AR~10989.  We already know that AR~10989 was not a very structured region and that it lay mostly in a region of open field outside the streamer belt.  Perhaps the relative magnetic helicity transported into the corona was not effectively trapped in the coronal field of AR~10989.

\section{Conclusion}
\label{s:conclusion}

We have calculated PFSS models for Carrington rotations 2067-9 and NLFFF models for the three active regions of the WHI, NOAA ARs~10987-9.  Together the models reveal the following properties of coronal magnetic field structure around the time of the WHI.

The global coronal field structure was very asymmetric around the time of the WHI with all significant activity crowded between about $180^{\circ}$ and $270^{\circ}$ of Carrington longitude.  This lopsided activity distribution, combined with relatively weak polar fields, resulted in a global coronal field structure that was far from the nearly axisymmetric, dipolar configuration expected during solar minimum.  There was a large warp in the streamer belt centered at around $240^{\circ}$ of Carrington longitude and a displacement of the streamer belt from the equator of 30 to 40$^{\circ}$ during all three rotations CR 2067-9 in both the modeled fields and the STEREO/SECCHI/COR1 streamer observations.  Furthermore, much of the flux connecting to the ecliptic plane originated from low-latitude coronal holes, an unusual state of affairs for a near-minimum corona.

During the WHI itself, CR 2068, the magnetic activity appeared to be particularly well organized in three active regions about 30$^{\circ}$ apart in longitude and approximately equal latitude of about $-10^{\circ}$.  All three were bipolar and had leading negative polarity, consistent with Hale's law, although their tilt angles appeared to be different.  There were further significant differences between the ARs.  The photospheric field of AR~10987 was dominated by two sunspots of approximately equal strength.  AR~10988 had a strong leading sunspot but widely distributed, less intense following flux while AR~10989 had no sunspot and was significantly weaker and more fragmented than the other two.  ARs~10987 and 10988 had comparable magnetic fluxes and AR~10989 had significantly less.  According to NLFFF models AR~10988 had the largest magnetic energy of the three regions although AR~10987 had approximately the same amount of photospheric magnetic flux and had proportionally more free magnetic energy.  AR~10987 also had the most relative magnetic helicity of the three regions and AR~10989 the least.  The models for ARs~10987 and 10988 indicate positive relative magnetic helicity, consistent with a well-known statistical tendency for southern-hemisphere active regions, while the relative magnetic helicity of AR~10989 was not significant and its sign was not determined by the models.  None of the regions had free magnetic energy or helicity normally associated with a flaring active region.   However, AR~10987 had a coherent S-shaped or sigmoidal structure connecting its two sunspots.  This was embedded in a bipolar loop system in the NLFFF model and had foot points in an intense current concentration in the positive-polarity sunspot.  This structure was also evident in the high-temperature STEREO/SECCHI/EUVI and Hinode/XRT images.  ARs~10988 and 10989 had no corresponding structure in the models or the observations.  The PFSS model for CR 2068 showed that the three regions were highly interconnected, and that ARs~10987 and 10988 were located mostly within the helmet streamer belt whereas most of AR~10989 lay outside the streamer belt among open fields.

According to \inlinecite{Webbetal2010} AR~10989 was the most CME- and flare-productive of the three regions.  While \inlinecite{WelschMcTiernan2010} argue that the flare activity may have been more balanced among the three regions, several predictors of flare activity that they calculated based on the photospheric magnetic field indicated that AR~10988 should have been the most active region of the three and AR~10989 the least.  Of the magnetic parameters they studied, the characteristic helicity, the cumulative relative magnetic helicity flux normalized by the square of the mean magnetic flux, was the only one for which AR~10989 had a higher value than the other two regions.  Our NLFFF models tell a similar story.  Of the three regions, AR~10989 had the lowest free magnetic energy and relative magnetic helicity.  In the present paper we have computed the two solutions each one corresponding to injection of electric current from one polarity. It would be interesting to extend this study using the new Optimization Grad Rubin Method (OGRM) described in Amari and Aly (2010) for the NLFF calculation to find a unique solution obtained by taking into account electric current from both polarities.

\inlinecite{Webbetal2010} and \inlinecite{WelschMcTiernan2010} discuss possible reasons why AR~10989 was relatively active.  According to helioseismological calculations, subsurface vortical flows were occurring beneath all three active regions but AR~10989 was distinguished in having strong neighboring vortical flows of opposite sign, a characteristic of flare-productive active regions \cite{Webbetal2010}.  Moreover the field of AR~10989 may have evolved between the time of the magnetic observations near disk-center and the time of the CME activity observed near AR~10989 at the limbs \cite{WelschMcTiernan2010}.    Our PFSS model indicates that AR~10989 was situated among open fields while our NLFFF models show how relatively weak and unstructured the field of AR~10989 was compared to the other two regions.  One tentative explanation is that the equilibrium of AR~10989's field was relatively unstable.  Perhaps the subsurface vortical motions and photospheric helicity flux easily destabilized the coronal field, whereupon magnetic energy and helicity readily escaped from the vicinity of AR~10989 via the ambient open fields.  What is more certain is that we have much to learn about the interacting processes of solar activity.

%

%

%
\begin{acks}
We thank Sarah Gibson, Pat McIntosh, Dave Webb and Brian Welsch for discussions and Janet Luhmann, Yan Li and Xuepu Zhao for contributing software to our PFSS modeling effort.  GP acknowledges funding from NASA Grant NNH08AH251.  AC and TA thank CNES for its support, as well as  NSO for financial support for AC's stay in Tucson.  SOLIS/VSM vector magnetograms are produced cooperatively by NSF/NSO and NASA/LWS.  The GONG program is managed by NSO.  NSO
is operated by AURA, Inc. under a cooperative agreement with the
NSF. The GONG data were acquired by instruments
operated by the Big Bear Solar Observatory, High Altitude Observatory,
Learmonth Solar Observatory, Udaipur Solar Observatory, Instituto de
Astrof\'{\i}sica de Canarias, and Cerro Tololo Interamerican
Observatory.
\end{acks}


\bibliographystyle{spr-mp-sola}

\bibliography{WHI_refs.bib} 

\end{article}

\end{document}